\documentclass[12pt]{article}
\usepackage{graphicx}
\usepackage{amsmath}
\usepackage{amssymb}
\usepackage{ulem}
\usepackage{enumerate}
\usepackage{cite}
\usepackage{latexsym}
\begin{document}
\begin{center}
{\bf {\large{Nilpotency Property, Physicality Criteria, Constraints and Standard BRST Algebra: A  4D Field-Theoretic System}}}

\vskip 3.4cm

{\sf  R. P. Malik$^{(a,b)}$}\\
$^{(a)}$ {\it Physics Department, Institute of Science,}\\
{\it Banaras Hindu University, Varanasi - 221 005, Uttar Pradesh, India}\\

\vskip 0.1cm

$^{(b)}$ {\it DST Center for Interdisciplinary Mathematical Sciences,}\\
{\it Institute of Science, Banaras Hindu University, Varanasi - 221 005, Uttar Pradesh,  India}\\
{\small {\sf {e-mails: rpmalik1995@gmail.com; malik@bhu.ac.in}}}
\end{center}

\vskip 3.0 cm

\noindent
{\bf Abstract:}
Within the framework of Becchi-Rouet-Stora-Tyutin (BRST) formalism, we discuss the off-shell nilpotent (anti-)BRST  and the bosonic ghost-scale
symmetries of a set of coupled (but equivalent) Lagrangian densities for the four (3 + 1)-dimensional (4D) {\it combined} field-theoretic  system
of the free Abelian 3-form and 1-form gauge theories. We demonstrate that the {\it Noether} (anti-)BRST  charges are
non-nilpotent due to the presence of a set of non-trivial Curci-Ferrari (CF) type restrictions on our theory. These CF-type restrictions
are derived and discussed from different theoretical angles in our present endeavor. In addition to it, we obtain the nilpotent
versions of the above (anti-)BRST charges from their  counterparts non-nilpotent versions and discuss the physicality criteria w.r.t. the 
{\it nilpotent} charges to show that the physical states (existing in the total quantum Hilbert space of states) 
are {\it those} that are annihilated by the operator forms of the first-class constraints of our classical 4D combined
field-theoretic system. The standard BRST algebra among the nilpotent (anti-)BRST charges
and the bosonic ghost charge is  derived, too.

\vskip 1.0cm
\noindent
PACS numbers: 11.15.-q, 12.20.-m, 03.70.+k \\

\vskip 0.5cm
\noindent
{\it {Keywords}}: Combined field-theoretic system of the 4D  Abelian 3-form and 1-form gauge theories; first-class constraints;
(anti-)BRST symmetries; non-nilpotent Noether (anti-)BRST charges; nilpotent versions of the  (anti-)BRST charges; physicality criteria

\newpage

\section {Introduction}

It is a well-known fact that, at the initial stages of the development of theoretical physics, the ideas and concepts from pure mathematics were
borrowed for the precise description of the natural phenomena (that were already observed and verified by the appropriate experiments). 
The modern developments in  the realm of theoretical high
energy physics (THEP) have brought together the ideas and concepts of pure mathematics {\it almost} on equal footing to the mathematical sophistication and rigor
required to explain the experimental observations/results in HEP.
In particular, the enormous theoretical progress in the domain of (super)string theories (see, e.g. [1-5] and references therein) has been able to
bring together a set of active practitioners in the realms of THEP and pure mathematics on an intellectual platform where both kinds of researchers have benefited from each-other. Such kinds of confluence and convergence  
between the tricks and techniques of pure mathematics and theoretical physics  
have been observed in the realm of quantum field theories, too. In this context, mention can be made of the topological field theories (see, e.g. [6-9] and references therein),  higher-spin gauge theories (see, e.g. [10,11] and references therein), supersymmetric Yang-Mills theories (see, e.g. [12,13]
and references therein), etc.

Against the backdrop of the above paragraph, it is worthwhile to mention that, during the past few years, we have devoted quite sometime on the study of the
{\it massless} and St{\" u}ckelberg-modified {\it massive} Abelian $p$-form ($ p = 1, 2, 3...$) gauge theories within the framework of Becchi-Rouet-Stora-Tyutin
(BRST) formalism [14-17] and established that {\it these} BRST-quantized theories are the field-theoretic examples for Hodge theory in $D = 2 p$ 
(D = 2, 4, 6) dimensions of the Minkowskian flat spacetime where it has been shown that there is convergence of ideas from the physical aspects
of the BRST formalism and mathematical ingredients 
that are associated with the de Rham cohomological operators\footnote{On a compact spacetime manifold without a boundary, we define a set of three
(i.e. $d, \, \delta, \Delta $) de Rham cohomological operators of differential geometry where the operator $d = \partial_\mu \, dx^\mu$ [with $d^2 = \frac{1}{2!}\,
(\partial_\mu \partial_\nu - \partial_\nu \partial_\mu)\, (dx^\mu \wedge dx^\nu) = 0$] is the exterior derivative, the operator 
$\delta = \pm\, *\, d\, *$ (with $\delta^2 = 0 $) is the co-exterior derivative and  $\Delta = (d + \delta)^2 = \{d, \,\delta \}$
corresponds to the Laplacian operator (which is always a positive definite quantity). In the mathematical relationship: 
$\delta = \pm\, *\, d\, *$, the symbol $*$ stands for the Hodge duality operator on the given spacetime manifold. These three operators obey the algebra: $d^2 =
\delta^2 = 0, \; \Delta = \{d, \; \delta\}, \; [\Delta, \; d] = 0, \;  [\Delta, \; \delta] = 0$ where (i) the symbols $\{\; ,\;\} $ 
and $[\; , \; ] $ correspond to the
definitions of the well-known (anti)commutators, and (ii) the Laplacian operator appears to be like the 
Casimir operator for this whole algebra. However, the commuting nature of 
the Laplacian operator, for {\it this} algebra, is {\it not}  in the sense of the more familiar Lie algebra(s). The algebra, obeyed by 
the above set of {\it three} de Rham  cohomological operators, is popularly known as 
the Hodge algebra.} of differential geometry (see, e.g. [18-23]) at the {\it algebraic} level.
To be precise, we have been able to provide the physical realizations of the cohomological operators of differential geometry in the terminology of the
discrete and continuous symmetries (and corresponding conserved charges) of the above 
{\it even} dimensional (i.e. D = 2, 4, 6) BRST-quantized theories at the {\it algebraic} level. In a  very recent set of works [24,25], we have shown the
odd-dimensional (i.e. D = 3) field-theoretic system of the Abelian 2-form and 1-form gauge  theories to 
provide a tractable field-theoretic example for Hodge theory. 
In addition, we have {\it also} proven some of the physically interesting set of (i) the 
BRST-quantized quantum mechanical models [26,27], and (ii) the  $\mathcal {N} = 2$ SUSY quantum mechanical 
systems (see, e.g. [28-30] and references therein)  to be the 1D tractable toy models of Hodge theory.

In the physical four (3 + 1)-dimensions of spacetime, we have considered the {\it combined} field-theoretic system of the free Abelian 3-form and 1-form gauge theories within the framework of BRST formalism [31] and shown the existence of (i) the nilpotent BRST and co-BRST 
{\it continuous} symmetry transformation operators, (ii)
a bosonic symmetry transformation operator that is derived from the anticommutator of the BRST and co-BRST symmetry transformation operators, and (iii) a
set of discrete symmetry transformation operators.  It has been explicitly shown that the bosonic symmetry transformation operator commutes with the nilpotent BRST and co-BRST symmetry transformation operators. We have also
established that the interplay between the discrete and nilpotent  (co-)BRST symmetry transformation operators provide the physical
realization(s) of the mathematical relationship that exists between the  (co-)exterior derivative operators of differential geometry. To
be precise, the discrete symmetry transformation operator of our BRST-quantized 4D field-theoretic system provides the analogue of the
Hodge duality $*$ operator of differential geometry and nilpotent (co-)BRST transformation operators provide the physical realization(s) of the
(co-)exterior derivatives $(\delta)d$ (with $d^2 = 0, \,\delta^2 = 0 $) of differential geometry in the mathematical relationship 
(i.e. $\delta = \pm\, *\, d\, * $) that exists between them.

In our earlier work [31], we have considered a 4D (co-)BRST invariant
Lagrangian density. However, we have {\it not} devoted time on the discussion of the anti-BRST and anti-co-BRST invariant Lagrangian density which happens to be 
a {\it coupled} (but equivalent) version of the (co-)BRST invariant Lagrangian density (of our earlier work [31]) due to the presence of a set
of {\it non-trivial} CF-type restrictions on our theory. In our present endeavor, we discuss the key features associated with the off-shell nilpotent
(anti-)BRST symmetries {\it only}\footnote{For our BRST-quantized 4D theory to be an example for Hodge theory, we expect that it should respect
the (anti-)BRST, (anti-)co-BRST, a unique bosonic and the ghost-scale symmetry transformations. However, for the discussion on the first-class
constraints at the {\it classical} as well as at the {\it quantum} level, we focus {\it only} on the nilpotent and absolutely anticommuting
(anti-)BRST symmetry transformations in our present investigation where the emphasis is laid on the derivations of (i) the Noether conserved
(but non-nilpotent) (anti-)BRST charges $Q_{(a)b}$,  (ii) the {\it modified} conserved and nilpotent ($Q_{(A)B}^2 = 0 $)
(anti-)BRST charges $Q_{(A)B}$, and (iii) the physicality criteria (i.e. $Q_{(A)B} \, |phys>\, = 0$) w.r.t. the nilpotent versions of the
conserved (anti-)BRST charges $Q_{(A)B}$ where the symbol $|phys>$ stands for the physical states.}. 
We derive (i) the non-nilpotent {\it Noether} conserved (anti-)BRST charges $Q_{(a)b}$, and (ii) the
off-shell nilpotent versions  of the (anti-)BRST charges  $Q_{(A)B}$ which are useful in the discussion
on the physicality criteria (cf. Secs. 5,4). In fact, we discuss the {\it latter}  in an elaborate fashion and 
demonstrate that the operator forms of the first-class constraints (cf. Appendix A) annihilate the {\it true} physical states (i.e. $|phys>$) 
by the requirements: $Q_{(A)B} \, |phys>\, = 0$ of the physicality criteria (within the framework of the BRST formalism). This observation is 
consistent with the Dirac quantization conditions for the physical systems that are endowed with constraints of any variety.

Our present investigation is essential and important on the following counts. First of all, our 4D BRST-quantized field-theoretic system of 
the {\it free} Abelian 3-form and 1-form gauge theories is characterized by the existence of the first-class constraints in the terminology
of Dirac's prescription for the classification scheme of constraints [32-36]. We have been able to show the existence of the {\it first-class} constraints
in an explicit fashion at the {\it classical} as well as at the {\it quantum } level within the framework of BRST formalism
(see, Appendix A and Secs. 4,5). It is interesting here to add that {\it this} BRST-quantized system has been proven to be a
field-theoretic example for Hodge theory  (see, e.g. [31]).
Second, we have demonstrated explicitly that the Noether conserved 
(anti-)BRST charges are non-nilpotent (cf. Secs. 5,4) due to the presence of 
the non-trivial CF-type restrictions (cf. Appendix B). Third, we have 
systematically (see, e.g. [37]) derived the off-shell nilpotent versions of the (anti-)BRST charges from
their non-nilpotent counterparts which have been utilized in (i) the physicality criteria (cf. Subsecs. 4.3 and 5.3) within the framework of
BRST formalism, and (ii) the derivation of the standard BRST algebra (cf. Appendix C). Finally, the existence of the 
(non-)trivial (anti-)BRST invariant
CF-type restrictions is the hallmark of a {\it properly} BRST-quantized theory. We have derived these {\it non-trivial} restrictions
from different theoretical angles in our present endeavor (cf. Appendix B).

The theoretical materials of our present endeavor are organized as follows. In Sec. 2, we recapitulate the bare
essentials of our earlier work [31] on the nilpotent BRST symmetry transformations  for an appropriate Lagrangian density that incorporates the
proper gauge-fixing and Faddeev-Popov (FP) ghost terms. Our Sec 3 is devoted to the 
discussion on the anti-BRST symmetry  transformations  for the coupled (but equivalent) Lagrangian density (corresponding the 
BRST invariant Lagrangian density of our Sec. 2). The theoretical content of our Sec. 4 concerns itself with the 
derivation of the Noether BRST conserved current and corresponding conserved (but non-nilpotent) BRST charge $Q_b$. We derive the 
nilpotent version of the BRST charge $Q_B$, too, from the non-nilpotent Noether BRST charge $Q_b$ and discuss the physicality criterion w.r.t. the nilpotent
version of the BRST charge  $Q_B$.
Our Sec. 5 deals with the derivation of the conserved (but non-nilpotent) Noether anti-BRST charge from the Noether anti-BRST conserved current.
We derive the nilpotent version of the anti-BRST charge $Q_{AB}$ from the Noether conserved charge $Q_{ab}$ and discuss the physicality criterion w.r.t. the 
{\it nilpotent} version of the anti-BRST charge $Q_{AB}$. Finally, in Sec. 6, we concisely summarize our key results, make a few crucial 
remarks and point out the future perspective and scope of our present investigation.

In our Appendix A, we demonstrate the existence of the first-class constraints on our {\it combined} 4D field-theoretic system at the {\it classical} level.
The subject matter of our Appendix B is to derive and discuss the (anti-)BRST invariant CF-type restrictions 
(on our BRST-quantized 4D theory) from different theoretical angles at the {\it quantum} level. Our Appendix C is devoted to the discussion on 
the {\it bosonic} ghost-scale symmetry transformations and the derivation of the conserved ghost charge which participates in the standard 
BRST algebra {\it together} with the off-shell nilpotent versions of the (anti-)BRST charges.\\

{\it Notations and Conventions:} We denote the off-shell nilpotent (anti-)BRST transformations 
by $s_{(a)b}$ and the corresponding non-nilpotent (i.e. $Q_{(a)b}^2 \neq 0 $) {\it Noether} conserved charges carry the symbols
$Q_{(a)b}$. The off-shell nilpotent (i.e. $Q_{(A)B}^2 = 0 $) versions of the (anti-)BRST charges are denoted by $Q_{(A)B}$.
We take the 4D flat Minkowskian metric tensor $\eta_{\mu\nu}$ as: $\eta_{\mu\nu}$ = 
diag $(+ 1, -1, -1, -1)$ so that the dot product between two {\it non-null} 4D vectors $S_\mu$ and $T_\mu$ is defined as: 
$S\cdot T = \eta_{\mu\nu} S^\mu\,T^\nu \equiv S_0\, T_0 - S_i\,T_i$ where the Greek indices 
$ \mu, \nu, \sigma ... = 0, 1, 2,  3$ denote the time and space directions and Latin indices $i, j, k... = 1, 2, 3$ stand for
the 3D space directions {\it only}. 
We adopt the convention of the left derivative w.r.t. all the {\it fermionic} fields of our theory. 
The 4D Levi-Civita tensor $\varepsilon_{\mu\nu\sigma\rho}$ is chosen such that:  $\varepsilon_{0123} 
= +1 = -\, \varepsilon^{0123}$ and, when two of them are contracted,
 they satisfy the standard relationships: $\varepsilon_{\mu\nu\eta\kappa} \varepsilon^{\mu\nu\eta\kappa}= - \,4!$,
$\varepsilon_{\mu\nu\eta\kappa} \varepsilon^{\mu\nu\eta\rho}= - \,3! \,\delta^\rho_\kappa, \; \varepsilon_{\mu\nu\eta\kappa} \varepsilon^{\mu\nu\sigma\rho}= - \,2!\,
\big(\delta^\sigma_\eta \delta^\rho_\kappa - \delta^\sigma_\kappa \delta^\rho_\eta\big)$, etc. 
We also adopt the convention: $(\delta A_{\mu\nu\sigma}/\delta A_{\alpha\beta\gamma}) = \frac{1}{3!}\,\big[ \delta^\alpha_\mu (\delta^\beta_\nu \,
\delta^\gamma_\sigma - \delta^\beta_\sigma \,\delta^\gamma_\nu) +  \delta^\alpha_\nu (\delta^\beta_\sigma \,
\delta^\gamma_\mu - \delta^\beta_\mu \,\delta^\gamma_\sigma) + \delta^\alpha_\sigma (\delta^\beta_\mu \,
\delta^\gamma_\nu - \delta^\beta_\nu \,\delta^\gamma_\mu)\big]$, etc., for the tensorial variation/diffferentiation
for various computational purposes.\\



\section{Preliminary: Nilpotent BRST Transformations}

We begin with the BRST invariant Lagrangian density ${\cal L}_{(B)} = {\cal L}_{(NG)} + {\cal L}_{(FP)}$ which incorporates
into it  (i) the non-ghost sector
(i.e. ${\cal L}_{(NG)} $), and (ii) the Faddeev-Popov (FP) ghost-sector (i.e. ${\cal L}_{(FP)} $). The properly gauge-fixed non-ghost sector of the 
BRST invariant Lagrangian density ${\cal L}_{(B)} $ is as follows (see, e.g. [31] for details) 
\begin{eqnarray}\label{1}
&&{\cal L}_{(NG)}  = \dfrac{1}{2}\, B_2 \big (\partial \cdot \phi \big ) - \dfrac{1}{4}\, B_2^2  + 
\dfrac{1}{2}\, B_3 \big (\partial \cdot \widetilde \phi \big ) - \dfrac{1}{4}\, B_3^2 
+ \dfrac{1}{2}\, B^2 - B \, (\partial \cdot A) \nonumber\\
&+&  \dfrac{1}{2} \, B_1^2 
+ \, B_1\, \Big(\dfrac{1}{3!}\,\varepsilon^{\mu\nu\sigma\rho} \,\partial_\mu A_{\nu\sigma\rho} \Big) 
 - \,\dfrac{1}{4} \, \big( B_{\mu\nu} \big)^2 + \dfrac{1}{2}\, B_{\mu\nu} \, \Big[\partial_\sigma A^{\sigma\mu\nu}
+ \dfrac{1}{2}\, \big (\partial^\mu \phi^\nu - \partial^\nu \phi^\mu \big) \Big ]  
   \nonumber\\
&-& \dfrac{1}{4}\, \big( {\cal B_{\mu\nu}} \big)^2 + \dfrac{1}{2} \, {\cal B_{\mu\nu}}\, \Big[ \varepsilon^{\mu\nu\sigma\rho} \,\partial_\sigma A_\rho 
+ \dfrac{1}{2}\, \big (\partial^\mu \widetilde \phi^\nu - \partial^\nu \widetilde \phi^\mu \big) \Big], 
 \end{eqnarray}
where $B,\, B_1,\, B_2,\, B_3,\, B_{\mu\nu},\, {\cal B_{\mu\nu}} $ are the {\it bosonic} Nakanishi-Lautrup type auxiliary fields that have been 
invoked for the purpose of the linearization of the quadratic terms that appear in the 
following {\it original} gauge-fixed Lagrangian density ${\cal L}_{(0)}$, namely; 
\begin{eqnarray}\label{2}
{\cal L}_{(0)}  &=& \dfrac{1}{4}\,  \big (\partial \cdot \phi \big )^2 + \dfrac{1}{4}\, \big (\partial \cdot \widetilde \phi \big )^2
- \dfrac{1}{2}\,(\partial \cdot A)^2  -  \dfrac{1}{2} \, \Big(- \dfrac{1}{3!}\,\varepsilon^{\mu\nu\sigma\rho} \,\partial_\mu A_{\nu\sigma\rho} \Big)^2  \nonumber\\
 &+& \,\dfrac{1}{4} \, \Big[\partial_\sigma A^{\sigma\mu\nu}
+ \dfrac{1}{2}\, \big (\partial^\mu \phi^\nu - \partial^\nu \phi^\mu \big) \Big ]^2  +    
 \dfrac{1}{4}\, \Big[ \varepsilon^{\mu\nu\sigma\rho} \,\partial_\sigma A_\rho 
+ \dfrac{1}{2}\, \big (\partial^\mu \widetilde \phi^\nu - \partial^\nu \widetilde \phi^\mu \big) \Big]^2.
\end{eqnarray}
In the above equations (1) and (2), the totally antisymmetric tensor Abelian gauge field $A_{\mu\nu\sigma}$ is derived from the Abelian 3-form:
$A^{(3)} = \frac{1}{3!}\,A_{\mu\nu\sigma}\,(d\,x^\mu \wedge d\,x^\nu \wedge d\, x^\sigma)$ and the Abelian vector field $A_\mu$ is obtained
from the Abelian 1-form:  $A^{(1)} = A_\mu\, dx^\mu $. The field-strength tensors: $H_{\mu \nu \sigma \rho} = \partial_\mu\, A_{\nu\sigma\rho} - \partial_\nu\,
 A_{\sigma\rho \mu } + \partial_\sigma\, A_{\rho \mu \nu }  - \partial_\rho \,A_{\mu \nu \sigma }$ and $F_{\mu\nu}  = \partial_\mu A_\nu - \partial_\nu A_\mu $
 are derived from the Abelian 4-form: $H^{(4)} = d\, A^{(3)} 
= \frac{1}{4!}\,\, H_{\mu \nu \sigma\rho}\, ( d\,x^\mu \wedge d\,x^\nu \wedge d\, x^\sigma \wedge d\, x^\rho)$ and the 
Abelian 2-form: $F^{(2)} = d\, A^{(1)} \equiv \frac{1}{2!}\, F_{\mu\nu} \big(dx^\mu \wedge dx^\nu \big)$, respectively. Here the symbol $d = \partial_\mu \, dx^\mu$ 
[with $d^2 = \frac{1}{2!}\, (\partial_\mu \partial_\nu - \partial_\nu \partial_\mu)\, (dx^\mu \wedge dx^\nu) = 0$]
 stands for the exterior derivative of differential geometry (see, e.g. [18-23] for details). It is the special feature of the 4D theory
 that the kinetic term [i.e. $\frac{1}{48}\, H^{\mu\nu\sigma\rho} H_{\mu\nu\sigma\rho}  = -\, \frac{1}{2} (H_{0123})^2$] for the 
Abelian 3-form gauge theory can be expressed in terms of the 4D Levi-Civita tensor as:  $-\, \frac{1}{2} (H_{0123})^2 = -\, \frac{1}{2}\,
(-\, \frac{1}{3!} \, \varepsilon^{\mu\nu\sigma\rho}\, \partial_\mu A_{\nu\sigma\rho} )^2$ because we note that the {\it single} existing
component of the field-strength tensor $H_{\mu \nu \sigma\rho}$ in 4D spacetime  can be expressed as: $H_{0123} 
= -\,\frac{1}{3!}\; \varepsilon^{\mu\nu\sigma\rho}\, \partial_\mu A_{\nu\sigma\rho}$. On the other hand, the standard kinetic term 
(i.e. $-\, \frac{1}{4}\, F^{\mu\nu} F_{\mu\nu} $) for the
Abelian 1-form gauge theory can be written in terms of the 4D Levi-Civita tensor as: $ -\, \frac{1}{4}\, F^{\mu\nu} F_{\mu\nu} = \frac{1}{4}\,
(\varepsilon^{\mu\nu\sigma\rho} \,\partial_\sigma A_\rho )^2$. The kinetic term for the Abelian 1-form gauge field and the 
gauge-fixing term for the Abelian 3-form gauge field can be generalized (due to the reducibility properties) 
by incorporating into them the (axial-) vector fields
$(\tilde \phi_\mu) \phi_\mu$ as very clearly expressed in equations (2) and (1). It is interesting to point out that a close look at the
equations (1) and (2) reveals that we have also taken into account the gauge-fixing terms [i.e. $\frac{1}{4}   (\partial \cdot \tilde \phi )^2, 
\; \frac{1}{4}  (\partial \cdot  \phi)^2 $] for the (axial-)vector fields $(\tilde \phi_\mu)\phi_\mu $ in our equations (2) and (1)
in their quadratic and linearized forms, respectively.

The FP-ghost part of our BRST invariant Lagrangian density ${\cal L}_{(B)}$ contains various kinds of (anti-)ghost fields 
which are needed for the validity of unitarity in our BRST-quantized 4D theory. This Lagrangian density ${\cal L}_{(FP)} $,  for our
combined 4D field-theoretic system of the free Abelian 3-form and 1-form gauge theories, is as follows [31]
\begin{eqnarray}\label{3}
{\cal L}_{(FP)}  &=&  \dfrac{1}{2}\, \Big [\big (\partial_\mu \bar C_{\nu\sigma} +  \partial_\nu \bar C_{\sigma\mu} 
+ \partial_\sigma \bar C_{\mu\nu} \big ) \big (\partial^\mu C^{\nu\sigma} \big ) +
\big (\partial_\mu \bar C^{\mu\nu}  + \partial^\nu \bar C_1 \big ) f_\nu \nonumber\\ 
&-& \big (\partial_\mu  C^{\mu\nu}  + \partial^\nu  C_1 \big ) \bar F_\nu 
 +  \big( \partial \cdot \bar \beta \big) \, B_4
 - \big( \partial \cdot  \beta \big) \, B_5 - B_4 \, B_5 - 2\, \bar F^\mu\, f_\mu \nonumber\\
&-& \big (\partial_\mu \bar \beta_\nu - \partial_\mu \bar \beta_\nu \big ) 
\big (\partial^\mu \beta^\nu \big )  
-\, \partial_\mu \bar C_2\, \partial^\mu C_2  \Big ] - \partial_\mu \bar C\, \partial^\mu C, 
\end{eqnarray}
where the Lorentz-scalar fermionic ($C^2 = 0, \, \bar C^2 = 0, C \, \bar C + \bar C \, C = 0$)
(anti-)ghost fields $(\bar C)C$, present in the last term, 
are associated with the Abelian 1-form gauge field $A_\mu$ and they carry the ghost numbers
(-1)+1, respectively. On the other hand, corresponding to our totally antisymmetric (i.e. $A_{\mu\nu\sigma} = - \,A_{\nu\mu\sigma} = -\,A_{\sigma\nu\mu}  $, etc.,)
Abelian 3-form gauge field $A_{\mu\nu\sigma} $, we have the antisymmetric
($\bar C_{\mu\nu} = -\, \bar C_{\nu\mu}, C_{\mu\nu} = -\,  C_{\nu\mu}$) tensor (anti-)ghost fields $(\bar C_{\mu\nu}) C_{\mu\nu}$ which are 
{\it also} fermionic ($C_{\mu\nu}^2 = 0, \, \bar C_{\mu\nu}^2 = 0,\, C_{\mu\nu}\, \bar C_{\sigma\rho} 
+ \bar C_{\sigma\rho}\,  C_{\mu\nu} = 0$, etc.) in nature and they are endowed 
with the ghost numbers (-1)+1, respectively. In our theory, we have 
the ghost-for-ghost {\it bosonic} (i.e. $\beta_\mu^2 = \neq 0,\, \bar \beta_\mu^2 = \neq 0, \, 
\beta_\mu \, \bar \beta_\nu - \bar \beta_\nu\, \beta_\mu = 0 $) vector (anti-)ghost fields $(\bar \beta_\mu)\beta_\mu$
and the ghost-for-ghost-for-ghost {\it fermionic} 
(i.e. $C_2^2 = 0, \, \bar C_2^2 = 0, C_2 \, \bar C_2 + \bar C_2 \, C_2 = 0 $) (anti-)ghost fields $(\bar C_2)C_2$ that carry the ghost numbers (-2)+2 and (-3)+3., respectively. The fermionic (i.e. $f_\mu^2 = 0, \, \bar F_\mu^2 = 0, f_\mu \, \bar F_\nu + \bar F_\nu \, f_\mu = 0 $, etc.)
auxiliary fields $(\bar F_\mu)f_\mu$ and bosonic (i.e. $B_5^2 \neq 0, \, B_4^2 \neq 0,\, B_5 \, B_4 = B_4 \, B_5 $)
auxiliary fields $(B_5)B_4$ of our theory carry the ghost numbers
(-1)+1 and (-2)+2, respectively, because we observe that: $B_5 = (\partial \cdot \bar \beta),\, B_4 = -\, (\partial \cdot \beta) $. 
The additional fermionic (i.e. $C_1^2 = 0, \, \bar C_1^2 = 0, C_1 \, \bar C_1 + \bar C_1 \, C_1 = 0 $)
(anti-)ghost fields $(\bar C_1)C_1$ are also endowed with the ghost numbers (-1)+1, respectively.

We concentrate now on the BRST symmetry transformations of the total Lagrangian density ${\cal L}_{(B)} = {\cal L}_{(NG)} + {\cal L}_{(FP)}$. 
In this connection, it is interesting
to point out that the following infinitesimal and off-shell nilpotent (i.e. $s_{b}^2 = 0 $) BRST transformations ($s_{b}$)
\begin{eqnarray}\label{4}
&&s_b A_{\mu\nu\sigma} = \partial_\mu C_{\nu\sigma} + \partial_\nu C_{\sigma\mu}
+ \partial_\sigma C_{\mu\nu}, \qquad  s_b C_{\mu\nu} = \partial_\mu \beta_\nu
- \partial_\nu \beta_\mu, \qquad s_b \bar C_{\mu\nu} = B_{\mu\nu}, \nonumber\\
&&s_b A_{\mu} = \partial_\mu C, \quad s_b \bar C = B, \quad  
s_b \bar \beta_\mu = \bar F_\mu, \quad
s_b \beta_\mu = \partial_\mu C_2,  \quad s_b \bar B_{\mu\nu} = \partial_\mu f_\nu - \partial_\nu f_\mu, \nonumber\\
&&s_b {\bar C}_2 = B_5, \qquad s_b C_1 = - B_4, \qquad  s_b \bar C_1 = B_2, \qquad s_b \phi_\mu = f_\mu, \qquad s_b F_\mu = -\, \partial_\mu B_4,
\nonumber\\
&& s_b \bar f_\mu = \partial_\mu B_2, \; s_b  \Bigl [ C_2,\, C,\, f_\mu,\, {\bar F}_\mu,\,\widetilde \phi_\mu, \, B,\, B_1, \, B_2,\, B_3, \,  B_4,\, B_5,\,
 B_{\mu\nu}, \, {\cal B_{\mu\nu}}, \bar {\cal B}_{\mu\nu}
\Bigl ] = 0,
\end{eqnarray}
leave the action integral, corresponding to the Lagrangian density ${\cal L}_{(B)}$, invariant because we observe that
{\it this} Lagrangian density transforms to the total spacetime derivative as: 
\begin{eqnarray}\label{5}
s_b\, {\cal L}_{(B)} &=& \dfrac{1}{2}\, \partial_\mu\, \Big[ (\partial^\mu\, C^{\nu\sigma} + \partial^\nu\, C^{\sigma\mu}
+ \partial^\sigma\, C^{\mu\nu}) \, B_{\nu\sigma}  + B^{\mu\nu}\, f_\nu - B_5\, \partial^\mu\, C_2  \nonumber\\
&+& B_2\, f^\mu + B_4\, \bar F^\mu - (\partial^\mu\,  \beta^\nu - \partial^\nu\, \beta^\mu)\,  \bar F_\nu \Big] 
- \partial_\mu \Big [B \, \partial^\mu  C \Big ].
\end{eqnarray}
Thus, we draw the conclusion that the infinitesimal, continuous and off-shell nilpotent BRST transformations [cf. Eq. (4)] are the {\it symmetry}
transformations [31] for our present combined 4D field-theoretic system  of the {\it free} Abelian 3-form and 1-form gauge theories.

We end this section with the following clinching remarks. First of all, the nilpotency (i.e. $s_b^2 = 0 $) 
property of the BRST symmetry transformation
operator ($s_b$) ensures that {\it this} operator is {\it fermionic} in nature. As a consequence, it transforms the bosonic field
of our theory into its counterpart fermionic field and vic{\`e}-versa. Second, a close look at the BRST symmetry transformations (4) makes it clear that 
the mass dimension (in the natural units: $\hbar = c = 1 $) of a field increases by one when it is operated upon by $s_b$. 
Third, the axial-vector field $\tilde \phi_\mu$ and its corresponding field-strength tensor
do {\it not} transform at all under the BRST symmetry transformations. Fourth, we note that
the field-strength tensors (e.g. $H_{\mu\nu\sigma\rho}, \, F_{\mu\nu} $), corresponding to the gauge fields $A_{\mu\nu\sigma}$ and $A_\mu$, respectively, 
remain invariant under the BRST transformations [cf. Eq. (4)]. As discussed earlier, it is  to be re-emphasized that 
the field strength tensors of the gauge fields owe their {\it mathematical} origin to the exterior derivative of differential geometry.
Fifth, the gauge-fixing terms (i.e. $\partial \cdot A $ and $\partial_\sigma A^{\sigma\mu\nu} $) owe their origin to the co-exterior derivative
(i.e. $\delta = - \, *\, d\, *$) as we note that: $\delta A^{(1)} = -\, *\, d\, * (A_\mu \, dx^\mu) = (\partial \cdot A) $ and
$\delta A^{(3)} = -\, *\, d\, * [\frac{1}{3!}\, A_{\mu\nu\sigma} \, (dx^\mu \wedge dx^\nu \wedge dx^\sigma)] = - \, \frac{1}{2}\,
(\partial^\sigma  A_{\sigma\mu\nu}) \, (dx^\mu \wedge dx^\nu) $. We have the freedom to add/subtract [cf. Eqs. (1),(8)]
a 2-form $\Phi^{(2)} = d \,\Phi^{(1)}$
to the gauge-fixing term (i.e. $\delta A^{(3)} = -\, *\,d\, *\, A^{(3)} $) of the Abelian 3-form gauge field
where the 1-form $\Phi^{(1)} = \phi_\mu\, dx^\mu $ defines a polar vector field $\phi_\mu$. This is how the vector field $\phi_\mu$ appears in 
the generalized gauge-fixing term of our Abelian 3-form gauge  field in our theory.
Finally, the axial-vector $\widetilde \phi_\mu$ appears in our theory in the generalization of the kinetic term for the Abelian 1-form gauge field
because we have expressed {\it its} kinetic term (i.e. $-\, \frac{1}{4} \, F^{\mu\nu}\, F_{\mu\nu} $) in terms of the
dual on the Abelian 2-form [i.e. $*\, F^{(2)} = \frac{1}{2!}\, \tilde F_{\mu\nu} \, (x^\mu \wedge dx^\nu) $] where we have
explicit expression for the dual field-strength tensor as: $\tilde F_{\mu\nu} = \frac{1}{2}\, \varepsilon_{\mu\nu\sigma\rho}\, F^{\sigma\rho} \equiv
\varepsilon_{\mu\nu\sigma\rho}\, \partial^\sigma A^\rho$. Thus, we have the freedom to add/subtract  a pseudo 2-form
$\widetilde \Phi^{(2)} = d \,\widetilde \Phi^{(1)}$ to the kinetic term of the 
Abelian 1-form gauge field where the pseudo  1-form $\widetilde \Phi^{(1)} = \widetilde \phi_\mu \, dx^\mu$ defines
the axial-vector field $\widetilde \phi_\mu$ which appears in its appropriate forms in the non-ghost sectors (1) and (8).\\


\section{Nilpotent Anti-BRST Symmetry Transformations}

As per the sacrosanct requirements of the BRST formalism, it is essential that, corresponding to a given infinitesimal and local gauge 
symmetry transformation at the {\it classical} level, we should have the {\it nilpotent} (i.e. $s_{(a)b}^2 = 0 $)
and absolutely anticommuting (i.e. $s_b \, s_{ab} + s_{ab}\, s_b = 0$) 
infinitesimal BRST (i.e. $s_b$) as well as the anti-BRST (i.e. $s_{ab}$) symmetry transformations at the {\it quantum} level
in a {\it properly}  BRST-quantized theory. In view of these requirements,
corresponding to the infinitesimal, continuous and off-shell nilpotent ($s_b^2 = 0$) BRST symmetry transformations (4), we have the following 
infinitesimal, continuous and off-shell nilpotent (i.e. $s_{ab}^2 = 0 $)
anti-BRST transformations ($s_{ab}$), namely;
\begin{eqnarray}\label{6}
&&s_{ab} A_{\mu\nu\sigma} = \partial_\mu \bar C_{\nu\sigma} + \partial_\nu \bar C_{\sigma\mu}
+ \partial_\sigma \bar C_{\mu\nu}, \quad  s_{ab} \bar C_{\mu\nu} = \partial_\mu \bar \beta_\nu
- \partial_\nu \bar \beta_\mu, \quad s_{ab}  C_{\mu\nu} = \bar B_{\mu\nu}, \nonumber\\
&&s_{ab} A_{\mu} = \partial_\mu \bar C, \quad s_{ab}  C = -\, B, \quad  
s_{ab}  \beta_\mu =  F_\mu, \quad
s_{ab} \bar \beta_\mu = \partial_\mu \bar C_2, \quad s_{ab} B_{\mu\nu} = \partial_\mu \bar f_\nu - \partial_\nu \bar f_\mu, \nonumber\\
&&s_{ab} {C}_2 = B_4, \quad s_{ab} C_1 = - B_2, \quad  s_{ab} \bar C_1 = -\, B_5, 
\quad s_{ab} \phi_\mu = \bar f_\mu, \quad s_{ab} \bar F_\mu =-\, \partial_\mu B_5,
\nonumber\\
&& s_{ab} f_\mu = -\, \partial_\mu B_2, \; s_{ab}  \Bigl [ \bar C_2,\, \bar C,\, \bar f_\mu,\, {F}_\mu,
\,\widetilde \phi_\mu, \, B,\, B_1, \, B_2,\, B_3, \,  B_4,\, B_5,\, \bar B_{\mu\nu}, \, 
\bar {\cal B}_{\mu\nu}, \, {\cal B}_{\mu\nu}\Bigl ]  = 0,
\end{eqnarray}
which transform the  Lagrangian density ${\cal L}_{(\bar B)}$ to the total spacetime derivative:
\begin{eqnarray}\label{7}
s_{ab}\, {\cal L}_{(\bar B)} &=& \dfrac{1}{2}\, \partial_\mu\, \Big[\bar B^{\mu\nu}\, \bar f_\nu - \, (\partial^\mu\, \bar C^{\nu\sigma} + \partial^\nu\, \bar C^{\sigma\mu}
+ \partial^\sigma\, \bar C^{\mu\nu}) \, \bar B_{\nu\sigma}   + B_4\, \partial^\mu\, \bar C_2  \nonumber\\
&-& B_5\, F^\mu + B_2\, \bar f^\mu - (\partial^\mu\,  \bar \beta^\nu - \partial^\nu\, \bar \beta^\mu)\,  F_\nu \Big] 
- \partial_\mu \Big [B \, \partial^\mu  \bar C \Big ].
\end{eqnarray}
The above observation renders the action integral, corresponding to the Lagrangian density  ${\cal L}_{(\bar B)}$, invariant due to
Gauss's divergence theorem where {\it all} the physical fields vanish off as $x \to \pm\, \infty $. This Lagrangian density 
 ${\cal L}_{(\bar B)} = {\cal L}_{(ng)} + {\cal L}_{(fp)}$ consists of 
the non-ghost part [i.e. ${\cal L}_{(ng)}$] and the FP-ghost part [i.e. ${\cal L}_{(fp)}$]
which are the analogues of the non-ghost part of the Lagrangian density ${\cal L}_{(NG)} $ [cf. Eq. (1)] and FP-ghost part of the
Lagrangian density ${\cal L}_{(FP)} $ [cf. Eq. (3)] of ${\cal L}_{(B)} $. First of
all, let us focus on the non-ghost sector of the Lagrangian density ${\cal L}_{(\bar B)} $
which is described by the following Lagrangian density, namely;
\begin{eqnarray}\label{8}
&&{\cal L}_{(ng)}  = \dfrac{1}{2}\, B^2 - B \, (\partial \cdot A) + \dfrac{1}{2}\, B_2 \,\big (\partial \cdot \phi \big ) - \dfrac{1}{4}\, B_2^2  + 
\dfrac{1}{2}\, B_3 \,\big (\partial \cdot \widetilde \phi \big ) - \dfrac{1}{4}\, B_3^2 
 \nonumber\\
&+&  \dfrac{1}{2} \, B_1^2 
+\, B_1\, \Big(\dfrac{1}{3!}\,\varepsilon^{\mu\nu\sigma\rho} \,\partial_\mu A_{\nu\sigma\rho} \Big) 
 - \,\dfrac{1}{4} \, \big( \bar B_{\mu\nu} \big)^2 - \dfrac{1}{2}\, \bar B_{\mu\nu} \, \Big[\partial_\sigma A^{\sigma\mu\nu}
- \dfrac{1}{2}\, \big (\partial^\mu \phi^\nu - \partial^\nu \phi^\mu \big) \Big ]  
   \nonumber\\
&-& \dfrac{1}{4}\, \big( {\bar {\cal B}_{\mu\nu}} \big)^2 - \dfrac{1}{2} \, {\bar {\cal B}_{\mu\nu}}\, 
\Big[ \varepsilon^{\mu\nu\sigma\rho} \,\partial_\sigma A_\rho 
- \dfrac{1}{2}\, \big (\partial^\mu \widetilde \phi^\nu - \partial^\nu \widetilde \phi^\mu \big) \Big], 
 \end{eqnarray}
where a {\it new} set of {\it bosonic} Nakanishi-Lautrup type auxiliary fields $\bar B_{\mu\nu}$ and $\bar {\cal B}_{\mu\nu}$ 
(with {\it zero} ghost numbers) have been invoked 
to linearize the gauge-fixing term for the Abelian 3-form gauge field and kinetic term for the Abelian 1-form field, respectively. 
A comparison between the Lagrangian density (1) and (8)
establishes that the {\it last  four} terms have some very crucial changes in the signs. The FP-ghost sector of the Lagrangian density 
${\cal L}_{(\bar B)}$ for our 4D  field-theoretic system is described by the following Lagrangian density [31]
\begin{eqnarray}\label{9}
{\cal L}_{(fp)}  &=&  \dfrac{1}{2}\, \Big [\big (\partial_\mu \bar C_{\nu\sigma} +  \partial_\nu \bar C_{\sigma\mu} 
+ \partial_\sigma \bar C_{\mu\nu} \big ) \big (\partial^\mu C^{\nu\sigma} \big ) -
\big (\partial_\mu \bar C^{\mu\nu}  - \partial^\nu \bar C_1 \big ) F_\nu \nonumber\\ 
&+& \big (\partial_\mu  C^{\mu\nu}  - \partial^\nu  C_1 \big ) \bar f_\nu 
 +  \big( \partial \cdot \bar \beta \big) \, B_4
 - \big( \partial \cdot  \beta \big) \, B_5 - B_4 \, B_5 - 2\, \bar f^\mu\, F_\mu \nonumber\\
&-& \big (\partial_\mu \bar \beta_\nu - \partial_\mu \bar \beta_\nu \big ) 
\big (\partial^\mu \beta^\nu \big )  
-\, \partial_\mu \bar C_2\, \partial^\mu C_2  \Big ] - \partial_\mu \bar C\, \partial^\mu C, 
\end{eqnarray}
which is {\it different} from our earlier FP-ghost Lagrangian  density [cf. Eq. (3)] that has 
been taken into account in the BRST invariant Lagrangian density ${\cal L}_{(B)}$.
It should be pointed out that we have (i) two {\it new} fermionic (anti-)ghost [i.e. $(\bar f_\mu)F_\mu $]
auxiliary fields in (9) that carry the ghost numbers (-1)+1, respectively, and (ii) the {\it second} and {\it third} terms of the FP-ghost parts of
the Lagrangian densities in (3) and (9) are {\it quite} different.

We  discuss concisely  the anticommutativity property of the nilpotent (anti-)BRST transformation operators $s_{(a)b}$ 
and lay emphasis on the CF-type restrictions (cf. Appendix B). 
It turns out that the absolute anticommutativity property (i.e. $\{s_b, s_{ab}\} = 0$) is satisfied for {\it all} the fields 
of our theory {\it except} fields $A_{\mu\nu\sigma}, C_{\mu\nu}$ and $\bar C_{\mu\nu}$ because we observe: 
 \begin{eqnarray}\label{10}
&& \{s_b, \,s_{ab}\}\, A_{\mu\nu\sigma} = \partial_\mu (B_{\nu\sigma} +  \bar B_{\nu\sigma}) + \partial_\nu (B_{\sigma\mu}
 +  \bar B_{\sigma\mu}) + \partial_\sigma (B_{\mu\nu} +  \bar B_{\mu\nu}), \nonumber\\
 && \{s_b, \,s_{ab}\}\, C_{\mu\nu} = \partial_\mu (f_\nu +  F_\nu) - \partial_\nu (f_\mu + F_\mu), \nonumber\\
  && \{s_b, \, s_{ab}\}\,\bar C_{\mu\nu} = \partial_\mu (\bar f_\nu +  \bar F_\nu) - \partial_\nu (\bar f_\mu + \bar F_\mu).
\end{eqnarray}  
However, we have the sanctity of the absolute  anticommutativity property (i.e.  $\{s_b, s_{ab}\} = 0$) of
the (anti-)BRST transformations  for the {\it above} fields, too, provided we use the following (anti-)BRST invariant CF-type restrictions
(cf. Appendix B)
which have been derived by using the superfield approach\footnote{It should be noted that the CF-type restriction: 
${\cal B}_{\mu\nu} + \bar {\cal B}_{\mu\nu} - (\partial_\mu \tilde \phi_\nu  - \partial_\nu \tilde \phi_\mu) = 0 $ [cf. Eq. (B.1)] which
is useful in the proof of the absolute anticommutativity between the nilpotent co-BRST and anti-co-BRST symmetry transformations
is {\it not} derived from the normal superfield approach to BRST formalism [38,39]. }
 to the D-dimensional free Abelian 3-form theory [38,39]:
\begin{eqnarray}\label{11}
B_{\mu\nu} + \bar B_{\mu\nu} = \partial_\mu \phi_\nu  - \partial_\nu \phi_\mu, \qquad 
f_\mu + F_\mu  = \partial_\mu C_1, \qquad \bar f_\mu + \bar F_\mu  = \partial_\mu \bar C_1.
\end{eqnarray} 
It is pertinent to point out that we have focused on the off-shell nilpotent (anti-)BRST symmetry transformation operators
and used the bosonic sector of the CF-type restriction of (11) to prove $\{s_b, \,s_{ab}\}\, A_{\mu\nu\sigma} = 0 $ and the fermionic sectors
of the CF-type restrictions: $f_\mu + F_\mu  = \partial_\mu C_1$ and $ \bar f_\mu + \bar F_\mu  = \partial_\mu \bar C_1 $ have been
exploited in proving $\{s_b, \,s_{ab}\}\, C_{\mu\nu} = 0 $ and $\{s_b, \, s_{ab}\}\,\bar C_{\mu\nu} = 0 $, respectively, in our equation (10).
Using the off-shell nilpotent versions of the (anti-)BRST transformations [cf. Eqs. (6),(4)], it can be checked that the CF-type restrictions [cf. Eq. (B.1)]
are (anti-)BRST invariant quantities, namely; 
\begin{eqnarray}\label{12}
&&s_{(a)b}[B_{\mu\nu} + \bar B_{\mu\nu} - (\partial_\mu \phi_\nu  - \partial_\nu \phi_\mu)] = 0, \quad 
s_{(a)b}[f_\mu + F_\mu  - \partial_\mu C_1] = 0,\nonumber\\
&& s_{(a)b}[\bar f_\mu + \bar F_\mu  - \partial_\mu \bar C_1] = 0, \quad 
s_{(a)b}[{\cal B}_{\mu\nu} + \bar {\cal B}_{\mu\nu} - (\partial_\mu \tilde \phi_\nu  - \partial_\nu \tilde \phi_\mu)] = 0,
\end{eqnarray} 
which corroborates our earlier claim [after equation (10)].
Hence, these constraints on our theory are {\it physical} and they are connected with the geometrical objects called gerbes [40,41]. At the far end of
our Appendix B, we have remarked about 
the existence of the CF-type restrictions on the basis of (i) the EL-EoMs w.r.t. the {\it auxiliary} fields of our
theory, and (ii) the proof of the {\it coupled} nature of the Lagrangian densities ${\cal L}_{(B)} $ and ${\cal L}_{(\bar B)} $.

For the sake of completeness, we {\it also} show the existence of the CF-type restrictions from the EL-EoMs w.r.t the {\it dynamical} fields of 
our theory. In this context, we observe that the FP-ghost sectors [cf. Eqs. (3),(9)] of the
{\it coupled} Lagrangian densities ${\cal L}_{(B)} $ and ${\cal L}_{(\bar B)} $ lead to the following EL-EoMs 
w.r.t. the fields $C_{\mu\nu}$ and $ \bar C_{\mu\nu}$, respectively: 
\begin{eqnarray}\label{13}
&&\partial_\mu\, (\partial^\mu\, \bar C^{\nu\lambda} + \partial^\nu\, \bar C^{\lambda\mu} + \partial^\lambda\, \bar C^{\mu\nu})
+ \frac{1}{2}\, (\partial^\nu\, \bar{F}^\lambda - \partial^\lambda \, \bar F^\nu) = 0, \nonumber\\
&&\partial_\mu\, (\partial^\mu\, \bar C^{\nu\lambda} + \partial^\nu\, \bar C^{\lambda\mu} + \partial^\lambda\, \bar C^{\mu\nu})
- \frac{1}{2}\, (\partial^\nu\, \bar{f}^\lambda - \partial^\lambda \, \bar f^\nu) = 0, \nonumber\\
&&\partial_\mu\, (\partial^\mu\,  C^{\nu\lambda} + \partial^\nu\,  C^{\lambda\mu} + \partial^\lambda\,  C^{\mu\nu})
+ \frac{1}{2}\, (\partial^\nu\, {f}^\lambda - \partial^\lambda \,  f^\nu) = 0, \nonumber\\
&&\partial_\mu\, (\partial^\mu\,  C^{\nu\lambda} + \partial^\nu\,  C^{\lambda\mu} + \partial^\lambda\,  C^{\mu\nu})
- \frac{1}{2}\, (\partial^\nu\, {F}^\lambda - \partial^\lambda \,  F^\nu) = 0. 
\end{eqnarray}
It is clear, from the above, that we obtain the following: 
\begin{eqnarray}\label{14}
\partial_\mu\, (\bar f_\nu + \bar F_\nu) - \partial_\nu\,(\bar f_\mu + \bar F_\mu) = 0, \;\qquad \;
\partial_\mu\, (f_\nu + F_\nu) - \partial_\nu\,(f_\mu + F_\mu) = 0.
\end{eqnarray}
At this juncture, we argue that the above equations in (13) lead to the derivations of the fermionic CF-type restrictions: 
$f_\mu + F_\mu = \partial_\mu\, C_1$ and $\bar f_\mu + \bar F_\mu = \partial_\mu\, \bar C_1$. 
To corroborate this claim, we note that, in the fermionic relationships  (14), the ghost numbers can be conserved iff we take into 
account the following {\it non-trivial} 
combinations of the pairs of the fermionic auxiliary fields $(\bar f_\mu, \, \bar F_\mu)$ and $(f_\mu, \,  F_\mu)$  and the 
(anti-)ghost fields $(\bar C_1)C_1$, namely; 
\begin{eqnarray}\label{15}
f_\mu + F_\mu =  \,\pm \, \partial_\mu\, C_1, \qquad  \bar f_\mu + \bar F_\mu =\,\pm \, \partial_\mu\, \bar C_1, 
\end{eqnarray}
to make the r.h.s. of equation (14) equal to zero. 
However, the requirement of the (anti-) BRST invariance of the
CF-type restrictions ensures that only the {\it positive} signs on the r.h.s. of (15) are permitted. 
We would like to lay emphasis on the fact  that the r.h.s. of (15) can {\it not} be anything other than the derivatives  
on the (anti-)ghost fields $(\bar C_1)C_1$. In other words, {\it no}  other independent {\it basic} (anti-)ghost 
fields are permitted on the r.h.s. of (15). Furthermore, we would like to add that 
there is a {\it trivial} solution where we can take, in a straightforward fashion, the combinations:
$f_\mu + F_\mu = 0$ and $\bar f_\mu + \bar F_\mu =0$. However, these combinations are {\it not} (anti-)BRST invariant
[cf. Eqs. (6),(4)]. Hence, they can not
be identified with the {\it physical} fermionic sectors of the CF-type restrictions on our theory.

Against the backdrop of the above discussions [cf. Eq. (15)], we would like to mention that the bosonic sectors of the CF-type restrictions (i.e.
$B_{\mu\nu} + \bar B_{\mu\nu} = \partial_\mu \phi_\nu - \partial_\nu \phi_\mu$ and
${\cal B}_{\mu\nu} + \bar {\cal B}_{\mu\nu} = \partial_\mu \tilde \phi_\nu - \partial_\nu \tilde \phi_\mu$) can {\it also} be derived
from the EL-EoMs w.r.t. the {\it basic} gauge fields of our theory where we shall concentrate on the non-ghost sectors of the
Lagrangian densities [cf. Eqs. (1),(8)]. For instance, it can be checked that the following EL-EoMs
\begin{eqnarray}\label{16}
\dfrac{1}{2}\, \varepsilon^{\mu\nu\sigma\rho} \, \partial_\nu {\cal B}_{\sigma\rho} + \partial^\mu B = 0, \qquad
\dfrac{1}{2}\, \varepsilon^{\mu\nu\sigma\rho} \, \partial_\nu B_{\sigma\rho} - \partial^\mu B_1 = 0,
\end{eqnarray}
emerge out from the Lagrangian density (1) w.r.t. the gauge fields $A_\mu$ and $A_{\mu\nu\sigma}$, respectively. On the other hand, the
non-ghost sector of the Lagrangian density (8) leads to the following EL-EoMs w.r.t. 
the basic gauge fields $A_\mu$ and $A_{\mu\nu\sigma}$, respectively:
\begin{eqnarray}\label{17}
\dfrac{1}{2}\, \varepsilon^{\mu\nu\sigma\rho} \, \partial_\nu \bar {\cal B}_{\sigma\rho} - \partial^\mu B = 0, \qquad
\dfrac{1}{2}\, \varepsilon^{\mu\nu\sigma\rho} \, \partial_\nu \bar B_{\sigma\rho} + \partial^\mu B_1 = 0.
\end{eqnarray}
 A close look at (16) and (17) implies that we have the following conditions
\begin{eqnarray}\label{18}
\dfrac{1}{2}\, \varepsilon^{\mu\nu\sigma\rho} \, \partial_\nu \big [ {\cal B}_{\sigma\rho} +  \bar {\cal B}_{\sigma\rho} \big ]= 0, \qquad
\dfrac{1}{2}\, \varepsilon^{\mu\nu\sigma\rho} \, \partial_\nu \big [ B_{\sigma\rho} + \bar B_{\sigma\rho} \big ] = 0,
\end{eqnarray}
on the Nakanishi-Lautrup auxiliary fields if we wish to maintain the {\it coupled} nature of the 
Lagrangian densities ${\cal L}_{(B)} $ and ${\cal L}_{(\bar B)} $.
The non-trivial solutions of the above conditions are:
\begin{eqnarray}\label{19}
B_{\mu\nu} + \bar B_{\mu\nu} = \pm \,\big (\partial_\mu \phi_\nu - \partial_\nu \phi_\mu \big ), \qquad
{\cal B}_{\mu\nu} + \bar {\cal B}_{\mu\nu} = \pm \,\big (\partial_\mu \tilde \phi_\nu - \partial_\nu \tilde \phi_\mu \big ).
\end{eqnarray}
However, the requirement of the (anti-)BRST invariance picks up only the {\it positive} sign from the r.h.s. of (19). Thus, we have derived the
bosonic sectors of the CF-type restrictions: $B_{\mu\nu} + \bar B_{\mu\nu} = \partial_\mu \phi_\nu - \partial_\nu \phi_\mu$ and
${\cal B}_{\mu\nu} + \bar {\cal B}_{\mu\nu} = \partial_\mu \tilde \phi_\nu - \partial_\nu \tilde \phi_\mu$ from the EL-EoMs
w.r.t. the {\it dynamical} basic Abelian 1-form and 3-form gauge fields (i.e. $A_\mu$ and $A_{\mu\nu\sigma}$). We would like to lay
emphasis on the fact that the {\it correct} relationships (i.e. $f_\mu + F_\mu =  \partial_\mu\, C_1,\;\bar f_\mu + \bar F_\mu = \partial_\mu\, \bar C_1,
B_{\mu\nu} + \bar B_{\mu\nu} = \partial_\mu \phi_\nu - \partial_\nu \phi_\mu, \; {\cal B}_{\mu\nu} + \bar {\cal B}_{\mu\nu} 
= \partial_\mu \tilde \phi_\nu - \partial_\nu \tilde \phi_\mu $) from (15) and (19) are the {\it physical} constraints on our theory and they are 
{\it not} the EL-EoMs because they have not been derived from a {\it single} Lagrangian density. Their existence on our theory {\it only} ensures that
the Lagrangian densities ${\cal L}_{(B)} $ and ${\cal L}_{(\bar B)} $ are the {\it correct} coupled Lagrangian densities.

We wrap-up  this section with a few concluding remarks. First of all, the coupled Lagrangian densities can yield the CF-type
restrictions from their EL-EoMs {\it but} the CF-type restrictions themselves are (i) {\it not} the
EL-EoMs in the true sense of the words, and (ii) responsible for the existence of the {\it coupled} 
Lagrangian densities ${\cal L}_{B}$ and ${\cal L}_{\bar B}$. Second, these Lagrangian densities are equivalent
from the point of view of the symmetry considerations (cf. Appendix B) as {\it both} of these respect the (anti-)BRST symmetry transformations 
on the submanifold of the quantum  fields where all the {\it three} (anti-)BRST invariant CF-type restrictions
(i.e. $B_{\mu\nu} + \bar B_{\mu\nu} = \partial_\mu\, \phi_\nu - \partial_\nu\, \phi_\mu, 
\,\;  f_\mu + F_\mu = \partial_\mu\, C_1, \; \bar f_\mu + \bar F_\mu = \partial_\mu\, \bar C_1$) 
are satisfied (see, e.g. [38,39] for details). Third, just like the BRST symmetry transformations (cf. Sec. 2), the field-strength tensors
(i.e. $ H_{\mu\nu\sigma\rho}, F_{\mu\nu}$) remain invariant under the nilpotent anti-BRST symmetry transformations, too. Fourth,
the anti-BRST symmetry operator (i) lowers the ghost number of a field by {\it one} on which it operates, and (ii) 
raises the mass dimension (in the natural unites: $\hbar = c = 1 $)
of the field by {\it one} on which it acts. Fifth, the BRST and anti-BRST symmetry
transformations are {\it fermionic} in nature and they transform the bosonic field into the fermionic field and vic{\`e}-versa.  
However, their absolute anticommutativity property {\it distinguishes} them from the $\mathcal {N} = 2$ SUSY transformations
which are also nilpotent but {\it not} absolutely anticommuting in nature. Sixth, the additional (axial-)vector fields $(\widetilde \phi_\mu)\phi_\mu$
appear in the perfectly anti-BRST invariant Lagrangian density ${\cal L}_{\bar B}$ with {\it negative} signs as against the perfectly BRST
invariant Lagrangian density ${\cal L}_{B}$ where they appear with {\it positive} signs at appropriate places in the kinetic term for the
Abelian 1-form field and the gauge-fixing term for the Abelian 3-form field, respectively. The mathematical origins for these additional fields
remain the {\it same} as we have discussed in the last paragraph of our Sec. 2.
Finally, the covariant canonical quantization of a
gauge theory is performed very elegantly within the framework of BRST formalism where the unitarity and 
quantum gauge (i.e. BRST) invariance are respected
at any arbitrary order of perturbative computations for an allowed physical process by the theory.\\


\section{Conserved BRST Current and  BRST Charge}

Our present section is divided into {\it three} parts. In subsection 4.1, we derive (i) the Noether conserved BRST current by exploiting the standard 
theoretical tricks of Noether's theorem, and (ii) the Noether conserved  BRST charge $Q_b$ where we demonstrate that it is non-nilpotent
(i.e. $Q_b^2 \neq 0$).
Our subsection 4.2 is devoted to the derivation of the nilpotent version of the BRST charge $Q_B $ from the 
non-nilpotent version of the Noether BRST charge $Q_b$.
In our subsection 4.3, we discuss the physicality criterion w.r.t. $Q_B$.\\

\vskip 0.7cm

\subsection{Noether  Conserved BRST Current}

We have seen that, under the continuous, infinitesimal and nilpotent BRST symmetry transformations (4), the Lagrangian density ${\cal L}_{(B)}$ transforms 
to a total spacetime derivative [cf. Eq. (5)]. Exploiting the standard theoretical techniques of Noether's theorem, we obtain the following
expression for the Noether BRST current ($J^\mu_{(b)} $), namely;
\begin{eqnarray}\label{20}
J^\mu_{(b)} &=&  s_b \Phi_i \; \dfrac {\partial {\cal L}_{(B)}}{\partial (\partial_\mu \Phi_i)} + B\, \partial^\mu C 
+ \dfrac{1}{2}\, \big [B_5 \, \partial^\mu C_2 + (\partial^\mu \beta^\nu - \partial^\nu \beta^\mu)\, \bar F_\nu \big ]
\nonumber\\
&-& \dfrac{1}{2}\, \big [(\partial^\mu C^{\nu\sigma} + \partial^\nu C^{\sigma\mu} + \partial^\sigma C^{\mu\nu})\, B_{\nu\sigma}
+ B^{\mu\nu}\, f_\nu + B_2\, f^\mu + B_4 \bar F^\mu \big],
\end{eqnarray}
where the generic field $\Phi_i$ stands for all the {\it dynamical} fields of our 4D combined system of the BRST-quantized Abelian 3-from and 1-form gauge theories
which is described by the perfectly BRST-invariant [cf. Eq. (5)] Lagrangian density ${\cal L}_{(B)}$.
In other words, we have: $\Phi_i \equiv A_{\mu\nu\sigma},  C_{\mu\nu}, \bar C_{\mu\nu}, A_\mu, \phi_\mu, \tilde \phi_\mu, \beta_\mu, \bar \beta_\mu, 
C_1, \bar C_1, C, \bar C $. The rest of the terms on the r.h.s. of (20) are the terms that are present in the sqaure bracket of equation (4) modulo
a sign factor as per the basic rules/ideas behind the Noether theorem. The first term on the r.h.s. of (20) can be explicitly computed from the non-ghost sector 
 [cf. Eq. (1)] and the ghost sector [cf. Eq. (3)] of the {\it total} BRST-invariant Lagrangian density  ${\cal L}_{(B)}$. It is worthwhile to 
 mention here that (i) the bosonic/fermionic auxiliary fields  of our theory, obviously, do {\it not} contribute 
anything in the computation of above Noether current, and
 (ii) the dynamical fields $\tilde \phi_\mu, C_2, C$ {\it also} do {\it not} contribute 
anything in the computation of $J^\mu_{(b)} $ because we observe that:
 $s_b C_2 = 0, \, s_b C = 0, \,s_b \widetilde \phi_\mu = 0$ [cf. Eq. (4)]. The final expression for the  Noether BRST current, due to
precise computation of (20),  is as follows: 
\begin{eqnarray}\label{21}
J^\mu_{(b)} &=&  \dfrac{1}{2}\, \Big [\varepsilon^{\mu\nu\sigma\rho}\, B_1 (\partial_\nu C_{\sigma\rho}) 
+  (\eta^{\mu\nu}  B^{\sigma\rho} + \eta^{\mu\sigma} B^{\rho\nu} + \eta^{\mu\rho} B^{\nu\sigma})\, (\partial_\nu C_{\sigma\rho})
 + B^{\mu\nu}\,f_\nu  
\nonumber\\
&+& \varepsilon^{\mu\nu\sigma\rho}\, (\partial_\nu C)\, {\cal B}_{\sigma\rho} - B_5 \, \partial^\mu C_2 - (\partial^\mu \bar \beta^\nu 
- \partial^\nu \bar \beta^\mu)\, \partial_\nu C_2 - (\partial^\mu \beta^\nu - \partial^\nu \beta^\mu)\, \bar F_\nu  \nonumber\\
&+& B_2\, f^\mu + B_4\, \bar F^\mu - (\partial^\mu \bar C^{\nu\sigma} + \partial^\nu \bar C^{\sigma\mu} + \partial^\sigma \bar C^{\mu\nu})\,(\partial_\nu \beta_\sigma 
- \partial_\sigma \beta_\nu) \Big ] - B \, \partial^\mu C.
\end{eqnarray}
The conservation law (i.e. $\partial_\mu J^\mu_{(b)} = 0$) can be proven by using the EL-EoMs (16) and (C.5) which have been derived 
from the non-ghost and ghost sectors of the {\it total} Lagrangian density ${\cal L}_{(B)}$. In addition, 
in the above proof, we have to exploit the following EL-EoMs
\begin{eqnarray}\label{22}
\big (\partial^\mu B^{\nu\sigma} + \partial^\nu B^{\sigma\mu}  + \partial^\sigma B^{\mu\nu} \big ) 
- \varepsilon^{\mu\nu\sigma\rho}\, \partial_\rho B_1 = 0, \qquad
\partial_\mu B^{\mu\nu} + \partial^\nu B_2 = 0,
\end{eqnarray}
which emerge out from the non-ghost sector [cf. Eq. (1)] of the perfectly BRST-invariant Lagrangian density ${\cal L}_{(B)}$ where the first
entry is nothing but equation (16)  which can {\it also} be expressed as: $\varepsilon^{\mu\nu\sigma\rho}\, \partial_\mu B_1
+ \big (\partial^\nu B^{\sigma\rho} + \partial^\sigma B^{\rho\nu}  + \partial^\rho B^{\nu\sigma} \big ) = 0 $. In addition, we have to 
make use of the relationship: $(\eta^{\mu\nu}  B^{\sigma\rho} + \eta^{\mu\sigma} B^{\rho\nu} + \eta^{\mu\rho} B^{\nu\sigma})\, (\partial_\nu C_{\sigma\rho})
= (\partial^\mu C^{\nu\sigma} + \partial^\nu C^{\sigma\mu} + \partial^\sigma C^{\mu\nu})\, B_{\nu\sigma} $ which {\it also} plays a key role in the 
conservation law (i.e. $\partial_\mu J^\mu_{(b)} = 0 $). In other words, depending on the situation, we can always exchange these
two terms at appropriate places. The above Noether conserved BRST current [cf. Eq. (21)] leads to the definition
of the Noether conserved BRST charge $Q_{b} = \int d^3 x\, J^0_{(b)}$ whose explicit expression is:
\begin{eqnarray}\label{23}
Q_b &=&   \int d^3 x\, \Big [\dfrac{1}{2}\,\Big \{\varepsilon^{0ijk}\, B_1 \,(\partial_i C_{jk}) 
+ (\partial^0 C^{ij} + \partial^i C^{j0} + \partial^j C^{0i})\, B_{ij}
 + B^{0i}\,f_i  \nonumber\\
&+& \varepsilon^{0ijk}\, (\partial_i C)\, {\cal B}_{jk} - B_5 \, \partial^0 C_2 - (\partial^0 \bar \beta^i
- \partial^i \bar \beta^0)\, \partial_i C_2 - (\partial^0 \beta^i - \partial^i \beta^0)\, \bar F_i  \nonumber\\
&+& B_2\, f^0 + B_4\, \bar F^0 - (\partial^0 \bar C^{ij} + \partial^i \bar C^{j0} + \partial^j \bar C^{0i})\,(\partial_i \beta_j
- \partial_j \beta_i) \Big \} - B \, \partial^0 C \Big ].
\end{eqnarray}
The above Noether conserved BRST charge turns out to be the generator for the infinitesimal,
continuous and off-shell nilpotent BRST symmetry transformations (4) provided we (i) express the BRST charge in terms of the
canonical conjugate momenta w.r.t. all the dynamical fields, and (ii) use the  canonical (anti)commutators for the dynamical fermionic/bosonic
fields of our theory.
This statement can be mathematically expressed as 
\begin{eqnarray}\label{24}
s_b \Phi_i = \pm\, i\, \Big [\Phi_i, \;\, Q_b \Big ]_{(\pm)},
\end{eqnarray}
where the subscripts $(\pm)$ on the square bracket stand for the (anti)commutator for the generic dynamical field $\Phi_i$ being
fermionic/bosonic in nature. The $(\pm)$ signs, in front of the square bracket, have to be chosen judiciously (cf. Appendix C for some examples).

We end his subsection with the remark that, even though the BRST symmetry transformations (4) are off-shell nilpotent, it turns out that the 
Noether BRST charge (23) is {\it not} nilpotent of order two (i.e. $Q_b^2 \neq 0 $). To corroborate this statement, we exploit the relationship in 
(24) which is very general and valid for the bosonic as well as the fermionic fields/quantities.  If we chose the generic field
to be $Q_b$ itself, we observe the following:
\begin{eqnarray}\label{25}
s_b \, Q_b = +\, i\, \big \{Q_b, \; Q_b \big \} &\equiv& -\, \dfrac{1}{2}\,\int d^3 x \Big [(\partial^0 B^{ij} 
+ \partial^i B^{j0} + \partial^j B^{0i})\,(\partial_i \beta_j- \partial_j \beta_i) \nonumber\\
&+& (\partial^0 \bar F^i - \partial^i \bar F^0)\, \partial_i C_2\Big ] \neq 0.
\end{eqnarray}
In the above, we have computed explicitly the l.h.s. by applying {\it directly} the 
nilpotent BRST symmetry transformations (4) on the expression for the Noether conserved BRST charge (23).
A close look at the above observation implies that the Noether BRST charge is {\it not} nilpotent of order two (i.e.
$+\,i\,\{Q_b. \; Q_b \} =  2\,i\, Q_b^2 \neq 0$). This happens because of the fact that there are non-trivial CF-type restrictions [cf. Eq. (B.1)] 
on our theory. An elaborate discussion on this issue has been performed in our earlier work (see, e.g. [37] for details). In this work [37], we have
also proposed a systematic theoretical method to obtain the nilpotent version of the 
conserved BRST charge from the non-nilpotent version of the Noether 
conserved BRST charge. It is worthwhile to mention, in passing, that the physicality criterion w.r.t. the Noether conserved 
(but non-nilpotent) BRST charge (i.e $Q_b \, |phys> = 0$) does {\it not} lead to the annihilation of the physical states (i.e $|phys> $) by
the operator forms of the {\it full} set of the 
 {\it classical} first-class constraints (cf. Appendix A). This observation is {\it not} consistent with the Dirac quantization 
conditions for the physical systems that are endowed with constraints [32,33]. Furthermore, the Noether non-nilpotent 
(i.e. $Q_b^2 \neq 0$) BRST charge $Q_b$
 is {\it not} suitable as far as the discussion on a few aspects of the BRST cohomology is concerned (see. e.g. [42-45] for details). At this juncture,
we would like to remark that the use of the appropriate forms of the components of the EL-EoMs (22) and (C.5) leads to the r.h.s. of (25) being
a 3D integral over the total {\it space} derivative terms which vanish off as $x \to \pm\,\infty$
which implies that the nilpotency of $Q_b$ is {\it true} only on the on-shell. However, we hasten to add that our observation
in (25) still implies that the Noether BRST charge $Q_b$ is a non-nilpotent (i.e. $Q_b^2 \neq 0 $) 
and BRST non-invariant (i.e. $s_b Q_b \neq 0 $)
object because we are dealing with the off-shell nilpotent
BRST transformations where the EL-EoMs do {\it not} play any role in the discussion on the nilpotency property.\\


\subsection{Nilpotent BRST Charge from Non-nilpotent Noether Charge }

The central theme of our present subsection is to modify 
the Noether conserved BRST charge $Q_b$ to $Q_B$ in such a manner that the {\it latter} remains conserved just like the {\it former} and 
we obtain the analogue of the relationship (25) such that $Q_B$ satisfies: $s_{b} \, Q_B = +\, i\, \big \{Q_B, \; Q_B \big \} = 0$
so that the modified BRST charge $Q_B$ obeys the nilpotency (i.e. $Q_B^2 = 0$) property. To accomplish this goal, we
exploit the key ideas and proposals of our earlier work [37].
As per the theoretical tricks and techniques that have been indicated in {\it this} work, we have to exploit (i) the 
partial integration followed by the Gauss divergence theorem, (ii) the appropriate EL-EoMs that
are derived from the Lagrangian density ${\cal L}_{(B)} $, and (ii)
the BRST symmetry transformations (4) at appropriate places, to derive the nilpotent (i.e. $Q_B^2 = 0$) version of the BRST charge $Q_B$ 
from the non-nilpotent (i.e. $Q_b^2 \neq 0$) version of the Noether BRST charge $Q_b$. In this context,  first of all, let us focus on the
following terms of the Noether BRST charge $Q_b$ [cf. Eq. (23)], namely;
\begin{eqnarray}\label{26}
 \int d^3 x\, \Big [\dfrac{1}{2}\,\varepsilon^{0ijk}\, (\partial_i C)\, {\cal B}_{jk} - B \, \partial^0 C  \Big ],
\end{eqnarray}
that are connected with the 4D BRST-quantized Abelian 1-form gauge theory.
Using the straightforward application of the partial integration followed by the 
Gauss divergence theorem, the above equation can be re-expressed in the following form 
\begin{eqnarray}\label{27}
\dfrac{1}{2}\, \int d^3 x\,\Big [ -\,\dfrac{1}{2}\, \varepsilon^{0ijk}\, C\, (\partial_i {\cal B}_{jk}) - B \, \dot C \Big ],
\end{eqnarray}
where we have dropped a total space derivative term because the physical fields are supposed to vanish off as $x \to \pm\, \infty$.
At this juncture, we substitute the following EL-EoM
\begin{eqnarray}\label{28}
\dfrac{1}{2}\, \varepsilon^{\mu\nu\sigma\rho} \, \partial_\nu {\cal B}_{\sigma\rho} + \partial^\mu B = 0,
\end{eqnarray}
which has been derived, in a straightforward fashion,  from the BRST-invariant Lagrangian density ${\cal L}_{(B)} $. 
The resulting integral from (27) is as follows
\begin{eqnarray}\label{29}
 \int d^3 x\,\big (\dot B \, C - B \, \dot C \big ),
\end{eqnarray}
which will be present in the nilpotent version of the BRST charge $Q_B$ because it is straightforward to check that the above integrand is the BRST invariant
(i.e. $s_b \big [\dot B \, C- B \, \dot C \big ] = 0$) quantity [cf. Eq. (4)]. 
Furthermore, we would like add that equation (29) is the {\it standard} expression for the
conserved and nilpotent BRST charge for the BRST-quantized Abelian 1-form gauge theory. Now we shall focus
on {\it only} the remaining terms of the Noether BRST charge $Q_b$ [cf. Eq. (23)]. However, we shall keep in our mind the observations [that have been
made in (26) and (29)] for the deduction of the off-shell nilpotent (i.e. $Q_B^2 = 0$)
version of the BRST charge $Q_B$ through the requirement: $s_b Q_B = 0$.

We begin with the {\it first}  term of $Q_b$ because it contains the derivative on the Abelian 3-form
gauge field (i.e. $A_{\mu\nu\sigma} $)
through the relationship: $B_1 = -\,\frac{1}{3!}\, \varepsilon^{\mu\nu\sigma\rho} \, \partial_\mu A_{\nu\sigma\rho} $. 
We perform the partial integration
and, after that,  apply the Gauss divergence theorem. These two successive operations lead to the following
\begin{eqnarray}\label{30}
\dfrac{1}{2}\, \int d^3 x\, \varepsilon^{0ijk}\, B_1 \,(\partial_i C_{jk}) \equiv - \, 
\dfrac{1}{2}\, \int d^3 x\, \varepsilon^{0ijk}\, (\partial_i B_1) \, C_{jk},
\end{eqnarray}
where we have dropped a total space derivative term. Using the following EL-EoM (that has been derived from ${\cal L}_{(B)}$) w.r.t.
the Abelian 3-form gauge field  $A_{\mu\nu\sigma}$, namely;
\begin{eqnarray}\label{31}
 \varepsilon^{\mu\nu\sigma\rho}\, \partial_\rho B_1
= \big (\partial^\mu B^{\nu\sigma} + \partial^\nu B^{\sigma\mu}  + \partial^\sigma B^{\mu\nu} \big ) \,\Longrightarrow \,
\varepsilon^{oijk}\, \partial_k B_1
= \big (\partial^0 B^{ij} + \partial^i B^{j0}  + \partial^j B^{0i} \big ),
\end{eqnarray}
the equation (30) can be re-expressed as follows:
\begin{eqnarray}\label{32} 
- \, \dfrac{1}{2}\, \int d^3 x\, \varepsilon^{0ijk}\, (\partial_i B_1) \, C_{jk}
= - \, \dfrac{1}{2}\, \int d^3 x\, \big (\partial^0 B^{ij} + \partial^i B^{j0}  + \partial^j B^{0i} \big ) \, C_{ij}.
\end{eqnarray}
This term would be present in the expression for the nilpotent version of the BRST charge $Q_B$. Toward our goal to obtain the 
desired result: $s_b Q_B = +i\, \{Q_B, \; Q_B \} = 0 $, we have to apply the BRST transformations (4) on the above expression
which leads to:
\begin{eqnarray}\label{33} 
- \, \dfrac{1}{2}\, \int d^3 x\, \big (\partial^0 B^{ij} + \partial^i B^{j0}  + \partial^j B^{0i} \big ) \, 
\big (\partial_i \beta_j - \partial_j \beta_i \big ).
\end{eqnarray}
We have to modify an appropriate term of the Noether conserved charge $Q_b$ [cf. Eq. (23)]
so that (33) can cancel out if we apply the BRST transformations (4) on it to accomplish
our central goal of obtaining: $s_b Q_B = 0$. Such a term is:
$-\, \frac{1}{2}\,(\partial^0 \bar C^{ij} + \partial^i \bar C^{j0} + \partial^j \bar C^{0i})\,(\partial_i \beta_j - \partial_j \beta_i) $.
However, the straightforward application of 
BRST transformations (4) on it does {\it not} lead to the desired result. Hence, we modify this term as follows:
\begin{eqnarray}\label{34} 
& + &\, \dfrac{1}{2}\, \int d^3 x\, \big (\partial^0 \bar C^{ij} + \partial^i \bar C^{j0} + \partial^j \bar C^{0i} \big )\,
\big (\partial_i \beta_j - \partial_j \beta_i \big )
\nonumber\\
& - &  \int d^3 x\, \big (\partial^0 \bar C^{ij} + \partial^i \bar C^{j0} + \partial^j \bar C^{0i} \big )\,
\big (\partial_i \beta_j - \partial_j \beta_i \big ).
\end{eqnarray}
It is now clear that if we apply the 
nilpotent BRST symmetry transformations (4) on the {\it first} term of the above equation, it will cancel out with (33). Thus, the
{\it first} term of (34) would be present in the expression for $Q_B$.  In other words, besides (29), 
we have obtained {\it two} terms of the nilpotent version of the
BRST charge $Q_B$ from the non-nilpotent version of the Noether BRST charge $Q_b$. These 
very {\it special} two terms are as follows:
\begin{eqnarray}\label{35} 
+  \dfrac{1}{2}\, \int d^3 x\, \Big [\big (\partial^0 \bar C^{ij} + \partial^i \bar C^{j0} + \partial^j \bar C^{0i} \big )\,
\big (\partial_i \beta_j - \partial_j \beta_i \big ) - \big (\partial^0 B^{ij} + \partial^i B^{j0}  + \partial^j B^{0i} \big ) \, C_{ij} \Big ].
\end{eqnarray}
It can be readily checked that the above combination of terms is BRST invariant. At this stage, we focus 
on the {\it second} term of (34)  which is nothing but the following:
\begin{eqnarray}\label{36} 
-\,2\,   \int d^3 x\, \big (\partial^0 \bar C^{ij} + \partial^i \bar C^{j0} + \partial^j \bar C^{0i} \big )\,
\big (\partial_i \beta_j \big ).
\end{eqnarray}
We carry out the partial integration 
in the above integral and exploit the theoretical strength of the Gauss divergence theorem which {\it together} imply
the following
\begin{eqnarray}\label{37} 
+\,2\,   \int d^3 x\, \partial_i \big (\partial^0 \bar C^{ij} + \partial^i \bar C^{j0} + \partial^j \bar C^{0i} \big )\,
\beta_j  = \int d^3 x\, \big (\partial^0 \bar F^j - \partial^j \bar F^0 \big )\, \beta_j,
\end{eqnarray}
where we have used the EL-EoM: $\partial_\mu \big (\partial^\mu \bar C^{\nu\sigma} + \partial^\nu \bar C^{\sigma\mu} 
+ \partial^\sigma \bar C^{\mu\nu} \big ) = -\, \frac{1}{2}\, (\partial^\nu \bar F^\sigma -  \partial^\sigma \bar F^\nu)$ that
has been derived from the Lagrangian density ${\cal L}_{(B)} $ [cf. Eq. (C.5)]. It is obvious that the above term would be present in the 
nilpotent version of the BRST charge $Q_B$. Thus, it is crystal clear that, so far, we have already obtained {\it three} integrals [i.e. (29), (35) and (37)]
which are part and parcel of the off-shell nilpotent version of the BRST charge $Q_B$.

To accomplish the goal of obtaining $s_b Q_B = 0$, we have to apply the BRST symmetry transformations (4) on the final form of (37) which leads to
the following
\begin{eqnarray}\label{38} 
s_b \Big [\int d^3 x\, \big (\partial^0 \bar F^i - \partial^i \bar F^0 \big )\, \beta_i \Big ] \equiv
-\, \int d^3 x\, \big (\partial^0 \bar F^i - \partial^i \bar F^0 \big )\,\partial_i C_2,
\end{eqnarray}
where we have taken into account the fermionic nature of $s_b$ and $\bar F_\mu$. This term [i.e. the final form of (38)] has to cancel out
with the modified form of the suitable term that is chosen from the expression for $Q_b$ [cf. Eq. (23)]. Such a term is : $- \, \frac{1}{2} \int\,
(\partial^0 \bar \beta^i - \partial^i \bar \beta^0)\, \partial_i C_2$. We modify this term by applyig the 
very simple algebraic trick, namely; 
\begin{eqnarray}\label{39} 
-\, \dfrac{1}{2}\, \int d^3 x\, \big (\partial^0 \bar \beta^i - \partial^i \bar \beta^0\big )\,\partial_i C_2 =
&+& \int d^3 x\, \big (\partial^0 \bar \beta^i- \partial^i \bar \beta^0\big )\,\partial_i C_2 
\nonumber\\ &-& \dfrac{3}{2}\,
\int d^3 x\, \big (\partial^0 \bar \beta^i- \partial^i \bar \beta^0\big )\,\partial_i C_2.
\end{eqnarray}
It is clear that when $s_b$ acts on the {\it first} term of (39), the ensuing result cancels out with the final form of (38). Hence, the
first term of (39) will be part and parcel of the {\it final} expression for the nilpotent version of the BRST charge $Q_B$. As a side remark, we
would like to add that, so far,  we have obtained {\it four} very crucial terms of the nilpotent version of the BRST charge $Q_B$ which
are nothing but the sum of (29), (35), (37) and the {\it first} term of (39). We concentrate now on the {\it second} term of (39) which becomes 
\begin{eqnarray}\label{40} 
+ \dfrac{3}{2}\, \int d^3 x\; \partial_i \big (\partial^0 \bar \beta^i - \partial^i \bar \beta^0\big )\, C_2 = + \dfrac{3}{2}\,
\int d^3 x\, \dot B_5\, C_2,
\end{eqnarray}
where we have  (i) carried out the partial integration followed by the Gauss divergence theorem, and (ii) used the EL-EoM:
$\partial_\mu (\partial^\mu \bar \beta^\nu - \partial^\nu \bar \beta^\mu) + \partial^\nu B_5 = 0 $. The above integral (40) is also a
BRST invariant quantity [i.e. $s_b (\dot B_5 \,C_2) = 0 $]. Hence, it will be present in the expression for the nilpotent BRST charge $Q_B$. The
{\it final} expression for the nilpotent (i.e. $Q_B^2 = 0 \Leftrightarrow s_b Q_B = i\, \{Q_B, \; Q_B \} = 0$) version of the BRST charge $Q_B$ is as follows
\begin{eqnarray}\label{41}
Q_B &=&   \int d^3 x\, \Big [\dfrac{1}{2}\, \Big \{(\partial^0 C^{ij} + \partial^i C^{j0} + \partial^j C^{0i})\, B_{ij} -
(\partial^0 B^{ij} + \partial^i B^{j0} + \partial^j B^{0i})\, C_{ij}
 + B^{0i}\,f_i \nonumber\\
 &+&  3\, \dot B_5 \, C_2 
- B_5 \, \dot C_2 - (\partial^0 \beta^i - \partial^i \beta^0)\, \bar F_i 
+ (\partial^0 \bar C^{ij} + \partial^i \bar C^{j0} + \partial^j \bar C^{0i})\,(\partial_i \beta_j
- \partial_j \beta_i )  \nonumber\\
&+& B_4\, \bar F^0 + B_2\, f^0 \Big \}
+  (\dot B \, C - B \, \dot C) + (\partial^0 \bar F^i - \partial^i \bar F^0)\, \beta_i  
+ (\partial^0 \bar \beta^i - \partial^i \bar \beta^0)\,\partial_i C_2 \Big ],
\end{eqnarray}
where, besides the integrals (29), (35), (37), the {\it first} term of (39) and (40), the rest of the terms 
(that exist in the above equation)  are the BRST invariant terms that are 
already present in the Noether conserved BRST charge $Q_b$ [cf. Eq. (23)].

We conclude this Subsec. with the following concluding remarks. First of all, we have exploited only the partial integration followed by the Gauss
divergence theorem and theoretical strength of the EL-EoMs that are derived from the BRST-invariant Lagrangian density ${\cal L}_{(B)}$. Hence,
the non-nilpotent Noether BRST charge $Q_b$ and the modified nilpotent version of the BRST charge $Q_B$ {\it both} are conserved quantities. Second,
the nilpotency (i.e. $s_b Q_B = + i\, \{Q_B, Q_B \} = 0 \Rightarrow Q_B^2 = 0 $) of the BRST charge $Q_B$ plays a key role in the discussions
on (i) the BRST cohomology (see, e.g. [42]), and (ii) the physicality criterion w.r.t. the BRST charge (cf. Subsec. 4.3 below). Finally, 
the conserved {\it Noether} BRST charge is the generator for the infinitesimal, continuous and nilpotent BRST symmetry transformations (4). Hence,
we note that {\it both} kinds of BRST charges (i.e. $Q_b$ and $Q_B$) 
have their own importance and relevance within the framework of BRST formalism. \\



\subsection{Physicality Criterion: Nilpotent BRST Charge}

We have seen that there are primary and secondary constraints on our theory at the {\it classical} level (cf. Appendix A) where the
D-dimensional Lagrangian density ${\cal L}_{(0)}^{(D)} $ [cf. Eq. (A.1)] for the combined field-theoretic system of the 
{\it free} Abelian 3-form and 1-form gauge 
theories has been considered. The 4D BRST-quantized version of this theory is described by the Lagrangian density: ${\cal L}_{(B)} 
=  {\cal L}_{(NG)} + {\cal L}_{(FP)}$ [cf. Eqs. (1),(3)] which does {\it not} have any kinds of constraints. To be precise, we note that the 
following canonical conjugate momenta (i.e. $\Pi^{\mu}_{(A1)}, \; \Pi^{\mu\nu\sigma}_{(A3)}$)
w.r.t. the Abelian 1-form  and 3-form gauge fields, namely;
\begin{eqnarray}\label{42}
\Pi^{\mu}_{(A1)} = \frac {\partial {\cal L}_{(B)}}{\partial (\partial_0 A_{\mu})} = 
\dfrac{1}{2}\, \varepsilon^{0\mu\nu\sigma}\, {\cal B}_{\nu\sigma} - \eta^{0\mu}\, B \;\;\Longrightarrow \;\; \Pi^{0}_{(A1)}  = -\, B, \qquad
\Pi^{i}_{(A1)} = \dfrac{1}{2}\, \varepsilon^{0ijk}\, {\cal B}_{jk}, 
\end{eqnarray}
\begin{eqnarray}\label{43}
\Pi^{\mu\nu\sigma}_{(A3)} = \frac {\partial {\cal L}_{(B)}}{\partial (\partial_0 A_{\mu\nu\sigma})} &=& \frac {1}{3!}\, 
\Big [ \varepsilon^{0\mu\nu\sigma}\, B_1 + \big (\eta^{0\mu} B^{\nu\sigma} + \eta^{0\nu} B^{\sigma\mu} + \eta^{0\sigma} B^{\mu\nu} \big ) \Big ]
\quad \Longrightarrow \quad 
\nonumber\\
&& \Pi^{0ij}_{(A3)}  = \frac {1}{3!}\, B^{ij} \equiv \frac {1}{3!}\, B_{ij}, \qquad \Pi^{ijk}_{(A3)}  = \frac {1}{3!}\,\varepsilon^{0ijk}\, B_1,
\end{eqnarray}
are {\it non-zero} which were {\it weakly} zero at the classical level (cf. Appendix A). Here we observe that 
(i) it is the non-ghost (i.e. ${\cal L}_{(NG)} $) part of the BRST-invariant Lagrangian density  ${\cal L}_{(B)}$
that contributes to the definitions of the canonical conjugate momenta [cf. Eqs. (42) and (43)], and (ii) the primary constraints (i.e. $\Pi^{0}_{(A1)}  \approx 0,\,
\Pi^{0ij}_{(A3)} \approx 0$) of the free {\it calssical} field-theoretic system of the Abelian 1-form and 3-form gauge theories have been traded with
the Nakanishi-Lautrup auxiliary fields (i.e. $\Pi^{0}_{(A1)}  = -\, B,\; \Pi^{0ij}_{(A3)}  = \frac {1}{3!}\, B_{ij}$) of our present BRST-quantized 4D theory.
The secondary constraints (i.e. $\partial_i \Pi ^{ijk}_{(A3)} \approx 0,\; \partial_i \Pi ^{i}_{(A1)} \approx 0$) of the {\it classical} 4D theory
(cf. Appendix A) appear in the EL-EoMs that are derived from the BRST-quantized version of the Lagrangian density ${\cal L}_{(B)}$. To corroborate this statement,
we note that the following EL-EoMs emerge out from the Lagrangian density ${\cal L}_{(B)}$, namely;
\begin{eqnarray}\label{44}
 \dfrac{1}{3!}\, \varepsilon^{\mu\nu\sigma\rho}\,\partial_\rho B_1 &=&  \dfrac{1}{3!}\, \big (\partial^\mu B^{\nu\sigma} + \partial^\nu B^{\sigma\mu} 
 + \partial^\sigma B^{\mu\nu} \big ) \nonumber\\
&\Longrightarrow& \;\partial_k \Pi^{kij}_{(A3)} = \dfrac{1}{3!}\, 
\big (\partial^0 B^{ij} + \partial^i B^{j0}  + \partial^j B^{0i} \big ), \nonumber\\
\dfrac{1}{2}\, \varepsilon^{\mu\nu\sigma\rho}\,\partial_\nu {\cal B}_{\sigma\rho} &=& \partial^\mu B \;\;\Longrightarrow \;\; \partial_i \Pi^i_{(A1)} = \dot B,
\end{eqnarray}
where we have chosen $\mu = 0, \, \nu = i, \, \sigma = j$ in the top entry and $\mu = 0$ in the bottom one.
It is clear that the above relations are connected with the secondary constraints of our 4D {\it classical} 
combined field-theoretic system of the Abelian 3-form and 1-form gauge theories (cf. Appendix A). In other words, the specific combinations of the
derivatives on the Nakanishi-Lautrup type auxiliary fields of the 4D BRST-quantized theory are connected with the secondary constraints 
of the 4D {\it classical} version of our theory. Within the framework of BRST formalism, we discuss the physicality criterion w.r.t. the nilpotent
(i.e. $Q_B^2 = 0$) version of the BRST charge $Q_B$ and show that the operator forms of {\it all} the primary and secondary constraints  of 
our 4D {\it classical} theory annihilate the physical states (i.e $|phys>$), existing in the {\it total} quantum Hilbert space of states, at the {\it quantum} level.

We observe that, right from the beginning, the (anti-)ghost fields of our 4D theory are {\it decoupled} from the rest of the 
physical fields of our theory (cf. coupled Lagrangian densities  ${\cal L}_{(B)}$ and  ${\cal L}_{(\bar B)}$). Hence, the 
quantum states (existing in the {\it total} quantum Hilbert space of states) are direct product [46] of the physical states (i.e $|phys>$) and the ghost states
(i.e $|ghost>$). We note that, in the expression for the nilpotent version of the BRST charge $Q_B$ [cf. Eq. (41)], we have the physical fields (with ghost number
equal to zero) and the basic as well as the auxiliary (anti-)ghost fields (with non-zero ghost number associated with everyone of them). When
the operator form of the specific ghost field acts on the ghost state (i.e.  $|ghost>$), we obtain a non-zero result. Thus, within the framework 
of the BRST formalism, the physicality criterion w.r.t. the nilpotent BRST charge $Q_B$ requires that: $Q_B \, |phys> = 0$ where the operator forms of the
physical fields (with the ghost number equal to zero) act on the physical states (i.e. $|phys>$). However, these 
specific physical fields {\it must} be associated
with the {\it basic} (anti-)ghost fields of our theory. Taking into account these inputs,
we observe that  $Q_B \, |phys> \,= 0$ implies 
\begin{eqnarray}\label{45}
B \, |phys> \,= 0, &\Longrightarrow& (\partial \cdot A) \,|phys>\, = 0 \qquad \Longrightarrow \qquad \Pi^{0}_{(A1)} |phys>\, = 0, \nonumber\\
\dot B \, |phys> \, = 0, &\Longrightarrow& \dfrac{1}{2}\, \varepsilon^{0ijk}\, \partial_i {\cal B}_{jk} \,|phys>\, = 0 \quad  \Longrightarrow \quad 
\partial_i \Pi^i_{(A1)} \, |phys>\, = 0,
\end{eqnarray}
as far as the BRST-quantized 4D Abelian 1-form gauge theory (hidden in the {\it combined} 
field-theoretic system of the BRST-quantized 4D Abelian 1-form and 3-form gauge theories) is concerned. In the above, we have used the definition of the canonical conjugate momenta [cf. Eq. (42)] and the EL-EoM for the Abelian 1-form field 
[cf. Eq. (44)] that are derived from the BRST-invariant Lagrangian density ${\cal L}_{(B)}$. Thus, we note that the 
{\it operator} forms of the primary and secondary first-class constraints of the {\it classical}  Abelian 1-form gauge theory (cf. Appendix A) annihilate
the physical states (i.e. $|phys>$) when we demand the physicality criterion w.r.t. the nilpotent version of the BRST charge $Q_B$. This observation is consistent
with the Dirac quantization conditions for the physical theories that are endowed with constraints of any variety [33]. In other words, within the framework of BRST 
formalism, we demand that the physical states (i.e. $|phys>$)
are {\it those} that are annihilated by the nilpotent version of the conserved BRST charge.

At this stage, we focus now on the portion of the nilpotent version of the BRST charge $Q_B$ which contains terms that are 
associated with the BRST-quantized
version of the Abelian 3-form gauge theory (that is hidden in our {\it combined} 4D field-theoretic system). It turns out that the physicality
criterion: $Q_B \, |phys> \,= 0$ implies the following
\begin{eqnarray}\label{46}
\dfrac{1}{2}\, B_{ij} \, |phys> \,= 0, \qquad &\Longrightarrow& \qquad  3\,\Pi^{0ij}_{(A3)} |phys>\, = 0, \nonumber\\
\dfrac{1}{2}\, \big (\partial^0 B^{ij} + (\partial^i B^{j0}  + (\partial^j B^{0i}  \big ) \, |phys> \, = 0, 
&\Longrightarrow& \dfrac{1}{2}\, \varepsilon^{0ijk}\, \partial_k B_1 \,|phys> = 0 \nonumber\\
 &\Longrightarrow& 3\,\partial_k \Pi^{kij}_{(A3)} \, |phys> = 0,
\end{eqnarray}
where we have used (i) the definitions of the canonical conjugate momenta [cf. Eq. (43)], and (ii) the EL-EoM w.r.t. the Abelian 3-form
gauge field $A_{\mu\nu\sigma}$ [cf. Eq. (44)]. Thus, we observe that the physicality criterion (i.e. $Q_B \, |phys>\, = 0$) w.r.t. the
nilpotent version of the BRST charge $Q_B$ [cf. Eq. (41)] leads to the annihilation of the physical states (i.e. $|phys>$) by the
operator forms of the primary and secondary first-class constraints of the {\it classical} free Abelian 3-form gauge theory
(cf. Appendix A) at the {\it quantum} level within the framework of BRST formalism. This specific observation is consistent
with the Dirac quantization conditions for the gauge theories that are endowed with the first-class constraints
(as is the case with our {\it combined} 4D field-theoretic system of gauge theories).

We end this subsection with a couple of remarks. First of all, we note that there is presence of the terms $B^{0i}\, f_i $ and $B_2 \, f^0$ 
in the expression for the 
nilpotent BRST charge $Q_B$ [cf. Eq. (41)] where the auxiliary fields
$B^{0i}$ and  $B_2$ carry the ghost numbers equal to zero. However, these are associated with
the specific components of the fermionic {\it auxiliary} vector field $f^\mu$ which is {\it not} the
{\it basic} ghost field of our theory. Thus, these auxiliary fields  
are {\it not} allowed to act on the physical states (i.e. $|phys>$). This is why, we do {\it not} have any 
conditions on the physical states (i.e. $|phys>$) from {\it these} terms\footnote{It is worthwhile to mention
that $B_2$ and $B^{0i}$ are the zeroth and space components of the canonical conjugate momenta (i.e. $\Pi^\mu_{(\phi)} $)  w.r.t. the 
vector field $\phi_\mu$. These are {\it not} the constraints on our 4D {\it classical} field-theoretic system
(cf. Appendix A). Hence, they are {\it not} supposed to put any condition on the physical states (i.e. $|phys>$) of our theory
unlike the conditions that emerge out in equations (45) and (46).}. Second, if we demand the physicality criterion 
(i.e. $Q_b \, |phys> = 0$) w.r.t. the Noether conserved BRST charge $Q_b$, we observe that {\it only} the operator forms of the primary
constraints (i.e. $B\, |phys> \, = 0, \; B_{ij}\, |phys> \, = 0 $) annihilate the physical states in a straightforward fashion
(without any application of the partial integration followed by the Gauss divergence theorem). Furthermore, the non-nilpotency 
(i.e. $Q_b^2 \neq 0 $) of the conserved Noether BRST charge $Q_b$ is {\it not} suitable as far as  the serious discussion on 
a few key aspects of the BRST cohomology is concerned (cf. e.g. [42] for details). \\


\section{Anti-BRST Current and Anti-BRST Charge}

Our present section contains {\it three} subsections. In subsection 5.1, we derive the Noether conserved anti-BRST current ($J^\mu_{(ab)}$)
using  the  off-shell nilpotent anti-BRST symmetry transformations (6) and obtain the conserved Noether anti-BRST charge $Q_{ab}$. We
show that the {\it latter} is {\it not} nilpotent of order two (i.e. $Q_{ab}^2 \neq 0$). Our subsection 5.2 is devoted to the derivation of the
off-shell nilpotent (i.e. $ Q_{AB}^2 = 0$) version of the anti-BRST charge $Q_{AB}$ from the non-nilpotent 
anti-BRST charge $Q_{ab}$. Finally, in our subsection 5.3, we {\it briefly} discuss the 
physicality criterion w.r.t. the nilpotent version of the anti-BRST charge $Q_{AB}$.\\

\subsection{Noether Conserved Anti-BRST Current and Charge}

We exploit here the theoretical tricks and techniques of the Noether theorem to derive the explicit expression for the 
Noether anti-BRST current due to our observation in equation (7). The general expression for the anti-BRST current, in this context,
is 
\begin{eqnarray}\label{47}
J^\mu_{(ab)} &=&  s_{ab} \Phi_i \; \dfrac {\partial {\cal L}_{(\bar B)}}{\partial (\partial_\mu \Phi_i)} + B\, \partial^\mu \bar C 
- \dfrac{1}{2}\, \big [B_4 \, \partial^\mu \bar C_2  + \bar B^{\mu\nu}\, \bar f_\nu + B_2\, \bar f^\mu \big ]
\nonumber\\
&+& \dfrac{1}{2}\, \big [(\partial^\mu C^{\nu\sigma} + \partial^\nu C^{\sigma\mu} + \partial^\sigma C^{\mu\nu})\, \bar B_{\nu\sigma}
+ (\partial^\mu \bar \beta^\nu - \partial^\nu \bar \beta^\mu)\,  F_\nu  + B_5  F^\mu
  \big],
\end{eqnarray}
where the symbol $s_{ab} \Phi_i $ stands for the infinitesimal, continuous and off-shell 
nilpotent anti-BRST symmetry transformations (6) on the generic {\it dynamical} field 
$\Phi_i \equiv A_{\mu\nu\sigma},  C_{\mu\nu}, \bar C_{\mu\nu}, A_\mu, \phi_\mu, \tilde \phi_\mu, \beta_\mu, \bar \beta_\mu, C_1, \bar C_1, C, \bar C $ 
and all the rest of the terms (in the above equation) have been picked up from our observation in (7). The first term in the above
equation can be computed explicitly from the Lagrangian density ${\cal L}_{(\bar B)} = 
{\cal L}_{(ng)} + {\cal L}_{(fp)} $ which is the sum of the 
Lagrangian densities that are present in the non-ghost sector [cf. Eq. (8)] and
the FP-ghost sector [cf. Eq. (9)], respectively. The final expression for the Noether anti-BRST current ($J^\mu_{(ab)}$) is as follows:
\begin{eqnarray}\label{48}
J^\mu_{(ab)} &=&  \dfrac{1}{2}\, \Big [\varepsilon^{\mu\nu\sigma\rho}\, B_1 (\partial_\nu \bar C_{\sigma\rho}) 
-  (\partial^\mu  \bar C^{\nu\sigma} + \partial^\nu  \bar C^{\sigma\mu} + \partial^\sigma \bar C^{\mu\nu})\, \bar B_{\nu\sigma}
 + \bar B^{\mu\nu}\,\bar f_\nu  
\nonumber\\
&-& \varepsilon^{\mu\nu\sigma\rho}\, (\partial_\nu \bar C)\, \bar {\cal B}_{\sigma\rho} + B_4 \, \partial^\mu \bar C_2 - (\partial^\mu  \beta^\nu 
- \partial^\nu \beta^\mu)\, \partial_\nu \bar C_2 - (\partial^\mu \bar \beta^\nu - \partial^\nu \bar \beta^\mu)\,  F_\nu  \nonumber\\
&-& B_5\,  F^\mu + B_2 \, \bar f^\mu +(\partial^\mu  C^{\nu\sigma} + \partial^\nu  C^{\sigma\mu} + \partial^\sigma C^{\mu\nu})\,(\partial_\nu \bar \beta_\sigma 
- \partial_\sigma \bar \beta_\nu) \Big ] - B \, \partial^\mu \bar C.
\end{eqnarray}
It is worthwhile to point out that (i) our observations in 
transformations (6) ensure that: $s_b \widetilde \phi_\mu = 0, s_b \bar C_2 = 0, \, s_b \bar C = 0$ which
imply that the dynamical fields $\widetilde \phi_\mu, \bar C_2, \bar C$ do {\it not} contribute anything to $J^\mu_{(ab)} $, (ii) all the
bosonic/fermionic auxiliary fields of our theory, for obvious reasons, do {\it not} contribute anything in the computation of $J^\mu_{(ab)} $
and (iii) for the algebraic convenience, the {\it second} term of the equation (48) has been chosen as:
$- \, (\partial^\mu  \bar C^{\nu\sigma} + \partial^\nu  \bar C^{\sigma\mu} + \partial^\sigma \bar C^{\mu\nu})\, \bar B_{\nu\sigma} $
instead of : $-\, (\eta^{\mu\nu}  \bar B^{\nu\sigma} + \eta^{\nu\sigma}  \bar B^{\sigma\mu} 
+ \eta^{\mu\rho} \bar B^{\mu\nu})\, (\partial _\nu \bar C_{\sigma\rho})$ which turns up in the actual computation.
It goes without saying that both these expressions are equal [i.e. $(\eta^{\mu\nu}  \bar B^{\sigma\rho} + \eta^{\mu\sigma} \bar B^{\rho\nu} 
+ \eta^{\mu\rho} \bar B^{\nu\sigma})\, (\partial_\nu \bar C_{\sigma\rho})
= (\partial^\mu \bar C^{\nu\sigma} + \partial^\nu \bar C^{\sigma\mu} + \partial^\sigma \bar C^{\mu\nu})\, \bar B_{\nu\sigma} $].
The conservation law (i.e. $\partial_\mu J^\mu_{(ab)} = 0$) of the Noether anti-BRST current can be proven by taking into account the appropriate EL-EoMs that
have been listed in (17), (C.5) and (C.6). It is worthwhile, at this stage, to point out that the following EL-EoMs that emerge out from the non-ghost sector [cf. Eq. (8)] of the perfectly anti-BRST invariant Lagrangian density ${\cal L}_{(\bar B)} $, namely;
\begin{eqnarray}\label{49}
\big (\partial^\mu \bar B^{\nu\sigma} + \partial^\nu \bar B^{\sigma\mu}  + \partial^\sigma \bar B^{\mu\nu} \big ) 
+ \varepsilon^{\mu\nu\sigma\rho}\, \partial_\rho B_1 = 0, \qquad
\partial_\mu \bar  B^{\mu\nu} + \partial^\nu B_2 = 0,
\end{eqnarray}
 are very useful in the above proof of the conservation law. As a side remark, we would like to state that the first entry in the equation  (49)
is nothing other than our equation (17) which can be also re-expressed as: $\varepsilon^{\mu\nu\sigma\rho}\, \partial_\mu B_1 
- \big (\partial^\nu \bar B^{\sigma\rho} + \partial^\sigma \bar B^{\rho\nu}  + \partial^\rho \bar B^{\nu\sigma} \big )  = 0$.

The above conserved Noether anti-BRST current current leads to the definition of the conserved Noether anti-BRST charge:
$Q_{ab} = \int d^3 x\, J^0_{(ab)}$ which can be expressed as follows:
 \begin{eqnarray}\label{50}
Q_{ab}  &=&   \int d^3 x\, \Big [\dfrac{1}{2}\, \Big \{\varepsilon^{0ijk}\, B_1 (\partial_i \bar C_{jk}) 
-  \big (\partial^0 \bar C^{ij} + \partial^i \bar C^{j0} + \partial^j \bar C^{0i} \big )\, \bar B_{ij} 
 + \bar B^{0i}\,\bar f_i  
\nonumber\\
&-& \varepsilon^{0ijk}\, (\partial_i \bar C)\, \bar {\cal B}_{jk} + B_4 \, \partial^0 \bar C_2 - (\partial^0  \beta^i
- \partial^i \beta^0)\, \partial_i \bar C_2 - (\partial^0 \bar \beta^i - \partial^i \bar \beta^0)\,  F_i  \nonumber\\
&-& B_5\,  F^0 + B_2 \, \bar f^0 +(\partial^0  C^{ij} + \partial^i  C^{j0} + \partial^j C^{0i})\,(\partial_i \bar \beta_j
- \partial_j \bar \beta_i) \Big \} - B \, \partial^0 \bar C \Big ].
\end{eqnarray}
This Noether conserved charge turns out to be the generator for the infinitesimal, continuous and nilpotent anti-BRST symmetry transformations
(6). For this purpose, we have to exploit the general relationship between the infinitesimal and continuous symmetry transformations and
their generator as the Noether conserved charge [cf. Eq. (24)]. In other words, to corroborate this statement, we have to replace:
$s_b \Rightarrow s_{ab}, \; Q_b \Rightarrow Q_{ab}$ in (24). It is very interesting to point out, at this juncture, that the following is true, namely; 
\begin{eqnarray}\label{51}
s_{ab} \, Q_{ab} = +\, i\, \big \{Q_{ab}, \; Q_{ab} \big \} &\equiv& \dfrac{1}{2}\,
\int d^3 x \Big [(\partial^0  \bar B^{ij} + \partial^i  \bar B^{j0} + \partial^j \bar B^{0i})\,(\partial_i \bar \beta_j - \partial_j \bar \beta_i) \nonumber\\
&-& (\partial^0  F^i - \partial^i F^0)\, \partial_i \bar C_2   \Big ] \neq 0.
\end{eqnarray}
In the above, we have computed explicitly the l.h.s. by {\it directly} applying the anti-BRST transformations (6) on expression for the Noether
anti-BRST charge [cf. Eq. (50)]. In other words, we observe that the Noether anti-BRST charge is {\it not} nilpotent of order two because of 
the fact that: $i\, \big \{Q_{ab}, \; Q_{ab} \big \} = 2\, i\, Q_{ab}^2 \neq 0 $. Thus, it is evident that, as far as the cohomological aspects
of our BRST-quantized theory is concerned (see, e.g. [42]), the Noether conserved anti-BRST charge is {\it not} suitable because it is non-nilpotent. 
As a side remark, we would like mention that if we use the proper components of the EL-EoMs (49) and (C.6), we can prove that the r.h.s. of
(51) is a 3D space integral on the total space derivative terms. Hence, the application of the Gauss divergence theorem will make
the r.h.s. of (51) equal to zero. However, the Noether anti-BRST charge $Q_{ab}$ would still remain the (i) non-anti-BRST invariant  
(i.e. $s_{ab} Q_{ab} \neq 0 $), and (ii) non-nilpotent (i.e. $Q_{ab}^2 \neq 0 $) quantity because we are dealing
with the off-shell nilpotent anti-BRST transformations [cf. Eq. (6)] where the nilpotency property does {\it not} depend on the use of EL-EoMs.

We wrap-up this subsection with a few concluding remarks. We have seen that the anti-BRST symmetry transformations are nilpotent of order two. These
are used in the computation of the conserved Noether current $J^\mu_{(ab)} $ from which the conserved Noether anti-BRST charge is derived. However,
it turns out that the Noether anti-BRST charge itself is {\it not} nilpotent of order two. The reason behind such an observation is the existence of the
non-trivial CF-type restrictions [cf. Eq. (B.1)] on our theory. A thorough discussion on this issue has been performed in our earlier work [37]
where we have also proposed a systematic theoretical method to obtain the {\it conserved} and nilpotent versions of the (anti-) BRST charges
from their counterpart conserved (but non-nilpotent) Noether (anti-)BRST charges. In the next subsection, we shall perform this exercise. Finally,
we would like to mention that the nilpotent version of the anti-BRST charge is essential in the context of (i) the physicality criterion w.r.t. to it
(cf. Subsec. 5.3 below),
and (ii) the discussion on the cohomologcal aspects of our theory in the language of it (see, e.g. [42] for details).\\

\subsection{Nilpotent Anti-BRST Charge $Q_{AB}$ from Noether Charge $Q_{ab}$ }

Against the backdrop of our thorough discussions in our subsection 4.2, we shall be {\it brief} here as the theoretical arguments are similar
as far as the derivation of the nilpotent (i.e. $Q_{AB}^2 = 0 $) version of the anti-BRST charge  $Q_{AB}$ from the non-nilpotent
(i.e. $Q_{ab}^2 \neq 0 $) version of the Noether anti-BRST charge $Q_{ab}$ is concerned. In this context, first of all, we concentrate on the
following terms of $Q_{ab}$ [cf. Eq. (50)] that are connected with the BRST-quantized Abelian 1-form gauge theory, namely;  
 \begin{eqnarray}\label{52}
\int d^3 x\, \Big [-\,\dfrac{1}{2}\, \varepsilon^{0ijk}\, (\partial_i \bar C)\, \bar {\cal B}_{jk} - B \, \partial^0 \bar C \Big ].
\end{eqnarray}
Using the partial integration and exploiting the theoretical strength of the Gauss divergence theorem, we have the following:
 \begin{eqnarray}\label{53}
\int d^3 x\, \Big [ +\,\dfrac{1}{2}\, \varepsilon^{0ijk}\,  \bar C\, (\partial_i \bar {\cal B}_{jk}) - B \, \partial^0 \bar C \Big ]
\equiv \int d^3 x\, \big (\dot B \, \bar C - B\, \dot {\bar C} \big ),
\end{eqnarray}
where we have the EL-EoM: $
\frac{1}{2}\, \varepsilon^{\mu\nu\sigma\rho} \, \partial_\nu \bar {\cal B}_{\sigma\rho} = \partial^\mu B$ 
which emerge out from the Lagrangian density ${\cal L}_{(\bar B)}$. The above form of the integral is the {\it standard} form of the BRST charge
for the BRST-quantized Abelian 1-form gauge theory. It is straightforward to note that the above integral is anti-BRST invariant quantity. Hence,
this integral will be present in the {\it final} expression for the nilpotent  (i.e. $Q_{AB}^2 = 0 $) version of the anti-BRST charge $Q_{AB}$.

We focus on the following term of the Noether anti-BRST charge $Q_{ab}$ [cf. Eq. (50)]
\begin{eqnarray}\label{54}
\dfrac{1}{2}\, \int d^3 x\, \varepsilon^{0ijk}\, B_1 (\partial_i \bar C_{jk}) = -\, 
\dfrac{1}{2}\, \int d^3 x\, \varepsilon^{0ijk}\, (\partial_i B_1) \, \bar C_{jk},
\end{eqnarray}
where we have performed the partial integration and taken into account the Gauss divergence theorem. Using the EL-EoM:
$ \varepsilon^{\mu\nu\sigma\rho}\, \partial_\mu B_1 
= -\, \big (\partial^\nu \bar B^{\sigma\rho} + \partial^\sigma \bar B^{\rho\nu}  + \partial^\rho \bar B^{\nu\sigma} \big )$, the above integral can be 
re-expressed as: 
\begin{eqnarray}\label{55}
-\,\dfrac{1}{2}\, \int d^3 x\, \varepsilon^{0ijk}\, (\partial_i B_1)  \bar C_{jk} = +\,
\dfrac{1}{2}\, \int d^3 x\,\big (\partial^0 \bar B^{ij} + \partial^i\bar B^{j0}  + \partial^j \bar B^{0i} \big )\, \bar C_{ij}.
\end{eqnarray}
This term would be present in the nilpotent version of the anti-BRST charge $Q_{AB}$. Toward our main goal to obtain $s_{ab} Q_{AB} = 0$, we have to
apply the anti-BRST symmetry transformations (6) on the above integral which leads to the following
\begin{eqnarray}\label{56}
\dfrac{1}{2}\, \int d^3 x\,\big (\partial^0 \bar B^{ij} + \partial^i\bar B^{j0}  + \partial^j \bar B^{0i} \big )\, 
\big (\partial _i \bar \beta_j - \partial _j \bar \beta_i\big ),
\end{eqnarray}
which must cancel out from the modified form of the specifically chosen term of the 
Noether anti-BRST charge $Q_{ab}$ [cf. Eq. (50)]. This {\it specific} term is:
$\frac{1}{2} \int d^3 x \, \big (\partial^0 C^{ij} + \partial^i C^{j0}  + \partial^j C^{0i} \big )\, 
\big (\partial _i \bar \beta_j - \partial _j \bar \beta_i\big ) $ which can be modified, using the very simple algrebraic trick, as:
\begin{eqnarray}\label{57}
&&- \dfrac{1}{2}\, \int d^3 x\,\big (\partial^0 C^{ij} + \partial^i C^{j0}  + \partial^j C^{0i} \big )\, 
\big (\partial _i \bar \beta_j - \partial _j \bar \beta_i\big ) \nonumber\\
&& +  
\int d^3 x\,\big (\partial^0 C^{ij} + \partial^i C^{j0}  + \partial^j C^{0i} \big )\, 
\big (\partial _i \bar \beta_j - \partial _j \bar \beta_i\big ).
\end{eqnarray}
It is straightforward to note that the application of the anti-BRST transformations (6) on the {\it first} integral of the above equation
cancels out with the integral in (56). Hence, the {\it first}  term of the above integral would be present in the {\it final} 
expression for the nilpotent $Q_{AB}$. We now concentrate on the {\it second} term of (57) which can be re-expressed as:
\begin{eqnarray}\label{58}
+ \,2\,  
\int d^3 x\,\big (\partial^0 C^{ij} + \partial^i C^{j0}  + \partial^j C^{0i} \big )\, 
\big (\partial_i \bar \beta_j \big ) = -\, 2\, \int d^3 x\,\partial _i \big (\partial^0 C^{ij} + \partial^i C^{j0}  + \partial^j C^{0i} \big )\, \bar \beta_j.
\end{eqnarray}
Using the EL-EoM: $\partial_i (\partial^0 C^{ij} + \partial^i C^{j0}  + \partial^j C^{0i}) = -\, \frac{1}{2}\, (\partial^0 F^j - \partial^j F^0)$  [cf. Eq. (C.6)], 
the above equation can be re-written as follows
\begin{eqnarray}\label{59}
+ \int d^3 x\,\big (\partial^0 F^i - \partial^i F^0\big )\, \bar \beta_i,
\end{eqnarray}
which would be present in the {\it final} expression for the nilpotent version of the anti-BRST charge $Q_{AB}$. Thus, in our attempt to obtain $s_{ab} Q_{AB} = 0$,
we have to apply the anti-BRST symmetry transformations (6) on the above integral which leads to the following
\begin{eqnarray}\label{60}
- \int d^3 x\,\big (\partial^0 F^i - \partial^i F^0 \big )\, \partial_i \bar C_2,
\end{eqnarray}
where we have taken into account the useful 
anticommuting property (i.e. fermionic nature) of $s_{ab}$ and $F_\mu$. The above integral {\it must} cancel out with the modified
form of the {\it specifically} chosen term of the Noether conserved anti-BRST charge $Q_{ab}$ [cf. Eq. (50)]. Such a term is: $ - \frac{1}{2}\int d^3 x\,
(\partial^0 \beta^i - \partial^i \beta^0)\, \partial_i \bar C_2 $ which can be modified to the following form, namely;
\begin{eqnarray}\label{61}
+ \int d^3 x\,\big (\partial^0 \beta^i - \partial^i \beta^0\big )\, \partial_i \bar C_2 - \dfrac{3}{2}\,
\int d^3 x\,\big (\partial^0 \beta^i - \partial^i \beta^0\big )\, \partial_i \bar C_2.
\end{eqnarray}
It can be readily checked that the application of the anti-BRST symmetry transformations (6) on the {\it first} term cancels out with the integral (60).
Hence, the {\it first} integral of the above equation would be present in the {\it final} expression for the nilpotent version of $Q_{AB}$. Let us focus on the
{\it second} term of the above equation and (i) perform the partial integration, and (ii) apply the Gauss divergence theorem. The final result
of these two operations is
\begin{eqnarray}\label{62}
+ \dfrac{3}{2}\,
\int d^3 x\,\partial_i \big (\partial^0 \beta^i - \partial^i \beta^0\big )\, \bar C_2 = -\, \dfrac{3}{2}\,\int d^3 x\, (\dot B_4\, \bar C_2),
\end{eqnarray}
where we have used the EL-EoM: $ \partial_i \big (\partial^0 \beta^i - \partial^i \beta^0\big ) = -\, \dot B_4$ [cf. Eq. (C.5)]. This
integral is an anti-BRST invariant quantity [i.e. $s_b (\dot B_4\, \bar C_2) = 0 $]
which can be readily checked by the application of the anti-BRST symmetry transformations (6). Hence,
this integral will be present in the {\it final} expression for the nilpotent version of the anti-BRST charge $Q_{AB}$.

The precise total  expression for $Q_{AB}$ is the sum of integrals (53), (55), (59), (62) and the {\it first} terms of (57) and (61). The rest of the
terms are anti-BRST invariant and they are present in the non-nilpotent version of the Noether conserved charge $Q_{ab}$ [cf. Eq. (50)]. 
Mathematically, the nilpotent version of the anti-BRST charge $Q_{AB}$ can be expressed as:
 \begin{eqnarray}\label{63}
Q_{AB}  &=&   \int d^3 x\, \Big [ \dfrac{1}{2}\, \Big \{B_4 \, \dot {\bar C}_2 -  3 \, \dot B_4\, \bar C_2 
-  \big (\partial^0 \bar C^{ij} + \partial^i \bar C^{j0} + \partial^j \bar C^{0i} \big )\, \bar B_{ij} 
- (\partial^0 \bar \beta^i - \partial^i \bar \beta^0)\,  F_i  \nonumber\\
&+& (\partial^0  \bar B^{ij} + \partial^i \bar B^{j0} + \partial^j \bar B^{0i})\, \bar C_{ij} 
- (\partial^0  C^{ij} + \partial^i  C^{j0} + \partial^j C^{0i})\,(\partial_i \bar \beta_j - \partial_j \bar \beta_i) - B_2 \, \bar f^0 \nonumber\\
&+& \bar B^{0i}\,\bar f_i  - B_5\,  F^0 \Big \} 
+  (\dot B \, \bar C - B \, \dot {\bar C}) +  (\partial^0 \bar \beta^i - \partial^i \bar \beta^0) \, \partial_i \bar C_2
+ (\partial^0 F^i - \partial^i F^0)\, \bar \beta_i \Big ].
\end{eqnarray}
It can be readily checked that $s_{ab} Q_{AB} = 0$ is satisfied when we apply {\it directly} the anti-BRST transformations (6) on the
above expression. In the next subsection, we shall see that it is the nilpotent version of the anti-BRST charge $Q_{AB}$ that is
used in the physicality criterion to yield the results that are consistent with Dirac's quantization conditions for the theories
that are endowed with  constraints of any variety. The conserved Noether anti-BRST charge $Q_{ab}$, even though non-nilpotent
(i.e. $Q_{ab}^2 \neq 0 $), is the generator for the infinitesimal, continuous and nilpotent anti-BRST symmetry transformations (6). Hence,
both {\it forms} the charges (i.e. $Q_{AB}$ and $Q_{ab}$) are important and useful in their own ways.\\

\subsection{Nilpotent Anti-BRST Charge and Physicality Criterion}

Against the backdrop of our thorough discussions in the context of the physicality criterion w.r.t. the 
conserved and nilpotent BRST charge $Q_B$ in our Subsec. 4.3, we 
shall be {\it brief} in our present discussions that are
connected with the physicality criterion w.r.t. the nilpotent version of the anti-BRST charge $Q_{AB}$ and, ultimately, we shall show that we find
the {\it same} conditions on the physical states (i.e. $|phys>$) as we have obtained in our Subsec. 4.3. In other words, we observe that the operator forms
of the primary and secondary first-class constraints (of the 4D {\it classical} field-theoretic system of the combination of the free Abelian 3-form and 1-form
gauge theories) annihilate the physical states (i.e. $|phys>$) when we demand that the physical states are {\it those} that are annihilated by the 
conserved and nilpotent version of the anti-BRST charge $Q_{AB}$. This observation, as emphasized earlier, is consistent with the Dirac quantization conditions for the systems that are endowed with constraints of any variety [33]. To proceed further, first of all, we note that the expressions for the canonical conjugate
momenta w.r.t. the Abelian 3-form and 1-form gauge fields (that emerge out from the anti-BRST invariant Lagrangian density ${\cal L}_{(\bar B)}$) are 
\begin{eqnarray}\label{64}
\Pi^{\mu\nu\sigma}_{(A3)} = \frac {\partial {\cal L}_{(\bar B)}}{\partial (\partial_0 A_{\mu\nu\sigma})} &=& \frac {1}{3!}\, 
\Big [ \varepsilon^{0\mu\nu\sigma}\, B_1 - \big (\eta^{0\mu} \bar B^{\nu\sigma} + \eta^{0\nu} \bar B^{\sigma\mu} 
+ \eta^{0\sigma} \bar B^{\mu\nu} \big ) \Big ]
\quad \Longrightarrow \quad 
\nonumber\\
&& \Pi^{0ij}_{(A3)}  = -\, \frac {1}{3!}\, \bar B^{ij} \equiv -\, \frac {1}{3!}\, \bar B_{ij}, \qquad \Pi^{ijk}_{(A3)}  = \frac {1}{3!}\,\varepsilon^{0ijk}\, B_1,
\end{eqnarray}
\begin{eqnarray}\label{65}
\Pi^{\mu}_{(A1)} = \frac {\partial {\cal L}_{(\bar B)}}{\partial (\partial_0 A_{\mu})} = -\,
\dfrac{1}{2}\, \varepsilon^{0\mu\nu\sigma}\, \bar {\cal B}_{\nu\sigma} - \eta^{0\mu}\, B \Longrightarrow  \Pi^{0}_{(A1)}  = -\, B, \quad
\Pi^{i}_{(A1)} = -\, \dfrac{1}{2}\, \varepsilon^{0ijk}\, \bar {\cal B}_{jk}, 
\end{eqnarray}
which demonstrate that all the components of the conjugate momenta are {\it non-zero}. Hence, within the framework of the BRST formalism, the primary
constraints (i.e. $\Pi^{0ij}_{(A3)} \approx 0, \;  \Pi^{0}_{(A1)}  \approx 0$) of our 4D classical 
field-theoretic system (cf. Appendix A) have been traded with the Nakanishi-Lautrup auxiliary fields of the anti-BRST invariant 
Lagrangian density ${\cal L}_{(\bar B)}$. In exactly similar fashion, we observe that the secondary constraints 
(i.e. $\partial_i \Pi ^{ijk}_{(A3)} \approx 0,\; \partial_i \Pi ^{i}_{(A1)} \approx 0$) of the {\it classical} 4D theory are found to be hidden in the
EL-EoMs that are derived from the anti-BRST invariant Lagrangian density ${\cal L}_{(\bar B)}$ w.r.t. the Abelian 3-form gauge
field $A_{\mu\nu\sigma}$ and the Abelian 1-form gauge field $A_\mu$, respectively. To be more precise,  we note the following
\begin{eqnarray}\label{66}
 \dfrac{1}{3!}\, \varepsilon^{\mu\nu\sigma\rho}\,\partial_\rho B_1 &=& -\, \dfrac{1}{3!}\, 
\big (\partial^\mu \bar B^{\nu\sigma} + \partial^\nu \bar B^{\sigma\mu} 
 + \partial^\sigma \bar B^{\mu\nu} \big ) \nonumber\\
&\Longrightarrow& \;\partial_k \Pi^{kij}_{(A3)} = -\, \dfrac{1}{3!}\, 
\big (\partial^0 \bar B^{ij} + \partial^i \bar B^{j0}  + \partial^j \bar B^{0i} \big ), \nonumber\\
\dfrac{1}{2}\, \varepsilon^{\mu\nu\sigma\rho}\,\partial_\nu \bar {\cal B}_{\sigma\rho} &=& -\; \partial^\mu B \;\;\Longrightarrow \;\; 
\partial_i \Pi^i_{(A1)} = +\, \dot B,
\end{eqnarray}
where we have used the definitions of the conjugate momenta that are present in equations (64) and (65).
Thus, we observe that the secondary constraints (cf. Appendix A) of the classical 4D field-theoretic system are nothing but the specific combinations of the
derivatives on the Nakansihi-Lautrup auxiliary fields for the BRST-quantized version of the Abelian 3-form gauge theory and a first-order
time derivative on the Nakanishi-Lautrup auxiliary field $B$ in the case of the BRST-quantized version of the Abelian 1-form gauge theory (which
are hidden in our {\it present} BRST-quantized version of the 4D {\it combined} field-theoretic system of the
free Abelian 3-form and 1-form gauge theories).

A close and careful look at the perfectly anti-BRST invariant Lagrangian density ${\cal L}_{(\bar B)} = {\cal L}_{(ng)} + {\cal L}_{(fp)}$ 
[cf. Eqs. (8),(9)] demonstrates that the ghost fields are decoupled from the rest of the theory that contains the
 physical fields (carrying the ghost number equal to zero). Hence, the quantum states (in the 
{\it total} quantum Hilbert space of states) are the direct
 product [46] of the physical states (i.e. $|phys>$) and the ghost states (i.e. $|ghost>$). We further note that every term of the 
 nilpotent version of the anti-BRST charge $Q_{AB}$ [cf. Eq. (63)] contains (i) the ghost fields, and  (ii) a few of them contain the 
 physical fields (with the ghost number equal to zero) along with the {\it basic} and/or {\it auxiliary} (anti-)ghost fields.  The physical fields that are
 associated with the {\it basic} ghost fields of our theory act on the physical states (i.e. $|phys>$) and the ghost fields act on the
 ghost states (i.e. $|ghost>$). The {\it latter} operation always yields the {\it non-zero} result. Hence, we demand that the physical fields
 (with ghost number equal to zero) should operate on the physical states (i.e. $|phys>$) provided they are associated with the 
 {\it basic} (anti-)ghost fields of our theory in the physicality criterion w.r.t. the nilpotent anti-BRST charge $Q_{AB}$. Thus, we
 observe that the physicality criterion (i.e. $Q_{AB}\, |phys>\, = 0 $) yields the following
\begin{eqnarray}\label{67}
&&B \, |phys> \,= 0 \;\Longrightarrow \; \Pi^{0}_{(A1)} |phys>\, = 0, \quad 
\dot B \, |phys> \, = 0 \; \Longrightarrow \; \partial_i \Pi^i_{(A1)} \, |phys>\, = 0,
\nonumber\\
&&\dfrac{1}{2}\, \big (\partial^0 \bar B^{ij} + \partial^i \bar B^{j0}  + \partial^j \bar B^{0i}  \big ) \, |phys> \, = 0 
\quad \Longrightarrow \quad -\, 3\,\partial_k \Pi^{kij}_{(A3)} \, |phys> = 0, \nonumber\\
&& -\, \dfrac{1}{2}\, \bar B_{ij} \, |phys> \,= 0 \qquad \Longrightarrow \qquad  3\,\Pi^{0ij}_{(A3)} |phys>\, = 0, 
\end{eqnarray}
which establishes that the operator forms of the primary and secondary constraints (modulo some constant numeral factors)
of our {\it classical } theory annihilate the physical states 
(i.e. $|phys>$) of our 4D BRST-quantized version of theory at the {\it quantum} level through the physicality criterion w.r.t. the nilpotent
version of the anti-BRST charge. Thus, we conclude that the physicality criteria w.r.t. the nilpotent (i.e. $Q_{(A)B}^2 = 0$)
versions of the (anti-)BRST charges
$Q_{(A)B}$ lead to the {\it same} consequences which are consistent with the Dirac quantization conditions for theories that are
endowed with constraints of any variety.

We end this Subsec. with a very crucial remark. We observe that the canonical conjugate momenta w.r.t. the vector field $\phi_\mu$, namely;
\begin{eqnarray}\label{68}
\Pi^{\mu}_{(\phi)} = \frac {\partial {\cal L}_{(\bar B)}}{\partial (\partial_0 \phi_{\mu})} = -\,
\dfrac{1}{2}\, \bar B^{0\mu} + \dfrac{1}{2}\,  \eta^{0\mu}\, B_2 \;\Longrightarrow\;  \Pi^{0}_{(\phi)}  = + \dfrac{1}{2}\, B_2, \qquad
\Pi^{i}_{(\phi)} = -\, \dfrac{1}{2}\, \bar  B^{0i}, 
\end{eqnarray}
are {\it not} a set of primary constraints on our {\it classical} theory (cf. Appendix A). In fact, the vector field $\phi_\mu$ has been introduced 
in the generalization of the gauge-fixing term
for the Abelian 3-form gauge field [cf. Eq. (8)] when we attempt to quantize the theory. We observe that terms $B_2\, \bar f^0 $ and $\bar B^{0i}\, \bar f_i $
are present in the expression for the nilpotent version of the anti-BRST charge $Q_{AB}$ [cf. Eq. (63)] where the auxiliary fields $B_2$
and $\bar B^{0i}$ carry the ghost number equal to zero. However, since they are associated with the components of the
{\it auxiliary} field $\bar f_\mu$, they are {\it not} allowed to act on the physical states (i.e. $|phys>$). Hence, they
do {\it not} lead to any conditions on the physical states like equation (67).\\

\vskip 1.0cm

\section{Summary and Future Perspective}

In our present endeavor, we have concentrated {\it only} on the infinitesimal, continuous and off-shell nilpotent 
(i.e. $s_{(a)b}^2 = 0 $) (anti-)BRST symmetry transformations $s_{(a)b}$ that are
respected by the action integrals corresponding to the 
coupled (but equivalent) Lagrangian densities. We have derived the Noether  conserved (anti-)BRST charges 
$Q_{(a)b}$ and have shown that they are
{\it not} nilpotent of order two (i.e. $Q_{(a)b}^2 \neq 0$). We have exploited the theoretical techniques of our earlier work [37]
to derive the off-shell nilpotent (i.e. $Q_{(A)B}^2 = 0$) versions of the (anti-)BRST charges $Q_{(A)B} $ from their
counterparts Noether (anti-)BRST charges $Q_{(a)b}$ which are non-nilpotent. We have laid a great deal of importance on 
the CF-type restrictions and we have derived them from different theoretical angles
(cf. Appendix B). The physicality criteria w.r.t. the conserved and 
nilpotent versions of the (anti-)BRST charges $Q_{(A)B} $ have been discussed in an elaborate manner 
(cf. Secs. 4 and 5) where we have demonstrated that the operator
forms of the first-class constraints (of our {\it classical} 4D theory) annihilate the physical states at the {\it quantum} level which are found
to be consistent with the Dirac quantization conditions (see, e.g. [32, 33]) for the physical systems that are endowed with constraints. We have also
devoted time on the derivation of the ghost charge which (i) generates the infinitesimal version of the  ghost-scale symmetry transformations
[cf. Eq. (C.2)], and (ii) participates in the derivation of the {\it standard}  form of the BRST algebra [cf. Eq. (C.17)]
along with the nilpotent versions of the 
(anti-)BRST charges $Q_{(A)B} $.

Some of the key and crucial observations of our present investigation are as follows. First of all, we note that the kinetic terms for the 
Abelian 3-form and 1-form gauge fields (owing their origin to the exterior derivative of differential geometry) remain invariant under the
off-shell nilpotent (anti-)BRST symmetry transformations [cf. Eqs. (6),(4)]. Second, the mass dimension
(in the natural units: $\hbar =  c = 1 $) of a specific  field of our theory increases by one when it is operated upon by the (anti-)BRST symmetry operators
[cf. Eqs. (6),(4)]. Third, the ghost number of a given field of our theory increases by one when it is operated upon by the 
BRST symmetry transformation operator [cf. Eq. (4)]. On the other hand, the operation of the anti-BRST symmetry transformation
operator [cf. Eq. (6)] decreases the ghost number of a given field by one. Fourth, the off-shell nilpotent (anti-)BRST symmetry
transformation operators  [cf. Eqs. (6),(4)] absolutely anticommute  (i.e. $\{ s_b, \; s_{ab} \} = 0$) with each-other only on the submanifold of the 
quantum fields where the CF-type restrictions (cf. Appendix B) are satisfied. Fifth, even though the (anti-)BRST charges $Q_{(a)b}$  are
derived from the infinitesimal, continuous and off-shell nilpotent (i.e. $s_{(a)b}^2 = 0 $) (anti-)BRST symmetry transformations $s_{(a)b}$,
they turn out to be non-nilpotent (i.e.  $Q_{(a)b}^2 \neq 0 $) because of the presence of the {\it non-trivial} CF-type restrictions (cf. Appendix B)
on our 4D BRST-quantized field-theoretic system [37]. Finally, the off-shell nilpotent (anti-)BRST symmetry transformation operators are 
inherently {\it different}
from their counterpart off-shell nilpotent $\mathcal {N} = 2 $ SUSY 
transformation operators because of {\it their} absolute anticommutativity (i.e.  $\{ s_b, \; s_{ab} \} = 0$)
property which  is certainly {\it not} satisfied by the off-shell nilpotent $\mathcal {N} = 2 $ SUSY transformation operators. Furthermore, there 
is {\it no} idea of the CF-type restrictions in the context of the $\mathcal {N} = 2 $ SUSY transformation operators which are the hallmarks
of the properly BRST-quantized theories (see, e.g. [40,41]).

As pointed out earlier (cf. Sec. 1), we have devoted quite a number of years in studying the BRST-quantized $D = 2\, p$ 
(i.e. $ D = 2, 4, 6 $) dimensional St{\" u}ckelberg-modified {\it massive} and {\it massless}
 Abelian $p$-form (i.e. $ p = 1, 2, 3$) gauge theories
and established that these {\it even} dimensional theories are 
the tractable field-theoretic examples for Hodge theory (see, e.g. [47-51] and references therein) 
where the symmetries and conserved charges 
have provided the physical realization(s) of the cohomologcal operators of differential geometry [18-23]. One of the upshots of this study
has been the observation that a tower of $p$-number of ``exotic'' {\it massive} as well as {\it massless} fields appear in these theories which carry
the negative kinetic terms\footnote{To be precise, the ``exotic'' massive fields appear in the BRST-quantized St{\" u}ckelberg-modified massive
gauge theories and massless ``exotic'' fields are seen in the context of the BRST-quantized {\it usual}  gauge theories which are massless by definition. In 
a very recent couple of papers [52,53], we have shown the existence of a {\it single} massive pseudo-scalar field as an ``exotic'' field
in the context of the BRST-quantized ($p + 1$)-dimensional  St{\" u}ckelberg-modified massive Abelian $p$-form (i.e. $p = 1, 2, 3$) gauge
theories. In particular, we have performed the explicit and elaborate computations for the BRST-quantized  St{\" u}ckelberg-modified massive Abelian 2-from 
gauge theory in the three (2 + 1)-dimensional (3D) spacetime [53].}. 
Such fields have become quite popular in the realm of the cyclic, bouncing and self-accelerated cosmological models of the Universe 
where these fields have been christened as the ``phantom'' and/or ``ghost'' fields (see, e.g. [54-59] and references therein). Such ``exotic'' fields
have been also proposed to be the possible candidates for dark matter/dark energy (see, e.g. [60,61] and references therein). One of the highlights of 
our {\it present} endeavor is the observation that the (axial-)vector fields $(\widetilde \phi_\mu)\phi_\mu$ (that have been introduced in the theory on mathematical
grounds) carry the {\it positive} kinetic terms. Hence, they are {\it not} the possible candidates for the ``phantom'' fields. This observation should
be contrasted with our earlier work [47] on the St{\" u}ckelberg-modified 4D {\it massive} Abelian 2-form gauge theory 
as an example for Hodge theory where we have shown the
existence of the two ``exotic'' fields which are nothing but (i) an axial-vector field, and (ii) a pseudo-scaler field. Both these fields 
carry the negative kinetic terms. However, they have been found to satisfy the {\it massive} Klein-Gordon equations of motion. The key observation of 
our present 4D example for Hodge theory (see, e.g. [31]) rules out the axial-vector field to be a {\it true} candidate for the phantom field
of the cosmological models. Thus, it seems to us that the pseudo-scalar field is the most fundamental field as a possible candidate
for the ``phantom'' field of the cosmology. This claim is backed by our observations (see, e.g. [48,24,25]) in the context of the 2D and 3D
field-theoretic examples for Hodge theory where the existence of a {\it single}  ``exotic'' pseudo-scalar field has been established in a very transparent manner.
The {\it massless} ``exotic'' pseudo-scalar field {\it might} be one of the possible candidates for the dark energy
which is responsible for the current observation of the accelerated expansion of the Universe (see, e.g. [62-64] and references therein).

In our earlier work [31], it has been amply indicated that our present 
BRST-quantized {\it combined} system of the 4D free Abelian 3-form and 1-form gauge theories is a possible field-theoretic example for Hodge theory
which is completely described by the coupled (but equivalent) Lagrangian densities that 
are expected to respect the infinitesimal and {\it continuous}
nilpotent (anti-)BRST, nilpotent (anti-)co-BRST, a unique bosonic and the ghost-scale symmetry transformations along with a couple if {\it discrete} duality 
symmetry transformations. In our future endeavor, we plan to devote time on the elaborate discussions on (i) the 
existence of the nilpotent (anti-)co-BRST
symmetries, (ii) the derivation of the non-nilpotent {\it Noether} (anti-)co-BRST charges, (iii) the deduction of the nilpotent versions
of the (anti-)co-BRST charges by exploiting the theoretical tricks of our earlier work [37], and (iv) the physicality criteria w.r.t. 
the nilpotent versions of the (anti-)co-BRST charges to show that the {\it true} physical states are {\it those} that are annihilated by
the operator forms of the {\it dual} versions of the first-class constraints\footnote{It is worthwhile to pinpoint the fact that the
physical states are annihilated by the first-class constraints due to the requirements of the physicality criteria w.r.t. the nilpotent versions of
the (anti-)BRST charges (cf. Secs. 4 and 5). However, for the BRST-quantized examples for Hodge theory, the {\it true} physical states
are annihilated by the off-shell nilpotent versions of the (anti-)BRST as well as the
(anti-)co-BRST charges because the {\it true} physical states of {\it such} theories are the (anti-)BRST as well as the (anti-)co-BRST invariant 
{\it harmonic} states of the Hodge decomposed quantum states in the Hilbert space of such a theory.}. In addition, we 
envisage to focus on the derivations of {\it all} the appropriate conserved charges
for this theory (corresponding to the set of {\it six} continuous symmetry transformations) and plan to devote time on the derivation of the
extended BRST algebra which would establish the mapping between the conserved charges and cohomological operators of differential geometry
at the {\it algebraic} level.

In addition to the above cited {\it future} directions, we wish to discuss our 4D field-theoretic system within the framework of the 
mathematically elegant and theoretically powerful  BV-formalism (see, e.g. [65-67] and references therein) which has reach and range that
encompasses in its folds (see, e.g. [68,69] and references therein)
the (BRST-)quantization of the higher $p$-form ($p = 2, 3, 4... $) gauge theories, topological field theories,
gravitational theory, string field theories, etc. In particular, it would be nice to recast  the discussions on (i) the coupled
(but equivalent) Lagrangian densities, (ii) the CF-type restrictions, (iii) the nilpotency property of the 
(anti-)BRST charges, etc, in the terminology of the BV-formalism. We believe, the powerful theoretical strength of the BV-formalism
would enable us to study the higher-dimensional (D = 5, 6, 7...) field-theoretic system of the higher $p$-form ($p = 2, 3, 4... $)
gauge theories. In particular, the 5D combined field-theoretic system of the free Abelian 2-form and 3-form  gauge theories 
would be interesting to study because it might turn out to be an example for Hodge theory in {\it odd} dimension. In our earlier works [40,41],
we have concentrated on the relationship between the geometrical objects 1-gerbs and 2-gerbes and the CF-type restrictions 
of the BRST-quantized D-dimensional (i) non-Abelian 1-form [41], and (ii) Abelian 2-form [40] as well as the Abeian 3-form [41] gauge theories. 
It would be nice future endeavor to study the cohomological and geometrical aspects of
our 4D combined field-theoretic system in the terminology of recent works [70,71] where it has been claimed that the  BRST cohomology is
the Lie algerboid cohomology. \\

\vskip 0.3cm

\noindent
{\bf Acknowledgment} \\

\noindent
The author owes a great deal to Prof. R. Rajaraman for his survival in the research areas 
that are connected with a few topics in the realm of theoretical high energy physics (THEP).
Prof. Rajaraman was one of the influential mentors of the THEP group at the Banaras Hindu University (BHU), Varanasi. He visited  BHU many times and
enlightened us with his very clear formal talks. His captivating  public lectures
were a treat in science communication to the general public.
He passed away on July 12, 2025. The author dedicates his present work, very
humbly and respectfully, to the memory of  Prof. Rajaraman. \\

\vskip 0.3cm

\noindent
{\bf Declarations} \\

\vskip 0.3cm

\noindent
{\bf Conflicts of interest}\\


\noindent
The author declares that there are no conflicts of interest.\\

\vskip 0.3cm

\noindent
{\bf Funding Statement}\\


\noindent
No funding was received for this research.\\

\vskip 0.3cm

\noindent
{\bf Data Availability Statement}\\


\noindent
No new data were created or analyzed in this study.\\

\vskip 0.7 cm 
\begin{center}
{\bf Appendix A: \bf On the First-Class Constraints}\\
\end{center}

\vskip 0.5 cm

\noindent
The central purpose of this Appendix is to show that our 4D field-theoretic system is endowed with the first-class constraints
in the terminology of the Dirac's prescription for the classification scheme of constraints (see, e.g. [32-36]). First of all, let us begin
with the D-dimensional {\it combined} field-theoretic system of the {\it free} Abelian 3-form and 1-form gauge theories which is 
described by the following Lagrangian density [i.e. ${\cal L}^{(D)}_{(0)} $], namely;
\[
{\cal L}^{(D)}_{(0)} = \dfrac{1}{48}\, H^{\mu\nu\sigma\rho}\, H_{\mu\nu\sigma\rho}  - \dfrac{1}{4}\, F^{\mu\nu}\, F_{\mu\nu},
\eqno (A.1)
\]
where the Greek indices $\mu, \, \nu,\, \sigma, \, \rho...=  0, 1, 2,....,(D-1) $ and field-strength tensors for the
Abelian 3-form and 1-form gauge fields have already been 
defined as: $H_{\mu \nu \sigma \rho} = \partial_\mu\, A_{\nu\sigma\rho} - \partial_\nu\,
 A_{\sigma\rho \mu } + \partial_\sigma\, A_{\rho \mu \nu }  - \partial_\rho \,A_{\mu \nu \sigma }$ and $F_{\mu\nu}  = \partial_\mu A_\nu - \partial_\nu A_\mu$,
respectively. In equation (A.1), the superscript $(D)$ stands for our chosen spacetime manifold to be the D-dimensional
flat Minkowskian in nature  and subscript $(0)$ denotes the starting free {\it original} Lagrangian density 
[i.e.  ${\cal L}^{(D)}_{(0)} $]. It is evident that the
canonical conjugate momenta w.r.t. to the Abelian 3-form and 1-form gauge fields, from the  Lagrangian density (A.1), are 
\[
\Pi^{\mu\nu\sigma}_{(A3)} = \frac {\partial {\cal L}^{(D)}_{(0)}}{\partial (\partial_0 A_{\mu\nu\sigma})} \;\; \equiv \;\;\frac {1}{3!}\, H^{0\mu\nu\sigma}
\quad \Longrightarrow \quad \Pi^{0ij}_{(A3)}  = \frac {1}{3!}\, H^{00ij} \approx 0,
\]
\[
\Pi^{\mu}_{(A1)} = \frac {\partial {\cal L}^{(D)}_{(0)}}{\partial (\partial_0 A_{\mu})} \;\; \equiv \;\;-\, F^{0\mu}
\qquad \;\Longrightarrow \quad \Pi^{0}_{(A1)}  = -\, F^{00} \approx 0,
\eqno(A.2)
\]
where the subscripts $(A3)$ and $(A1)$, associated with the definitions of the conjugate momenta, denote the momenta w.r.t the totally antisymmetry
3-form gauge field $A_{\mu\nu\sigma}$ and the Lorentz-vector Abelian 1-form field $A_\mu$, respectively. 
In the above equation (A.2), the quantities $\Pi^{0ij}_{(A3)} \approx 0$ and $\Pi^{0}_{(A1)}  \approx 0 $ 
are the primary constraints (PCs) on the theory where the Dirac notation of the symbol for the idea of {\it weakly} zero has been taken into account. 
It is straightforward to note that the Euler-Lagrange (EL) equations of motion (EoM)
\[
\partial_\mu H^{\mu\nu\sigma\rho} = 0 \qquad \Longrightarrow \quad \partial _0 H^{00jk} + \partial_i H^{i0jk} = 0,
\]
\[
\partial_\mu F^{\mu\nu} = 0 \quad \;\Longrightarrow \quad \partial _0 F^{00} + \partial_i F^{i0} = 0,
\eqno(A.3)
\]
are {\it true} where the choices:  (i) $\nu = 0, \;\sigma = j$ and $\rho = k$,  and (ii) $\nu = 0$ have been taken into considerations
in the top and bottom EL-EoMs, respectively. The requirements of the time-evolution invariance of these PCs lead to the derivations
of the secondary constraints (SCs).
It is clear from equations (A.2) and (A.3)  that we have the time derivatives (in the natural units: $\hbar = c = 1$, $\partial_0 = \partial_t$) 
on the PCs: $\Pi^{0ij}_{(A3)}  \approx 0$  and $\Pi^{0}_{(A1)}  \approx 0 $, respectively, as 
\[
\frac {\partial \Pi ^{0jk}_{(A3)}} {\partial t}\,  = \frac {1}{3!}\,\partial_i H^{0ijk} \approx 0 \quad \Longrightarrow 
\quad \frac {\partial \Pi ^{0jk}_{(A3)}}{\partial t} \equiv \partial_i \Pi ^{ijk}_{(A3)} \approx 0,
\]
\[
\frac {\partial \Pi ^{0}_{(A1)}} {\partial t}\,  = \,\partial_i F^{i0} \approx 0 \qquad \;\;\Longrightarrow 
\quad \frac {\partial \Pi ^{0}_{(A1)}}{\partial t} \equiv \partial_i \Pi ^{i}_{(A1)} \approx 0,
\eqno(A.4)
\]
where $\Pi ^{ijk}_{(A3)} = \frac {1}{3!}\, H^{0ijk}$  and $\Pi ^{i}_{(A1)} = F^{i0} $  
are the expressions for the {\it purely}  space components  of the conjugate momenta (A.2). 
 There are {\it no} further  constraints on the theory after the derivations of the 
secondary constraints (i.e. $\partial_i \Pi ^{ijk} \approx 0,\; \partial_i \Pi ^{i}_{(A1)} \approx 0$) in (A.4). 
It is interesting to point out that {\it both}  
the primary as well as the secondary constraints are expressed in terms of  
the components  of the canonical conjugate momenta [defined in Eq. (A.2) w.r.t. the gauge fields $A_{\mu\nu\lambda}$ and $A_\mu$]. 
Hence, they  commute among themselves leading to their characterization  as the first-class constraints in the terminology 
of Dirac's prescription for the classification scheme of constraints (see, e.g. [32-36] for details). Thus, we note that we have: (i) a set of two PCs:
$\Pi ^{0ij}_{(A3)} \approx 0, \;\Pi^{0}_{(A1)}  \approx 0$, and 
(ii) another set of two SCs: $\partial_i \Pi ^{ijk}_{(A3)} \approx 0, \; \partial_i \Pi ^{i}_{(A1)} \approx 0 $. 
As pointed out earlier, these PCs and SCs are {\it first-class} in the terminology of Dirac's prescription for the classification scheme
of constraints and they generate (see, e.g. [72] and references therein) the infinitesimal, local and continuous gauge symmetry transformations: 
$\delta_g A_{\mu\nu\sigma} =  \partial_\mu \Lambda_{\nu\sigma} + \partial_\nu \Lambda_{\sigma\mu} 
+ \partial_\sigma \Lambda_{\mu\nu}, \; \delta_g A_\mu = \partial_\mu\Lambda $ under which the field-strength tensors
for {\it both} the gauge fields remain invariant (and, hence, the
Lagrangian density (A.1), too). Here the antisymmetric (i.e. $\Lambda_{\mu\nu} = -\, \Lambda_{\nu\mu} $) tensor 
$\Lambda_{\mu\nu} $ and Lorentz-scalar $\Lambda$ 
are the infinitesimal local gauge symmetry transformation parameters. This statement is true 
in any arbitrary D-dimension of spacetime (including the physical
four (3 + 1)-dimensions of spacetime). It is worthwhile to point out that theories, endowed with the second-class constraints,  can be 
converted into {\it gauge} theories by introducing {\it new} set of fields in such a manner that the 
given set of second-class constraints are converted into
the appropriate set of first-class constraints (see, e.g. [72,73] and references therein).

We end our present Appendix with the following remarks. First of all, we note that the Lagrangian density (A.1) takes the following interesting form
for the {\it special} case of the physical four (3 + 1)-dimensions of spacetime, namely;
\[
{\cal L}^{(D = 4)}_{(0)} = 
-\, \dfrac{1}{2}\,\Big(- \dfrac{1}{3!}\varepsilon^{\mu\nu\sigma\rho} \,\partial_\mu A_{\nu\sigma\rho} \Big)^2
+ \dfrac{1}{4}\, \Big( \varepsilon^{\mu\nu\sigma\rho} \,\partial_\sigma A_\rho \Big)^2.
\eqno(A.5)
\]
All our discussions [that have been covered from equation (A.2) to (A.4)] remain {\it true}
for our present {\it four} (3 + 1)-dimensional (4D) theory as well. However, we note that the expressions for the
conjugate momenta [cf. Eq. (A.2)] take the following forms, namely;
\[
\Pi^{\mu\nu\sigma}_{(A3)} = \frac {\partial {\cal L}^{(D = 4)}_{(0)}}{\partial (\partial_0 A_{\mu\nu\sigma})} \;\; \equiv \;\;\frac {1}{3!}\, 
\varepsilon^{0\mu\nu\sigma}\, (H_{0123})
\quad \Longrightarrow \quad \Pi^{0ij}_{(A3)}  = \frac {1}{3!}\, \varepsilon^{00ij} (H_{0123})\approx 0,
\]
\[
\Pi^{\mu}_{(A1)} = \frac {\partial {\cal L}^{(D = 4)}_{(0)}}{\partial (\partial_0 A_{\mu})} \;\; \equiv \;\;-\, F^{0\mu}
\qquad\; \Longrightarrow \quad \Pi^{0}_{(A1)}  = -\, F^{00} \approx 0.
\eqno(A.6)
\]
The above equation demonstrates that we have, once again, a set of two PCs. However, their forms have changed in the sense that, even for
the bottom entry in (A.6), we have the following conjugate momenta for our 4D field-theoretic system, namely;
\[
\Pi^{\mu}_{(A1)} = \frac {\partial {\cal L}^{(D = 4)}_{(0)}}{\partial (\partial_0 A_{\mu})} 
= \dfrac{1}{2}\, \varepsilon^{\alpha\beta 0 \mu}\, \varepsilon_{\alpha\beta\sigma\rho} \, \partial^\sigma A^\rho \;\; \equiv \;\;-\, F^{0\mu},
\eqno(A.7)
\]
where we have used the standard relationship: $\varepsilon^{\mu\nu\eta\kappa} \,\varepsilon_{\mu\nu\sigma\rho}= - \,2!\,
(\delta_\sigma^\eta \delta_\rho^\kappa - \delta_\sigma^\kappa \delta_\rho^\eta )$. Thus, we still have (i) a set of two PCs [cf. Eq. (A.6)], and
(ii) a set of two SCs as: $\partial_i \Pi ^{ijk}_{(A3)} \approx 0, \; \partial_i \Pi ^{i}_{(A1)} \approx 0 $ where the expressions for the space
components of conjugate momenta are:
\[
\Pi^{ijk}_{(A3)}  = \frac {1}{3!}\, \varepsilon^{0ijk} (H_{0123}), \qquad \Pi^{i}_{(A1)}  = -\, F^{0i} \equiv F^{i0}.
\eqno(A.8)
\]
All the above constraints are a set of {\it first-class} constraints as they commute among themselves. The operator forms of these
constraints annihilate the physical states of our theory due to the requirements of the physicality criteria (cf. Subsecs. 4.3, 5.3).\\

\vskip 0.7 cm 
\begin{center}
{\bf Appendix B: \bf On the CF-Type Restrictions}\\
\end{center}

\vskip 0.5cm

\noindent
The derivations of the CF-type restrictions are the central themes of our present Appendix from  the requirements of (i) the direct equality
(i.e. ${\cal L}_{(B)} -  {\cal L}_{(\bar B)} = 0$), and (ii) the {\it symmetry equivalence}  of the 
coupled Lagrangian densities ${\cal L}_{(B)} $ and ${\cal L}_{(\bar B)} $ (under the off-shell nilpotent 
(anti-)BRST symmetry transformations [cf. Eqs. (6),(4)]). The explicit forms of the CF-type restrictions,
on our present {\it combined} 4D field-theoretic system, are: \\
\[
B_{\mu\nu} + \bar B_{\mu\nu} = \partial_\mu \phi_\nu - \partial_\nu \phi_\mu, \qquad f_\mu + F_\mu = \partial_\mu C_1,
\]
\[
{\cal B}_{\mu\nu} + \bar {\cal B}_{\mu\nu} = \partial_\mu \tilde \phi_\nu 
- \partial_\nu \tilde \phi_\mu, \qquad \bar f_\mu + \bar F_\mu = \partial_\mu \bar C_1.
\eqno(B.1)
\]
It is straightforward to note that there are {\it bosonic} sectors (i.e. $B_{\mu\nu} + \bar B_{\mu\nu} = \partial_\mu \phi_\nu - \partial_\nu \phi_\mu,\;
{\cal B}_{\mu\nu} + \bar {\cal B}_{\mu\nu} = \partial_\mu \tilde \phi_\nu - \partial_\nu \tilde \phi_\mu $) as well as the {\it fermionic}
sectors (i.e. $f_\mu + F_\mu = \partial_\mu C_1,\; \bar f_\mu + \bar F_\mu = \partial_\mu \bar C_1 $) in the above CF-type restrictions (B.1). 
These restrictions  are found to be (anti-)BRST invariant quantities that can be checked explicitly by applying the off-shell nilpotent (anti-)BRST symmetry
transformations [cf. Eqs. (6),(4)] {\it directly} on them.  Hence, these CF-type restrictions are the {\it physical} restrictions on our theory.

Let us, first of all, concentrate on the {\it direct} equality (i.e. ${\cal L}_{(B)} -  {\cal L}_{(\bar B)} = 0$) of the coupled Lagrangian densities
${\cal L}_{(B)}$ and ${\cal L}_{(\bar B)} $. It is straightforward to note that, in the above difference, the {\it difference}
in the FP-ghost sectors [cf. Eqs. (3),(9)] is: 
\[
{\cal L}_{(FP)} -  {\cal L}_{(fp)} = \big (\partial_\mu \bar C^{\mu\nu}\big )\, (f_\nu + F_\nu) 
-  \big (\partial_\mu  C^{\mu\nu} \big )\, (\bar f_\nu + \bar F_{\nu}) + 2\, \bar f^\mu\, F_\mu 
\]
\[
+ \big (\partial^\mu \bar C_1 \big)\, (f_\mu - F_\mu) + \big (\partial^\mu  C_1 \big)\, (\bar f_\mu - \bar F_\mu) -  2\, \bar F^\mu\, f_\mu.
\eqno(B.2)
\]
We would like to point out that the {\it first} two terms can be expressed
(modulo a couple of total spacetime derivative terms)  in the terminology of the fermionic CF-type restrictions
(i.e. $f_\mu + F_\mu = \partial_\mu C_1,\; \bar f_\mu + \bar F_\mu = \partial_\mu \bar C_1 $) as follows
\[
\big (\partial_\mu \bar C^{\mu\nu}\big )\, (f_\nu + F_\nu - \partial_\nu C_1) + \partial_\mu \big [\bar C^{\mu\nu} (\partial_\nu C_1) \big ]
\]
\[
-  \big (\partial_\mu  C^{\mu\nu} \big )\, (\bar f_\nu + \bar F_{\nu} - \partial_\nu \bar C_1) - \partial_\mu \big [C^{\mu\nu} (\partial_\nu \bar C_1) \big ],
\eqno(B.3)
\]
where the total spacetime derivative terms do {\it not} play any significant role in the description of the  {\it dynamics} of our theory
in the Lagrangian formulation. Hence, these terms will be ignored in our further discussions. The left-over terms 
in (B.2) can be re-arranged, with a bit of {\it involved} algebraic exercise, as follows
\[
\big  (2\,f^\mu - \partial^\mu C_1 \big ) \, [\bar f_\mu + \bar F_{\mu} - \partial_\mu \bar C_1]
+ \big  (2\,\bar f^\mu - \partial^\mu \bar C_1 \big ) \, [f_\mu +  F_{\mu} - \partial_\mu C_1],
\eqno(B.4)
\]
where, as is obvious, the CF-type restrictions: $\bar f_\mu + \bar F_\mu = \partial_\mu \bar C_1, \; f_\mu + F_\mu = \partial_\mu C_1  $ are present in
{\it both} the terms, respectively. Taking the sum of our expressions in (B.3) and (B.4), we note that the difference in the ghost-sectors (B.2) can be,
ultimately, expressed as:
\[
{\cal L}_{(FP)} -  {\cal L}_{(fp)} = 
\big  (\partial_\mu \bar C^{\mu\nu} + 2\,\bar f^\nu - \partial^\nu \bar C_1 \big ) \, [\bar f_\nu + \bar F_{\nu} - \partial_\nu \bar C_1]
\]
\[
-\,\big  (\partial_\mu C^{\mu\nu} - 2\,f^\nu + \partial^\nu C_1 \big ) \, [\bar f_\nu + \bar F_{\nu} - \partial_\nu \bar C_1].
\eqno(B.5)
\]
Thus, we draw the conclusion that the FP-ghost sectors [cf. Eqs. (3),(9)] of the coupled Lagrangian densities ${\cal L}_{(B)}$ and ${\cal L}_{(\bar B)} $
are equal modulo (i) the total spacetime derivative terms: $\partial_\mu [  \bar C^{\mu\nu} \, 
\partial_\nu C_1 - C^{\mu\nu}\, \partial_\nu \bar C_1] $, and (ii) the validity of the fermionic CF-type 
restrictions (i.e. $f_\mu + F_\mu = \partial_\mu C_1,\; \bar f_\mu + \bar F_\mu = \partial_\mu \bar C_1 $). In other words, the
direct equality requirement of the ghost-sectors of the Lagrangian densities [cf. Eqs. (3),(9)]  leads to the derivations of the (anti-)BRST invariant 
fermionic  sectors (i.e. $f_\mu + F_\mu = \partial_\mu C_1,\; \bar f_\mu + \bar F_\mu = \partial_\mu \bar C_1 $) of the CF-type 
restrictions that have been listed completely in (B.1).

We focus now on the contribution from the non-ghost sectors [cf. Eqs. (1),(8)] in the direct equality (i.e. ${\cal L}_{(B)} -  {\cal L}_{(\bar B)} = 0$) of
the coupled Lagrangian densities ${\cal L}_{(B)}$ and ${\cal L}_{(\bar B)} $ of our theory. In explicit form, this contribution is:
\[
{\cal L}_{(NG)} -  {\cal L}_{(ng)} = \dfrac{1}{4}\, \big [(\bar B_{\mu\nu})^2 - (B_{\mu\nu})^2 \big ] + \dfrac{1}{2}\, \big(\partial_\sigma A^{\sigma\mu\nu} \big )
\,\big (\bar B_{\mu\nu} +  B_{\mu\nu} \big ) + \dfrac{1}{4}\, \big (\partial^\mu \phi^\nu - \partial^\nu \phi^\mu \big )\, \big [B_{\mu\nu} - \bar B_{\mu\nu} \big ]
\]
\[
+ \dfrac{1}{4}\, \big [(\bar {\cal B}_{\mu\nu})^2 - ({\cal B}_{\mu\nu})^2 \big ] + \dfrac{1}{2}\, \big(\varepsilon^{\mu\nu\sigma\rho} \,\partial_\sigma A_\rho \big )
\ \big [{\cal B}_{\mu\nu} - \bar {\cal B}_{\mu\nu} \big ] + \dfrac{1}{4}\, 
\big (\partial^\mu \tilde \phi^\nu - \partial^\nu \tilde \phi^\mu \big )\, \big [{\cal B}_{\mu\nu} - \bar {\cal B}_{\mu\nu} \big ].
\eqno(B.6)
\]
Using the simple trick of factorization [e.g. $(\bar B_{\mu\nu})^2 - (B_{\mu\nu})^2 = (\bar B_{\mu\nu} + B_{\mu\nu}) \, (\bar B^{\mu\nu} - B^{\mu\nu})$, etc.]
and the re-arrangements of terms lead to the following equation from (B.6), namely;
\[
{\cal L}_{(NG)} -  {\cal L}_{(ng)} = \partial_\mu  \Big [ \big(\partial_\sigma A^{\sigma\mu\nu} \big )\, \phi_\nu
+ \varepsilon^{\mu\nu\sigma\rho} \big (\partial_\sigma A_\rho \big ) \, \tilde \phi_\nu \Big ] 
\]
\[
+ \dfrac{1}{2} \,\Big [\partial_\sigma A^{\sigma\mu\nu} + \dfrac{1}{2} \, \big (\bar B^{\mu\nu} - B^{\mu\nu} \big ) \Big ]\,
\Big [  {B}_{\mu\nu} + \bar {B}_{\mu\nu} - \big (\partial_\mu \phi_\nu - \partial_\nu \phi_\mu \big ) \Big ]
\]
\[
+ \dfrac{1}{2}\, \Big [\varepsilon^{\mu\nu\sigma\rho} \,\partial_\sigma A_\rho   + \dfrac{1}{2} \, \big (\bar {\cal B}^{\mu\nu} - 
{\cal B}^{\mu\nu} \big ) \Big ]\,
\Big [  {\cal B}_{\mu\nu} + \bar {\cal B}_{\mu\nu} - \big (\partial_\mu \tilde \phi_\nu - \partial_\nu \tilde \phi_\mu \big ) \Big ].
\eqno(B.7)
\]
A close look at (B.7) demonstrates that modulo (i) the total spacetime derivative terms (i.e. $\partial_\mu   [ \big(\partial_\sigma A^{\sigma\mu\nu} \big )\, \phi_\nu
+ \varepsilon^{\mu\nu\sigma\rho} \big (\partial_\sigma A_\rho \big ) \, \tilde \phi_\nu ]  $), and (ii) the validity of the bosonic sectors of the
CF-type restrictions [i.e. $\bar {B}_{\mu\nu} +  {B}_{\mu\nu} - \big (\partial_\mu \phi_\nu - \partial_\nu \phi_\mu \big ) = 0, \;
\bar {\cal B}_{\mu\nu} +  {\cal B}_{\mu\nu} - \big (\partial_\mu \tilde \phi_\nu - \partial_\nu \tilde \phi_\mu \big ) = 0 $], the {non-ghost
sectors of the 
Lagrangian densities ${\cal L}_{(NG)}$ and ${\cal L}_{(ng)} $ 
[cf. Eqs. (1),(8)] are {\it also} equivalent. In other words, the direct equality 
(i.e. ${\cal L}_{(NG)} -  {\cal L}_{(ng)} = 0 $) of the non-ghost sectors of the Lagrangian densities [cf. Eqs. (1),(8)]  leads to the
derivations of the bosonic sectors [i.e. $\bar {B}_{\mu\nu} +  {B}_{\mu\nu} - \big (\partial_\mu \phi_\nu - \partial_\nu \phi_\mu \big ) = 0, \;
\bar {\cal B}_{\mu\nu} +  {\cal B}_{\mu\nu} - \big (\partial_\mu \tilde \phi_\nu - \partial_\nu \tilde \phi_\mu \big ) = 0 $]
of the (anti-)BRST invariant CF-type restrictions that are present in (B.1).

At this stage, we devote time on the derivations of the CF-type restrictions (B.1) from the point of view of the {\it symmetry}
considerations. In other words, we prove that {\it both} the coupled Lagrangian densities ${\cal L}_{(B)}$ and ${\cal L}_{(\bar B)} $ of our theory
respect BRST as well as the anti-BRST symmetry transformations [cf. Eqs. (4),(6)] on the submanifold of the quantum fields where the CF-type restrictions 
(B.1) are satisfied. We have already shown that the Lagrangian density ${\cal L}_{(B)} $ transforms to the total spacetime derivative [cf. Eq. (5)] under
the BRST symmetry transformations (4). Similarly, the Lagrangian density ${\cal L}_{(\bar B)} $ transforms to the total spacetime derivative [cf. Eq. (7)]
under the anti-BRST symmetry transformations (6). Let us concentrate {\it now} on the application of the anti-BRST symmetry
transformations (6) on the Lagrangian density ${\cal L}_{(B)} $. This operation, ultimately, leads to the following explicit 
transformation of the Lagrangian density ${\cal L}_{(B)} $:
\[
s_{ab} {\cal L}_{(B}) = \dfrac{1}{2}\, \partial_\mu \,\Big [B_2 \, \bar f^\mu  +  \big (\partial^\mu \bar C^{\nu\sigma} +  \partial^\nu \bar C^{\sigma\mu} 
+ \partial^\sigma \bar C^{\mu\nu} \big )\, B_{\nu\sigma} + \big (\partial^\mu \bar \beta^\nu - \partial^\nu \bar \beta^\mu \big )\, f_\nu
\]
\[
- \bar B^{\mu\nu}\, \bar F_\nu - C^{\mu\nu}\, \partial_\nu B_5 + \bar C^{\mu\nu}\, \partial_\nu B_2- F^\mu\,B_5 - B_4\, \partial^\mu \bar C_2 \Big ]
+ \;\partial_\mu \, \Big [(\partial_\sigma A^{\sigma\mu\nu})\, \bar f_\nu - B\, \partial^\mu \bar C \Big]
\]
\[ 
+ \dfrac{1}{2} \, \Big [ \bar B^{\mu\nu}\, \partial_\mu [\bar f_\nu + \bar F_\nu - \partial_\nu \bar C_1 ] + \big (f^\mu + F^\mu - \partial^\mu C_1 \big )\,
\partial_\mu B_5 - \big (\bar f^\mu + \bar F^\mu - \partial^\mu \bar C_1 \big )\,\partial_\mu B_2 \Big ]
\]
\[
-\, \dfrac{1}{2}\, \Big [\big (\partial^\mu \bar \beta^\nu - \partial^\nu \bar \beta^\mu \big )\,\partial_\mu \big [f_\nu + F_\nu - \partial_\nu C_1 \big ]
+ \big \{B^{\mu\nu} + \bar B^{\mu\nu} - (\partial^\mu \phi^\nu - \partial^\nu \phi^\mu) \big \}\, \partial_\mu \bar f_\nu \Big ]
\]
\[
- \, \dfrac{1}{2}\,  \big (\partial^\mu \bar C^{\nu\sigma} +  \partial^\nu \bar C^{\sigma\mu} 
+ \partial^\sigma \bar C^{\mu\nu} \big )\, \partial_\mu \big [B_{\nu\sigma} + \bar B_{\nu\sigma} 
- (\partial_\nu \phi_\sigma - \partial_\sigma \phi_\nu)  \big ].
\eqno(B.8)
\]
We would like to mention, in passing, that the algebraic exercise is slightly {\it involved} in the above derivation.
Our observation in (B.8) shows that the action integral corresponding to the  Lagrangian density ${\cal L}_{(B)} $ respects (i) the BRST symmetry 
transformations (4), and (ii) the anti-BRST symmetry transformations (6) provided we take into account the validity of the (anti-)BRST
invariant CF-type restrictions: $B_{\mu\nu} + \bar B_{\mu\nu} = \partial_\mu \phi_\nu - \partial_\nu \phi_\mu, \; f_\mu + F_\mu = \partial_\mu C_1, 
\; \bar f_\mu + \bar F_\mu = \partial_\mu \bar C_1$. This observation amounts to the derivations of the above CF-type restrictions by
the requirements of the {\it symmetry equivalence} of the Lagrangian density ${\cal L}_{(B)} $ w.r.t. the BRST as well as the anti-BRST symmetry
transformations [cf. Eqs. (4),(6)].

Against the backdrop of the above paragraph, we apply now the BRST symmetry transformations (4) on the Lagrangian density ${\cal L}_{(\bar B)} $
which leads to:
\[
s_{b} {\cal L}_{(\bar B)} = \dfrac{1}{2}\, \partial_\mu \,\Big [B_2 \,  f^\mu  -  \big (\partial^\mu C^{\nu\sigma} +  \partial^\nu  C^{\sigma\mu} 
+ \partial^\sigma  C^{\mu\nu} \big )\, \bar B_{\nu\sigma} +\big (\partial^\mu  \beta^\nu - \partial^\nu  \beta^\mu \big )\, \bar f_\nu
\]
\[
-  B^{\mu\nu}\,  F_\nu - \bar C^{\mu\nu}\, \partial_\nu B_4 - C^{\mu\nu}\, \partial_\nu B_2 + \bar F^\mu\,B_4 - B_5\, \partial^\mu C_2 \Big ]
- \;\partial_\mu \, \Big [(\partial_\sigma A^{\sigma\mu\nu})\, f_\nu + B\, \partial^\mu  C \Big]
\]
\[ 
+ \dfrac{1}{2} \, \Big [ B^{\mu\nu}\, \partial_\mu [f_\nu +  F_\nu - \partial_\nu  C_1 ] - \big (f^\mu + F^\mu - \partial^\mu C_1 \big )\,
\partial_\mu B_2 - \big (\bar f^\mu + \bar F^\mu - \partial^\mu \bar C_1 \big )\,\partial_\mu B_4\Big ]
\]
\[
-\, \dfrac{1}{2}\, \Big [\big (\partial^\mu  \beta^\nu - \partial^\nu  \beta^\mu \big )\,\partial_\mu \big [\bar f_\nu + \bar F_\nu - \partial_\nu \bar C_1 \big ]
+ \big \{B^{\mu\nu} + \bar B^{\mu\nu} - (\partial^\mu \phi^\nu - \partial^\nu \phi^\mu) \big \}\, \partial_\mu f_\nu \Big ]
\]
\[
+ \, \dfrac{1}{2}\,  \big (\partial^\mu  C^{\nu\sigma} +  \partial^\nu  C^{\sigma\mu} 
+ \partial^\sigma C^{\mu\nu} \big )\, \partial_\mu \big [B_{\nu\sigma} + \bar B_{\nu\sigma} 
- (\partial_\nu \phi_\sigma - \partial_\sigma \phi_\nu)  \big ].
\eqno(B.9)
\]
In the above derivation, the algebraic exercise is slightly {\it involved}.
This establishes the fact that the Lagrangian density ${\cal L}_{(\bar B)} $ respects (i) the off-shell nilpotent anti-BRST symmetry transformations
(6), and (ii) the BRST transformations [cf. Eq. (B.9)] provided we take into account the validity of the appropriate CF-type restrictions
from (B.1). In other words, the requirements of the {\it symmetry equivalence} w.r.t. the nilpotent (anti-)BRST  transformations  for the 
coupled (but equivalent) Lagrangian densities ${\cal L}_{(B)}$ and ${\cal L}_{(\bar B)} $,
respectively,  lead to the derivations of the (anti-)BRST invariant 
CF-type restrictions (B.1).

We end this Appendix with a few {\it  side remarks}. First of all, we note that the EL-EoMs from the FP-Lagrangian densities (3) and (9) w.r.t. the 
pair of fermionic {\it auxiliary} fields
($\bar F_\mu, \, f_\mu$) and ($ F_\mu, \,\bar f_\mu $), respectively, lead to the following relationships:
\[
f_\mu = \dfrac{1}{2} \, \big (\partial^\nu C_{\nu\mu} + \partial_\mu C_1 \big ), \qquad 
\bar F_\mu = \dfrac{1}{2} \, \big (\partial^\nu \bar C_{\nu\mu} + \partial_\mu \bar C_1 \big ),
\]
\[
\bar f_\mu = -\, \dfrac{1}{2} \, \big (\partial^\nu \bar C_{\nu\mu} - \partial_\mu \bar C_1 \big ), \qquad 
F_\mu = -\,\dfrac{1}{2} \, \big (\partial^\nu  C_{\nu\mu} - \partial_\mu  C_1 \big ).
\eqno(B.10)
\]
A close look at the above relationships implies that we have
\[
f_\mu + F_\mu = \partial_\mu C_1, \qquad \bar f_\mu + \bar F_\mu = \partial_\mu \bar C_1,
\eqno(B.11)
\]
which  are nothing but the (anti-)BRST invariant {\it fermionic} sectors of the CF-type restrictions on our theory
that have been mentioned in (B.1). On the other hand, the EL-EoMs w.r.t. to the pair of Nakanishi-Lautrup
auxiliary fields ($B_{\mu\nu}, {\cal B}_{\mu\nu} $) and ($\bar B_{\mu\nu}, \bar {\cal B}_{\mu\nu} $) from the non-ghost sectors of
the Lagrangian densities (1) and (8), respectively, yield the following 
\[
B_{\mu\nu} = \partial^\sigma A_{\sigma\mu\nu} + \dfrac{1}{2} \, \big (\partial_\mu \phi_\nu - \partial_\nu \phi_\mu \big ),
\qquad
{\cal B}_{\mu\nu} = \varepsilon_{\mu\nu\sigma\rho} \,\partial^\sigma A^\rho
+ \dfrac{1}{2} \, \big (\partial_\mu \tilde \phi_\nu - \partial_\nu \tilde \phi_\mu \big ),
\]
\[
\bar B_{\mu\nu} = -\, \partial^\sigma A_{\sigma\mu\nu} + \dfrac{1}{2} \, \big (\partial_\mu \phi_\nu - \partial_\nu \phi_\mu \big ),
\quad \;
\bar {\cal B}_{\mu\nu} = -\,  \varepsilon_{\mu\nu\sigma\rho}\, \partial^\sigma A^\rho
+ \dfrac{1}{2} \, \big (\partial_\mu \tilde \phi_\nu - \partial_\nu \tilde \phi_\mu \big ),
\eqno(B.12)
\]
which imply that we have the validity of the relationships: $B_{\mu\nu} + \bar B_{\mu\nu} = \partial_\mu \phi_\nu - \partial_\nu \phi_\mu, \;
{\cal B}_{\mu\nu} + \bar {\cal B}_{\mu\nu} = \partial_\mu \tilde \phi_\nu - \partial_\nu \tilde \phi_\mu$ which are nothing but the 
bosonic sectors of the CF-type restrictions [cf. Eq. (B.1)]. We would like to lay emphasis on the fact that our key observations
in (B.10), (B.11) and (B.12) are {\it not} the {\it formal} proof of the derivations the CF-type restrictions (B.1). However, these
observations are {\it crucial} in the proof that the Lagrangian densities ${\cal L}_{(B)} $ and ${\cal L}_{(\bar B)} $ are
{\it coupled} Lagrangian  densities\footnote{It is pertinent to point out that, for the first time,
 the ideas behind (i) the {\it original} CF-condition [74], and
(ii) the coupled (but equivalent) Lagrangian densities appeared  in the context of the BRST
approach to the D-dimensional non-Abelian gauge theory (see. e.g. [45] for details).}. In other words, the CF-type restrictions (B.1) are the root-cause behind the
existence of the {\it coupled} Lagrangian densities (i.e. ${\cal L}_{(B)} $ and ${\cal L}_{(\bar B)} $). Furthermore, we would like to add here that the 
bosonic as well as the fermionic sectors of the CF-type restrictions [cf. Eq. (B.1)] have not been derived from a {\it single} Lagrangian density. Hence, these
(anti-)BRST invariant relationships [cf. Eq. (12)] are the {\it physical} constraints on our theory and {\it not} the EL-EoMs.
Finally, in all our discussions (connected with the off-shell nilpotent (anti-)BRST symmetry transformations and coupled 
Lagrangian densities ${\cal L}_{(B)}$ and ${\cal L}_{(\bar B)} $), we observe that the CF-type restriction:
${\cal B}_{\mu\nu} + \bar {\cal B}_{\mu\nu} = \partial_\mu \tilde \phi_\nu - \partial_\nu \tilde \phi_\mu $ does {\it not} play any crucial role.
This is due to the fact that {\it this} CF-type restriction is useful in the context of the nilpotent (anti-)co-BRST symmetry transformations
which we plan to discuss in our future endeavor.  The combined set of the key CF-type restrictions 
(i.e. $B_{\mu\nu} + \bar B_{\mu\nu} = \partial_\mu \phi_\nu - \partial_\nu \phi_\mu, \; f_\mu + F_\mu = \partial_\mu C_1, 
\; \bar f_\mu + \bar F_\mu = \partial_\mu \bar C_1$) can be {\it also} derived from the requirement of the absolute anticommutativty 
(i.e. $\{ Q_B, \; Q_{AB} \} \equiv Q_B\, Q_{AB} + Q_{AB} \, Q_B = 0$) of the 
conserved and {\it nilpotent} versions of the (anti-)BRST charges $Q_{(A)B}$. However,
we envisage to perform this exercise in our forthcoming research paper.\\

\vskip 0.7 cm 
\begin{center}
{\bf Appendix C: \bf On the Ghost Charge and Standard BRST Algebra}\\
\end{center}

\vskip 0.5 cm 

\noindent
It is an elementary exercise to observe that the  {\it coupled} (but equivalent) Lagrangian densities 
${\cal L}_{(B)} $ and ${\cal L}_{(\bar B)} $ of our 4D field-theoretic system remain invariant under the following ghost-scale symmetry
transformations, namely;
\[
C_{\mu\nu} \rightarrow e^{+\Sigma} C_{\mu\nu}, \quad \bar{C}_{\mu\nu} \rightarrow e^{-\Sigma} \bar{C}_{\mu\nu}, \quad 
\beta_\mu \rightarrow e^{+2 \Sigma} \beta_\mu, \quad \bar{\beta}_\mu \rightarrow e^{-2 \Sigma} \bar{\beta}_\mu, 
\]
\[
f_\mu \rightarrow e^{+ \Sigma} f_\mu, \quad \bar{f}_\mu \rightarrow e^{- \Sigma} \bar{f}_\mu, \quad
F_\mu \rightarrow e^{+ \Sigma} F_\mu, \quad \bar{F}_\mu \rightarrow e^{- \Sigma} \bar{F}_\mu,
\]
\[
C_2 \rightarrow e^{+3\,\Sigma} C_2, \quad \bar{C_2} \rightarrow e^{- 3\,\Sigma} \bar{C_2}, \quad
C \rightarrow e^{+\Sigma} C, \quad \bar{C} \rightarrow e^{-\Sigma} \bar{C}, 
\]
\[
C_1 \rightarrow e^{+\Sigma} C_1, \quad \bar{C_1} \rightarrow e^{-\Sigma} \bar{C_1}, 
\quad B_4 \rightarrow e^{+ 2\, \Sigma} B_4, \quad B_5 \rightarrow e^{-2\,\Sigma} B_5, 
\]
\[ 
\Phi_i \rightarrow e^{0} \Phi_i \quad\left(\Phi_i = A_{\mu\nu\sigma}, B_{\mu \nu}, {\cal B}_{\mu\nu}, \bar B_{\mu \nu}, \bar {\cal B}_{\mu\nu}, A_\mu,
B, B_1, B_2, B_3, \phi_{\mu}, \tilde \phi_\mu \right),
\eqno(C.1)
\]
where $\Sigma$ is a spacetime-independent (i.e. global) scale transformation parameter and the numerals in the exponents correspond to the ghost numbers for the fields. It is clear that {\it all} the fields of the non-ghost sectors 
(i.e. $A_{\mu\nu\sigma}, B_{\mu \nu}, {\cal B}_{\mu\nu}, \bar B_{\mu \nu}, \bar {\cal B}_{\mu\nu}, A_\mu, , \phi_{\mu}, \tilde \phi_\mu, B, B_1, B_2, B_3$) 
carry the ghost number equal to zero. Hence, in the exponent, corresponding to the ghost-scale symmetry transformation on the generic field $\Phi_i$, we have 
{\it zero} as the numeral. For the sake of brevity, we set the global
scale transformation parameter  $\Sigma=1$ so that the infinitesimal version $\left(s_{g}\right)$ 
of the above ghost-scale symmetry transformations is
\[
s_{g} C_{\mu\nu}= + \,C_{\mu\nu}, \quad s_{g} \bar{C}_{\mu\nu}= -\, \bar{C}_{\mu\nu}, \quad s_{g} \beta_\mu = + 2 \,\beta_\mu, 
\quad s_{g} \bar{\beta}_\mu = - 2\, \bar{\beta}_\mu, \]
\[
 s_{g} f_\mu = +  \,f_\mu, \quad s_{g} \bar f_\mu = - \, \bar f_\mu, \quad s_{g} F_\mu = +  \,F_\mu, 
\quad s_{g} \bar F_\mu = - \, \bar F_\mu,
\]
\[
s_{g} C_2 = + 3\, C_2, \quad s_{g} \bar{C}_2= - 3\, \bar{C}_2, \quad s_{g} C = +\, C, 
\quad s_{g} \bar{C}= - \, \bar{C}, \]
\[ 
s_{g} C_1 = +\, C_1, \quad s_{g} \bar{C}_1= - \, \bar{C}_1, \quad s_{g} B_4 = + 2\, B_4, \quad s_{g} B_5= - 2 \, B_5, \quad s_{g} \Phi=0,
\eqno(C.2)
\]
where it can be readily checked that $s_{g}$ is bosonic (i.e. $s_{g}^{2} \neq 0$) in nature. As a consequence, we observe that the bosonic
fields of the coupled (but equivalent) Lagrangian densities ${\cal L}_{(B)} $ and ${\cal L}_{(\bar B)} $ of our 4D field-theoretic system
transform to the bosonic fields and the fermionic fields transform to the fermionic fields {\it without} any change in the ghost numbers.
Under these infinitesimal transformations, the {\it perfectly} (anti-)BRST invariant Lagrangian densities remain invariant 
(i.e. $s_{g} \mathcal{L}_{(B)}=0,\; s_{g} \mathcal{L}_{(\bar{B})}=0$). As a consequence, according to Noether's theorem, we have the following expression for the ghost current $J_{(g)}^{\mu}$, namely;
\[
J_{(g)}^{\mu}= \dfrac{1}{2}\, \Big [(\partial^\mu C^{\nu\sigma} + \partial^\nu C^{\sigma\mu}  + \partial^\sigma C^{\mu\nu} \big )\, \bar C_{\nu\sigma}
+ \big (\partial^\mu \bar C^{\nu\sigma} + \partial^\nu \bar C^{\sigma\mu}  + \partial^\sigma \bar C^{\mu\nu} \big )\,  C_{\nu\sigma}
\]
\[ 
- 2\, \left(\partial^{\mu} \bar \beta^{\nu}-\partial^{\nu} \bar \beta^{\mu}\right)\, \beta_{\nu}
+ 2\, \left(\partial^{\mu} \beta^{\nu}-\partial^{\nu} \beta^{\mu}\right)\, \bar \beta_{\nu} - C^{\mu\nu} \, \bar F_\nu - \bar C^{\mu\nu} \, f_\nu
- C_1 \, \bar F^{\mu}\] 
\[
-  \,\bar C_1 \, f^{\mu} 
+ 3\, C_2\, \partial^{\mu} \bar{C}_2 + 3\, \bar{C}_2 \, \partial^{\mu}  C_2  - 2\, \beta^\mu\, B_5 - 2\, \bar \beta^\mu\, B_4 \Big ]
+ C\, \partial^{\mu} \bar{C} + \bar{C} \, \partial^{\mu}  C,
\eqno(C.3)
\]
derived from the Lagrangian density ${\cal L}_{(FP)}$ [cf. Eq. (3)]. On the other hand, the Noether ghost current $J_{(\bar g)}^{\mu}$ (that emerges out
from the Lagrangian density corresponding to the FP-ghost term ${\cal L}_{(fp)}$ [cf. Eq. (9)]) is as follows:
\[
J_{(\bar g)}^{\mu}= \dfrac{1}{2}\,
\Big  [(\partial^\mu C^{\nu\sigma} + \partial^\nu C^{\sigma\mu}  + \partial^\sigma C^{\mu\nu} \big )\, \bar C_{\nu\sigma}
+ \big (\partial^\mu \bar C^{\nu\sigma} + \partial^\nu \bar C^{\sigma\mu}  + \partial^\sigma \bar C^{\mu\nu} \big )\,  C_{\nu\sigma}
\]
\[ 
- 2\, \left(\partial^{\mu} \bar \beta^{\nu}-\partial^{\nu} \bar \beta^{\mu}\right)\, \beta_{\nu}
+ 2\, \left(\partial^{\mu} \beta^{\nu}-\partial^{\nu} \beta^{\mu}\right)\, \bar \beta_{\nu} + C^{\mu\nu} \, \bar f_\nu + \bar C^{\mu\nu} \, F_\nu
- C_1 \, \bar f^{\mu} 
\]
\[
-  \,\bar C_1 \, F^{\mu}  
+ 3\, C_2\, \partial^{\mu} \bar{C}_2 + 3\, \bar{C}_2 \, \partial^{\mu}  C_2  - 2\, \beta^\mu\, B_5 - 2\, \bar \beta^\mu\, B_4 \Big ]
+ C\, \partial^{\mu} \bar{C} + \bar{C} \, \partial^{\mu}  C.
\eqno(C.4)
\]
It is obvious that (i) only the {\it dynamical} (anti-)ghost fields from the FP-ghost parts 
[cf. Eqs. (3),(9)] of the (anti-)BRST invariant Lagrangian  densities
contribute in the computation of $J_{(g)}^{\mu} $ and $J_{(\bar g)}^{\mu} $, and (ii) the {\it auxiliary} (anti-)ghost fields as well as the generic field
$\Phi_i$ of the non-ghost sectors [cf. Eqs. (1),(8)] of the coupled (but equivalent) (anti-)BRST invariant Lagrangian  
densities ${\cal L}_{(B)} $ and ${\cal L}_{(\bar B)} $ also do {\it not} contribute anything to (C.3) and (C.4).

The conservation law ($\partial_{\mu} J_{(g)}^{\mu}=0$) can be proven by exploiting the EL-EoMs from the
ghost-sector [cf. Eq. (3)] 
of the Lagrangian density (i.e. ${\cal L}_{(FP)}$), namely; 
\[
\Box C = 0, \qquad \Box \bar C = 0, \qquad  \Box C_2 = 0, \qquad \Box \bar C_2 = 0, \qquad (\partial \cdot \bar F) = 0, \qquad (\partial \cdot f) = 0, 
\]
\[
B_4 = -\, (\partial \cdot \beta),  \quad 
B_5 = (\partial \cdot \bar \beta), \quad f_\mu = \dfrac{1}{2} \, \big (\partial^\nu C_{\nu\mu} + \partial_\mu C_1 \big ), \quad 
\bar F_\mu = \dfrac{1}{2} \, \big (\partial^\nu \bar C_{\nu\mu} + \partial_\mu \bar C_1 \big ), 
\]
\[
\partial_\mu \big (\partial^\mu \beta^\nu - \partial^\nu \beta^\mu \big) - \partial^\nu B_4 = 0, \qquad
\partial_\mu \big (\partial^\mu \bar \beta^\nu - \partial^\nu \bar \beta^\mu \big) + \partial^\nu B_5 = 0,
\]
\[ 
\partial_\mu \big (\partial^\mu \bar C^{\nu\sigma} + \partial^\nu \bar C^{\sigma\mu} + \partial^\sigma \bar C^{\mu\nu} \big )
+ \dfrac{1}{2}\, \big (\partial^\nu \bar F^\sigma - \partial^\sigma \bar F^\nu \big ) = 0,
\] 
\[ 
\partial_\mu \big (\partial^\mu  C^{\nu\sigma} + \partial^\nu C^{\sigma\mu} + \partial^\sigma  C^{\mu\nu} \big )
+ \dfrac{1}{2}\, \big (\partial^\nu f^\sigma - \partial^\sigma f^\nu \big ) = 0.
\eqno(C.5)
\] 
In exactly similar fashion, the conservation law ($\partial_{\mu} J_{(\bar g)}^{\mu}=0$) can be proven by exploiting the EL-EoMs that are
derived from the ghost-sector of the Lagrangian density ${\cal L}_{(fp)}$ [cf. Eq. (9)]. These EL-EoMs are {\it same} as (C.5) {\it except}
the following
\[
F_\mu = -\, \dfrac{1}{2} \, \big (\partial^\nu C_{\nu\mu} - \partial_\mu C_1 \big ), \quad 
\bar f_\mu = -\, \dfrac{1}{2} \, \big (\partial^\nu \bar C_{\nu\mu} - \partial_\mu \bar C_1 \big ), \quad  (\partial \cdot  F) = 0, \quad 
(\partial \cdot \bar f) = 0,
\]  
\[ 
\partial_\mu \big (\partial^\mu \bar C^{\nu\sigma} + \partial^\nu \bar C^{\sigma\mu} + \partial^\sigma \bar C^{\mu\nu} \big )
- \dfrac{1}{2}\, \big (\partial^\nu \bar f^\sigma - \partial^\sigma \bar f^\nu \big ) = 0,
\] 
\[ 
\partial_\mu \big (\partial^\mu  C^{\nu\sigma} + \partial^\nu C^{\sigma\mu} + \partial^\sigma  C^{\mu\nu} \big )
- \dfrac{1}{2}\, \big (\partial^\nu F^\sigma - \partial^\sigma F^\nu \big ) = 0,
\eqno(C.6)
\] 
which are {\it different} from their analogues in (C.5). The expressions for the ghost charges $Q_{(g)}=\int d^{3} x J_{(g)}^{0}$ 
and $Q_{(\bar g)}=\int d^{3} x J_{(\bar g)}^{0}$ are as follows:
\[
Q_{(g)}=\int d^{3} x\, \Big [  \frac{1}{2}\, \Big \{ \big (\partial^0 \bar C^{ij} + \partial^i \bar C^{j0} + \partial^j \bar C^{0i} \big )\, C_{ij}
+ \big (\partial^0 C^{ij} + \partial^i  C^{j0} + \partial^j  C^{0i} \big )\, \bar C_{ij}
\]
\[
- 2\, \big (\partial^0 \bar \beta^i -  \partial^i \bar \beta^0 \big )\, \beta_i + 2\, \big (\partial^0 \beta^i -  \partial^i  \beta^0 \big )\, \bar \beta_i
- \frac{1}{2}\, C^{0i}\, \bar F_i + \frac{1}{2} \,  C^{i0}\, \bar F_i- \frac{1}{2}\, \bar C^{0i} \, f_i + \frac{1}{2}\, \bar C^{i0} \, f_i
\]
\[
- C_1\, \bar F^0 - \bar C_1\, f^0
+ 3\, C_2 \, \dot {\bar C_2} + 3\, \bar C_2 \, \dot { C_2} - 2\, \beta^0\, B_5 - 2\, \bar \beta^0\, B_4 \Big \} + C\, \dot {\bar C} + \bar C \, \dot C
\Big ],
\eqno(C.7)
\]
\[
Q_{(\bar g)}=\int d^{3} x\, \Big [ \frac{1}{2}\, \Big \{ \big (\partial^0 \bar C^{ij} + \partial^i \bar C^{j0} + \partial^j \bar C^{0i} \big )\, C_{ij}
+ \big (\partial^0 C^{ij} + \partial^i  C^{j0} + \partial^j  C^{0i} \big )\, \bar C_{ij}
\]
\[
- 2\, \big (\partial^0 \bar \beta^i -  \partial^i \bar \beta^0 \big )\, \beta_i + 2\, \big (\partial^0 \beta^i -  \partial^i  \beta^0 \big )\, \bar \beta_i
+ \frac{1}{2}\, C^{0i}\,\bar  f_i  - \frac{1}{2}\, C^{io}\,\bar  f_i + \frac{1}{2}\, \bar C^{0i} \, F_i - \frac{1}{2}\, \bar C^{i0} \, F_i
\]
\[
- C_1\, \bar f^0 - \bar C_1\, F^0 
+ 3\, C_2 \, \dot {\bar C}_2
 + 3\, \bar C_2 \, \dot { C_2} - 2\, \beta^0\, B_5 - 2\, \bar \beta^0\, B_4 \Big \} + C\, \dot {\bar C} + \bar C \, \dot C
\Big ].
\eqno(C.8)
\]
The above conserved charges $Q_{(g)}$  and $Q_{(\bar g)}$ are  the {\it equivalent} generators for the infinitesimal ghost-scale
symmetry  transformations (C.2) if we express the ghost charges in terms of the 
 canonical conjugate  momenta w.r.t. the {\it basic} dynamical (anti-)ghost fields of the ghost-sectors of the Lagrangian  densities
${\cal L}_{(FP)} $ and ${\cal L}_{(fp)} $ [cf. Eqs. (3),(9)].

To corroborate the above statement, we note that the following conjugate
momenta are {\it common} for both the FP-Lagrangian densities (3) and (9), namely;
\[
\Pi^{\mu}_{(\beta)}  = \dfrac{\partial{\cal L}_{(FP, fp)}}{\partial (\partial_0 \beta_{\mu})} =  -\,\dfrac{1}{2}\,
 \big(\partial^0 \bar \beta^\mu -  \partial^\mu \bar \beta^{0} \big ) -\frac{1}{2} \,\big (\eta^{0\mu}\, B_5 \big )  \Longrightarrow 
\]
\[
\Pi^{0}_{(\beta)} = -\, \frac{1}{2}\, B_5, \quad
\Pi^{i}_{(\beta)}  =   -\, \dfrac{1}{2}\, 
\big(\partial^0 \bar \beta^\mu -  \partial^\mu \bar \beta^{0} \big ),  \quad  
\Pi_{(C)}  = \dfrac{\partial{\cal L}_{(FP, fp)}}{\partial (\partial_0 C)} =  \dot {\bar C},                                 
\]
\[
\Pi^{\mu}_{(\bar \beta)}  = \dfrac{\partial{\cal L}_{(FP, fp)}}{\partial (\partial_0 \bar \beta_{\mu})} =  -\,\dfrac{1}{2}\,
 \big(\partial^0  \beta^\mu -  \partial^\mu  \beta^{0} \big ) + \frac{1}{2} \,\big (\eta^{0\mu}\, B_4 \big )  \Longrightarrow 
\]
\[
\Pi^{0}_{(\bar \beta)} = +\, \frac{1}{2}\, B_4, \quad
\Pi^{i}_{(\bar \beta)}  =   -\, \dfrac{1}{2}\, 
\big(\partial^0 \bar \beta^\mu -  \partial^\mu \bar \beta^{0} \big ),  \quad
  \Pi_{(\bar C )}  = \dfrac{\partial{\cal L}_{(FP, fp)}}{\partial (\partial_0 \bar C)} =  -\, \dot C, 
\]                                 
\[
\Pi_{(\bar C_2)}  = \dfrac{\partial{\cal L}_{(FP, fp)}}{\partial (\partial_0 {\bar C}_2)} =  -\,\dfrac{1}{2}\,\dot { C_2}, \qquad 
\Pi_{(C_2)}  = \dfrac{\partial{\cal L}_{(FP, fp)}}{\partial (\partial_0 C_2)} =  \,\dfrac{1}{2}\,\dot {\bar  C}_2, 
\eqno(C.9)
\]
where the subscripts $(FP,fp)$ on the Lagrangian density imply that the above conjugate momenta have been derived from {\it both} the
Lagrangian densities in the ghost-sectors [cf. Eqs. (3),(9)]. In addition, we have the following expressions for the conjugate momenta which are
derived {\it only} from the Lagrangian density  ${\cal L}_{(FP)} $ [cf. Eq. (3)], namely;
\[
\Pi^{\mu\nu}_{(C)} (FP) = \dfrac{\partial{\cal L}_{(FP)}}{\partial (\partial_0 C_{\mu\nu})} = -\, \dfrac{1}{2}\,
\Big[ \big(\partial^0 \bar C^{\mu\nu} +  \partial^\mu \bar C^{\nu 0} + \partial^\nu \bar C^{0\mu} \big ) + \frac{1}{2} \,\big (\eta^{0\mu}\, \bar F^\nu - 
\eta^{0\nu}\, \bar F^\mu \big ) \Big ] \Longrightarrow 
\]
\[
\Pi^{ij}_{(C)} (FP) = - \frac{1}{2}\, \big(\partial^0 \bar C^{ij} +  \partial^i \bar C^{j0} + \partial^j \bar C^{i0} \big ), \quad
\Pi^{0i}_{(C)} (FP) =   \dfrac{1}{4} \bar F_{i}, \quad  \Pi^{io}_{(C)} (FP) =  -\, \dfrac{1}{4} \bar F_{i},                                     
\]
\[
\Pi^{\mu\nu}_{(\bar C)} (FP) = \dfrac{\partial{\cal L}_{(FP)}}{\partial (\partial_0 \bar C_{\mu\nu})} =  \dfrac{1}{2}\,
\Big[ \big(\partial^0 C^{\mu\nu} +  \partial^\mu  C^{\nu 0} + \partial^\nu  C^{0\mu} \big ) + \frac{1}{2} \,\big (\eta^{0\mu}\, f^\nu - 
\eta^{0\nu}\, f^\mu \big ) \Big ] \Longrightarrow 
\]
\[
\Pi^{ij}_{(\bar C)} (FP)=  \frac{1}{2}\, \big(\partial^0  C^{ij} +  \partial^i  C^{j0} + \partial^j  C^{i0} \big ), \quad
\Pi^{0i}_{(\bar C)}  (FP) = -\, \dfrac{1}{4} f_{i}, \quad  \Pi^{io}_{(\bar C)} (FP)  = \dfrac{1}{4} f_{i},                                     
\]
\[
\Pi_{(\bar C_1)} (FP) = \dfrac{\partial{\cal L}_{(FP)}}{\partial (\partial_0 \bar C_{1})}  = +\,\dfrac{1}{2}\, f^0, \qquad
\Pi_{(C_1)} (FP) = \dfrac{\partial{\cal L}_{(FP)}}{\partial (\partial_0 C_{1})} = -\, \dfrac{1}{2}\, \bar F^0,
\eqno(C.10)                                   
\]
where $(FP)$, in the parenthesis of all the conjugate momenta, denotes the fact that they have been derived from 
the Lagrangian density (3). Analogous to  (C.10), we have the conjugate momenta from the Lagrangian density (9), namely;
\[
\Pi^{\mu\nu}_{(C)} (fp) = \dfrac{\partial{\cal L}_{(fp)}}{\partial (\partial_0 C_{\mu\nu})} = -\, \dfrac{1}{2}\,
\Big[ \big(\partial^0 \bar C^{\mu\nu} +  \partial^\mu \bar C^{\nu 0} + \partial^\nu \bar C^{0\mu} \big ) - \frac{1}{2} \,\big (\eta^{0\mu}\, \bar f^\nu - 
\eta^{0\nu}\, \bar f^\mu \big ) \Big ] \Longrightarrow 
\]
\[
\Pi^{ij}_{(C)} (fp) = - \frac{1}{2}\, \big(\partial^0 \bar C^{ij} +  \partial^i \bar C^{j0} + \partial^j \bar C^{i0} \big ), \quad
\Pi^{0i}_{(C)} (fp) =  -\, \dfrac{1}{4} \,\bar f_{i}, \quad  \Pi^{io}_{(C)} (fp) =  +\, \dfrac{1}{4}\, \bar f_{i},                                     
\]
\[
\Pi^{\mu\nu}_{(\bar C)} (fp) = \dfrac{\partial{\cal L}_{(fp)}}{\partial (\partial_0 \bar C_{\mu\nu})} =  \dfrac{1}{2}\,
\Big[ \big(\partial^0 C^{\mu\nu} +  \partial^\mu  C^{\nu 0} + \partial^\nu  C^{0\mu} \big ) - \frac{1}{2} \,\big (\eta^{0\mu}\, F^\nu - 
\eta^{0\nu}\, F^\mu \big ) \Big ] \Longrightarrow 
\]
\[
\Pi^{ij}_{(\bar C)} (fp)=  \frac{1}{2}\, \big(\partial^0  C^{ij} +  \partial^i  C^{j0} + \partial^j  C^{i0} \big ), \quad
\Pi^{0i}_{(\bar C)}  (fp) = +\, \dfrac{1}{4} F_{i}, \quad  \Pi^{io}_{(\bar C)} (fp)  = -\,\dfrac{1}{4} F_{i},                                     
\]
\[
\Pi_{({\bar C}_1)} (fp) = \dfrac{\partial{\cal L}_{(fp)}}{\partial (\partial_0 \bar C_{1})}  = \dfrac{1}{2}\, F^0, \qquad
\Pi_{(C_1)} (fp) = \dfrac{\partial{\cal L}_{(FP)}}{\partial (\partial_0 C_{1})} = -\, \dfrac{1}{2}\, \bar f^0,
\eqno(C.11)                                   
\]
where we have chosen the symbol $(fp)$, in the parenthesis of all the conjugate momenta, to denote that they have been
derived from the Lagrangian density ${\cal L}_{(fp)} $ [cf. Eq. (9)].

Taking into account the inputs from (C.9), (C.10) and (C.11), we can 
succinctly express the conserved ghost charges in (C.7) and (C.8) as:
\[
Q_{(g)}=\int d^{3} x\, \Big \{ -\, \big [\Pi^{ij}_{(C)} (FP)\big ]\, C_{ij}
+ \big [\Pi^{ij}_{(\bar C)} (FP) \big ]\, \bar C_{ij} + 2\, \big [\Pi^i_{(\beta)} \big ]\, \beta_i 
- 2\, \big [\Pi^i_{(\bar \beta)}\big ]\, \bar \beta_i
\]
\[
+ \big [\Pi^{0i}_{(C)} (FP) \big ]\; C_{0i} + \big [\Pi^{i0}_{(C)} (FP)\big ]\; C_{i0} 
+ \big [ \Pi^{0i}_{(\bar C)} (FP) \big ]\;\bar  C_{0i} - \big [\Pi^{i0}_{(\bar C)} (FP) \big ]\;\bar  C_{i0}
\]
\[
- \big [\Pi_{(C_1)} (FP) \big ] \, C_1 + \big [\Pi_{(\bar C_1)} (FP) \big ] \,  \bar C_1 
- 3\,\big [\Pi_{(C_2)} \big ] \,C_2  + 3\, \big [\Pi_{(\bar C_2)} \big ]\,\bar C_2  
\]
\[
+ 2\, \big [\Pi^0_{(\beta)} \big ] \, \beta_0 - 2\, \big [\Pi^0_{(\bar \beta)} \big ]\,\bar \beta_0 -  \big [\Pi_{(C)} \big ]
C + \big [\Pi_{(\bar C)} \big ] \, \bar C \Big \},
\eqno(C.12)
\]
\[
Q_{(\bar g)}=\int d^{3} x\, \Big \{ -\, \big [\Pi^{ij}_{(C)} (fp)\big ]\, C_{ij}
+ \big [\Pi^{ij}_{(\bar C)} (fp) \big ]\, \bar C_{ij} + 2\, \big [\Pi^i_{(\beta)} \big ]\, \beta_i 
-\, 2\, \big [\Pi^i_{(\bar \beta)}\big ]\, \bar \beta_i
\]
\[
+ \big [\Pi^{0i}_{(C)} (fp) \big ]\; C_{0i} + \big [\Pi^{i0}_{(C)} (fp)\big ]\; C_{i0} 
+ \big [ \Pi^{0i}_{(\bar C)} (fp) \big ]\;\bar  C_{0i} - \big [\Pi^{i0}_{(\bar C)} (fp) \big ]\;\bar  C_{i0}
\]
\[
- \big [\Pi_{(C_1)} (fp) \big ] \, C_1 + \big [\Pi_{(\bar C_1)} (fp) \big ] \,  \bar C_1 
- 3\,\big [\Pi_{(C_2)} \big ] \,C_2  + 3\, \big [\Pi_{(\bar C_2)} \big ]\,\bar C_2  
\]
\[
+ \,2\, \big [\Pi^0_{(\beta)} \big ] \, \beta_0 - 2\, \big [\Pi^0_{(\bar \beta)} \big ]\,\bar \beta_0 -  \big [\Pi_{(C)} \big ]
C + \big [\Pi_{(\bar C)} \big ] \, \bar C \Big \}.
\eqno(C.13)
\]
A close look at the above expressions demonstrate that {\it both}  the charges are of the
{\it same} form and they lead to the derivations of the infinitesimal versions
of the ghost-scale symmetry transformations (C.2) if we use the canonical brackets between the {\it dynamical} (anti-)ghost fields that are
present in the Lagrangian densities ${\cal L}_{(FP, fp)}$ [cf. Eqs. (3),(9)] and corresponding conjugate momenta that have been listed in (C.9), (C.10)
and (C.11). In other words, the infinitesimal ghost-scale symmetry transformations 
(C.2) are exactly {\it same} for the Lagrangian  densities  ${\cal L}_{(FP)}$ and  ${\cal L}_{(fp)}$. For readers' convenience, we work out a couple of 
examples to establish that {\it both} the Noether charges $Q_{(g)}$ and $Q_{(\bar g)}$ generate (C.2). First of all, let us 
take the fermionic field $C_{ij}$  and show that: $s_g C_{ij} = +\, C_{ij}$
is generated 
by the conserved ghost charge $Q_{(g)}$ as well as $Q_{(\bar g)}$. Toward this goal in mind, we note that\footnote{For the sake of brevity, we have chosen
the notation $Q_{(g, \,\bar g)} $ for the ghost charges $Q_{(g)}$ and $Q_{(\bar g)}$. To be precise, in the computation of (C.14) and (C.15), 
{\it either} of the ghost charges can be taken into account.}
\[
s_g C_{ij} (\vec {x}, t) = +\,i \,\big [C_{ij} (\vec {x}, t), \; Q_{(g, \,\bar g)} \big ] = -\, i\,
\int d^3 y \big [C_{ij} (\vec {x}, t), \; \Pi^{kl}_{(C)} (\vec {y}, t)\; C_{kl} (\vec {y}, t) \big ]
\]
\[
\equiv -\, i\, \int d^3 y \big \{ C_{ij} (\vec {x}, t), \; \Pi^{kl}_{(C)} (\vec {y}, t) \big \} \, C_{kl} (\vec {y}, t)
 + \, i\, \int d^3 y \, \Pi^{kl}_{(C)} (\vec {y}, t)\, \big \{ C_{ij} (\vec {x}, t), \;   C_{kl} (\vec {y}, t) \big \},
 \eqno(C.14)
\]
where we have used (i) the relationship between the infinitesimal transformations (C.2) and their generators as: ($Q_{(g)}, \,Q_{(\bar g)}$), and (ii)
the commutator relationship: $ [A, \; B\, C ] = \{A, \; B \}\, C - B\, \{A, \; C \}$ for 
{\it all} the generic operators $A, B$ and $C$ being {\it fermionic} in nature.
Now it is straightforward to observe that if 
we use the canonical anticommutator: $\big \{ C_{ij} (\vec {x}, t), \; \Pi^{kl}_{(C)} (\vec {y}, t) \big \} = \frac{i}{2}\, \big (
\delta_i^k \, \delta_j^l  - \delta_i^l \, \delta_j^k \big )\, \delta^{(3)} \big (\vec{x} - \vec{y} \big )$, we obatin the result:  $s_g C_{ij} = +\, C_{ij}$
provided we also take into account the canonical anticommutator:
$\big \{ C_{ij} (\vec {x}, t), \;   C_{kl} (\vec {y}, t) \big \} = 0 $. For the sake of completeness, let us take into consideration a bosonic field
$\beta_i$ and establish that: $s_g \beta_i = + 2\, \beta_i$ is generated by $Q_{(g)}$ and $Q_{(\bar g)}$. To accomplish this goal, let us focus on:
\[
s_g \beta_{i} (\vec {x}, t) = -\,i \,\big [\beta_{i} (\vec {x}, t), \; Q_{(g, \,\bar g)} \big ] = -\, i\,
\int d^3 y \big [\beta_{i} (\vec {x}, t), \; 2\,\Pi^{j}_{(\beta)} (\vec {y}, t)\; \beta_{j} (\vec {y}, t) \big ]
\]
\[
\equiv -\,2\, i\, \int d^3 y \big [ \beta_{i} (\vec {x}, t), \; \Pi^{j}_{(\beta)} (\vec {y}, t) \big ] \, \beta_{j} (\vec {y}, t)
 -\,2\, i\, \int d^3 y \, \Pi^{j}_{(\beta)}  (\vec {y}, t)\, \big [ \beta_{i} (\vec {x}, t), \;   \beta_{j} (\vec {y}, t) \big ].
 \eqno(C.15)
\]
It is crystal clear, from the above equation,  that if we take into account the canonical commutators: 
$\big [ \beta_{i} (\vec {x}, t), \; \Pi^{j}_{(\beta)} (\vec {y}, t) \big ] = i\, \delta_i^j \, \delta^{(3)} \big (\vec{x} - \vec{y} \big )$ and 
$\big [ \beta_{i} (\vec {x}, t), \;   \beta_{j} (\vec {y}, t) \big ] = 0 $, we obtain the desired result: $s_g \beta_i = + 2\, \beta_i$
in a straightforward fashion from (C.15).

At this juncture, we choose  the ghost charge $Q_{g}$ as the generator for the infinitesimal ghost-scale transformations $s_{g}$ and 
observe that the following relationships between the ghost charge and the non-nilpotent  (i.e. $Q_{(a) b}^{2} \neq 0$) versions
and nilpotent   (i.e. $ Q_{(A) B}^{2}=0$) versions of the (anti-)BRST charges (i.e. $Q_{(a) b},\; Q_{(A) B}$) are true, namely;
\[
s_{g} Q_{b}=-i\left[Q_{b}, Q_{g}\right]= + \,Q_{b}  \quad \Longrightarrow \quad  i\left[Q_{g}, Q_{b}\right]= + \,Q_{b}, 
\]
\[
s_{g} Q_{a b}=-i\left[Q_{a b}, Q_{g}\right]=-\,Q_{a b}  \quad \Longrightarrow \quad   i\left[Q_{g}, Q_{a b}\right]=- \,Q_{a b},
 \]
 \[
s_{g} Q_{B}=-i\left[Q_{B}, Q_{g}\right]= + \,Q_{B}  \quad \Longrightarrow \quad   i\left[Q_{g}, Q_{B}\right]=+ \,Q_{B},
\]
\[
s_{g} Q_{A B}=-i\left[Q_{A B}, Q_{g}\right]= -\, Q_{A B}  \quad \Longrightarrow \quad   i\left[Q_{g}, Q_{A B}\right]=- \,Q_{A B},
\eqno(C.16)
\]
which demonstrate that the Noether non-nilpotent (anti-)BRST charges $Q_{(a) b}$ and the nilpotent versions of the (anti-)BRST charges $Q_{(A) B}$ obey the 
{\it same} kinds of algebras with the conserved ghost charge $Q_{g}$. However, we know that the off-shell 
nilpotency property of the (anti-)BRST charges is very important from the point of view of (i) the BRST cohomology (see, e.g. [42]), 
and (ii) the physicality criteria and their consistency with the Dirac quantization conditions for the systems with constraints [33]. 
Thus, the standard BRST algebra is obeyed amongst the nilpotent  (i.e. $Q_{(A) B}^{2}=0$) 
versions of the (anti-)BRST charges $Q_{(A) B}$ and the conserved ghost charge $Q_{g}$. This well-known algebra is:
\[
Q_{B}^{2}=0, \qquad Q_{A B}^{2}=0, \qquad i\left[Q_{g}, Q_{B}\right]= +\, Q_{B}, \qquad i\left[Q_{g}, Q_{A B}\right]= - \,Q_{A B}.
\eqno(C.17)
\]
The above algebra encodes the fact that the ghost numbers of the (anti-)BRST charges are $(-1)+1$, respectively. In other words, the BRST transformation raises
 the ghost number of a field by one [cf. Eq. (4)]. On the other hand, the ghost number is lowered
 by one  for a field which is acted upon by the anti-BRST transformation operator [cf. Eq. (6)].\\


\vskip 1.0cm


\begin{thebibliography}{99}
\bibitem{RPM1}   M. B. Green, J. H. Schwarz, E. Witten, Superstring Theory \\(Cambridge University Press, Cambridge, 1987) 
\bibitem{RPM2}   J. Polchinski, String Theory (Cambridge University Press, Cambridge, 1998)    
\bibitem{RPM3}   D. Lust, S. Theisen, Lectures in String Theory (Springer-Verlag, New York, 1989)    
\bibitem{RPM4}   K. Becker, M. Becker, J.H. Schwarz, String Theory and M-Theory \\(Cambridge University Press, Cambridge, 2007)   
\bibitem{RPM5}   D. Rickles, A Brief History of String Theory From Dual Models to M-Theory \\(Springer, Germany, 2014)
\bibitem{RPM6}   E. Witten, Topological quantum field theory. Commun. Math. Phys. 117, 353 (1988)    
\bibitem{RPM7}   A. S. Schwarz, On quantum fluctuations of instantons. Lett. Math. Phys. 2, 217 (1978)
\bibitem{RPM8}   D. Birmingham, Matthias Blau, Mark Rakowski, George Thompson, Topological field theory. Physics Reports 209, 129 (1991)
\bibitem{RPM9}   R. P. Malik, New topological field theories in two dimensions. \\J. Phys. A Math. Gen. 34, 4167 (2001)
\bibitem{Hari10} M. A. Vasiliev, Higher spin gauge theories in any dimension. \\Comptes Rendus Physique 5, 1101 (2004)
\bibitem{Hari11} C. Sleight, M. Taronna, Higher-spin gauge theories and bulk locality.\\ Phys. Rev. Lett. 121, 171604 (2018)
\bibitem{Hari12} Eric D'Hoker, D. H. Phong, {\it Lectures on Supersymmetric Yang-Mills Theory and Integrable Systems}, arXiv: 9912271 [hep-th]
\bibitem{Hari13} Lars Brink, John H. Schwarz, J. Scherk, Supersymmetric Yang-Mills theories. \\Nucl. Phys. B 121, 77 (1977)
\bibitem{Hari14} C. Becchi, A. Rouet, R. Stora, The Abelian Higgs-Kibble model: unitarity of the S-operator. Phys. Lett. B 52, 344 (1974)
\bibitem{SKP15}  C. Becchi, A. Rouet, R. Stora, Renormalization of the Abelian Higgs-Kibble model. Comm. Math. Phys. 42, 127 (1975)
\bibitem{SKP16}  C. Becchi, A. Rouet, R. Stora, Renormalization of gauge theories. \\ Annals of  Physics
 (N. Y.) 98, 287 (1976)
\bibitem{SKP17}  I. V. Tyutin, Gauge invariance in field theory and statistical physics in operator formalism, in Lebedev Institute Preprint, 
                 Report Number: FIAN-39 (1975) (unpublished), arXiv: 0812.0580 [hep-th]
\bibitem{SKP18} T. Eguchi, P. B. Gilkey, A. Hanson, Gravitation, gauge theories and differential geometry. Phys. Rep. 66, 213 (1980)
\bibitem{SKP19} S. Mukhi, N. Mukunda, Introduction to Topology Differential Geometry and Group Theory for Physicists 
                (Wiley Eastern Private Limited, New Delhi, 1990)
\bibitem{SKP20} M. G{\" o}ckeler, T. Sch{\" u}cker, Differential Geometry, Gauge Theories and Gravity\\ (Cambridge University Press, Cambridge, 1987)
\bibitem{SKP21} J. W. van Holten, The BRST complex and the cohomology of compact Lie algebras.
                Phys. Rev. Lett. 64, 2863 (1990) 
\bibitem{SKP22} K. Nishijima, The Casimir operator in the representations of BRS algebra.\\ Prog. Theor. Phys. 80, 897 (1988) 
\bibitem{RPM23} Joel W. Robbin, Dietmar A. Salamon, Introduction to Differential Geometry \\(Springer Spektrum Berlin, Heidelberg, 2022)
\bibitem{RPM24} R. Kumar, R. P. Malik,  A 3D field-theoretic model: discrete duality symmetry.\\ Annals of Physics 481, 170188 (2025)
\bibitem{RPM25} R. Kumar, R. P. Malik, Symmetries of a 3D field-theoretic model, \\arXiv: 2412.10852  [hep-th]
\bibitem{RPM26}  Saurabh Gupta, R. P. Malik, Rigid rotor as a toy model for Hodge theory.\\ Eur. Phys. J. C 68, 325 (2010)
\bibitem{RPM27}  Shri Krishna, R. P. Malik, A quantum mechanical example for Hodge theory.\\
                 Annals of Physics 464, 169657 (2024)
\bibitem{RPM28}  S. Krishna, R. P. Malik, A free ${\mathcal N} = 2$  supersymmetric system: novel symmetries.\\
                 Euro. Phys. Lett. (EPL) 109, 31001  (2015)                
\bibitem{RPM29} S. Krishna, A. Shukla, R. P. Malik, General ${\mathcal N} = 2$  supersymmetric quantum mechanical model: supervariable approach to 
                its off-shell nilpotent symmetries. \\Annals of Physics. 351,  558 (2014)
\bibitem{RPM30} S. Krishna, R. P. Malik, ${\mathcal N} = 2$ SUSY symmetries for a moving charged particle
                under influence of a magnetic field: supervariable approach.\\ Annals of Physics 355, 204 (2015)                
\bibitem{RPM31} R. P. Malik, Continuous and discrete symmetries in a  4D field-theoretic system:  symmetry operators and their algebraic structures.\\
                Euro. Phys. Lett. (EPL) 151, 12004 (2025)  
\bibitem{Hari32} P. A. M. Dirac, Lectures on Quantum Mechanics, Belfer Graduate School of Science (Yeshiva University Press, New York, 1964)
\bibitem{Hari33} K. Sundermeyer, Constraint Dynamics, Lecture Notes in Physics \\(Springer-Verlag, Berlin, 1982)
\bibitem{SKP34} E. C. G. Sudarshan, N. Mukunda, Classical Dynamics: A Modern Perspective \\(Wiley, New York, 1972)
\bibitem{SKP35} D. M. Gitman, I. V. Tyutin, Quantization of Fields with Constraints \\(Springer-Verlag, Berlin, Heidelberg, 1990)
\bibitem{skp36} H. J. Rothe. K. D. Rothe, Classical and Quantum Dynamics of Constrained Hamiltonian Systems: World Scientific Lecture Notes in Physics, Vol. 81
                (World Scientific, Singapore, 2010)  
\bibitem{RPM37} A. K. Rao, A. Tripathi, B. Chauhan, R. P. Malik, Noether theorem and nilpotency property of the (anti-)BRST charges in 
                the BRST formalism: a brief review.\\ Universe 8, 566 (2022)
\bibitem{RPM38}     R. P. Malik,  Abelian 3-Form Gauge Theory: Superfield Approach, \\
                    Physics of Particles and Nuclei {\bf 43}, 669 (2012)                
\bibitem{RPM39}      R. P. Malik, Abelian 2-Form Gauge Theory: Superfield Approach, \\ Eur. Phys. J. C {\bf 60}, 457 (2009) (see Appendix)
\bibitem{Hari40} L. Bonora, R. P. Malik, BRST, anti-BRST and gerbes. Phys. Lett. B 655, 75 (2007)
\bibitem{Hari41} L. Bonora, R. P. Malik, BRST, anti-BRST and their geometry. \\J. Phys. A: Math. Theor. 43, 375403 (2010)
\bibitem{SKP42} S. Weinberg, The Quantum Theory of Fields: Modern Applications, Vol. 2 (Cambridge University Press, Cambridge, 1996)

\bibitem{RPM43} M. Henneaux, C. Teitelboim, Quantization of Gauge Systems \\(Princeton University, New Jersey, 1992)
\bibitem{RPM44} N. Nakanishi, I. Ojima, Covariant Operator Formalism of Gauge Theories and Quantum Gravity (World Scientific, Singapore, 1996)
\bibitem{RPM45} K. Nishijima, B. R. S. invariance, asymptotic freedom and color confinement. Czechoslov. J. Phys. 46, 140 (1996)
\bibitem{Hari46} T. Kugo, I. Ojima, Local covariant operator formalism of non-Abelian gauge theories and quark confinement problem.
                 Prog. Theo. Phys. (Suppl) 66, 1 (1979) 
\bibitem{Hari47} S. Krishna, R. Kumar, R. P. Malik, A massive field-theoretic model for Hodge theory. Annals of  Physics 414, 168087 (2020)
\bibitem{SKP48}  B. Chauhan, S. Kumar, A. Tripathi, R. P. Malik, Modified 2D Proca theory: revisited under BRST and (anti-)chiral superfield formalisms. \\
                Advances in High Energy Physics 2020, 3495168 (2020)
\bibitem{Hari49} Saurabh Gupta, R. P. Malik, A field-theoretic model for Hodge theory. \\Eur. Phys. J. C 58, 517 (2008)                
\bibitem{Hari50} R. Kumar, S. Krishna, A. Shukla, R. P. Malik, Abelian $p$-form ($p = 1, 2, 3$) gauge theories as the field theoretic models for the Hodge theory.\\
                 Int. J. Mod. Phys. A 29, 1450135 (2014)
\bibitem{Hari51} E. Harikumar, R. P. Malik, M. Sivakumar, Hodge decomposition theorem for Abelian two-form gauge theory.
                 J. Phys. A: Math. Gen. 33,  7149 (2000)
\bibitem{RPM52} E. Harikumar, R. P. Malik, Pseudo-scalar field as a possible candidate for phantom field.
                 Advances in High Energy Physics 2025, 6687988 (2025) 
\bibitem{Hari53}  S. K. Panja, E. Harikumar, R. P. Malik, Modified 3D massive Abelian 2-from theory with a single 
                  pseudo-scalar field as a phantom field: BRST approach. \\Advances in High Energy Physics 2025, 7924160  (2025)
\bibitem{RPM54} P. J. Steinhardt, N. Turok, A cyclic model of the universe. Science 296, 1436 (2002)
\bibitem{rpm55}  J. L. Lehners, Ekpyrotic and cyclic cosmology. Phys. Rep. 465, 223 (2008)                  
\bibitem{rpm56}   S. Alexander, S. Cormack, M. Gleiser, A cyclic universe approach to fine tuning.\\ Phys. Lett. B 757, 147 (2016) 
\bibitem{RPM57} Y. F. Cai, A. Marcian, D.-G. Wang, E. Wilson-Ewing, Bouncing cosmologies with dark matter and dark energy. Universe 3, 1 (2017)
\bibitem{rpm58}   M. Novello, S.E.P. Bergliaffa, Bouncing cosmologies. Phys. Rep. 463, 127 (2008) 
\bibitem{RPM59} K. Koyama, Ghost in self-accelerating universe.\\ Classical and  Quantum Gravity 24, R231 (2007)
\bibitem{rpm60} V. M. Zhuravlev, D. A. Kornilov, E. P. Savelova, The scalar fields with negative kinetic energy, dark matter and dark energy.
                Gen. Relat. Gravity 36, 1736 (2004) 
\bibitem{Hari61} Y. Aharonov, S. Popescu, D. Rohrlich, L. Vaidman, Measurements, errors, and negative kinetic energy. Phys. Rev. A 48, 4084 (1993)
\bibitem{HARI62}  B. P. Schmidt,  Nobel Lecture: Accelerating expansion of the Universe through observations of distant supernovae.
                  Rev. Mod. Phys. 84, 1151 (2012)                     
\bibitem{HARI63}  P. Astier, R. Pain, Observational evidence of the accelerated expansion of the Universe,
                  arXiv: 1204.5493 [astro-ph.CO] 
\bibitem{HARI64}  Y. Gong, A. Wang, Energy conditions and current acceleration of the Universe.\\ Phys. Lett. B 652, 63 (2007) 
\bibitem{HARI65}  I. A. Batalin, G. A. Vilkovisky, Gauge algebra and quantization.\\ Phys. Lett. B 102, 27 (1981)
\bibitem{HARI66}  I. A. Batalin, G. A. Vilkovisky, Quantization of gauge theories with linearly dependent generators. Phys. Rev. D 28, 2567 (1983)
\bibitem{HARI67}  I. A. Batalin, G. A. Vilkovisky, Closure of the gauge algebra, generalized Lie equations and Feynman rules. Nucl. Phys. B 234, 106 (1984)
\bibitem{HARI68}  L. Baulieu, E. Bergshoeff, E. Sezgin., Open BRST algebras, ghost unification and string field theory.  Nucl. Phys. B 307, 348 (1988)
\bibitem{HARI69}  L. Baulieu, B-V quantization and field-anti-field duality for p-form gauge fields, topological field theories and 2-D gravity. 
                  Nucl. Phys. B 478, 431 (1996)

\bibitem{HARI70}  L. Ciambelli, R. G. Leigh, Lie algebroids and the geometry of off-shell BRST. \\Nucl. Phys. B 972, 115553 (2021)
\bibitem{HARI71}  W. Jia, M. S. Klinger, R. G. Leigh, BRST cohomology is Lie algebroid cohomology. Nucl. Phys. B 994, 116317 (2023) 
\bibitem{SKP72} P. Mitra, R. Rajaraman, Gauge-invariant reformulation of theories with second-class constraints. Annals of  Physics 203, 157 (1990)                 
\bibitem{SKP73} P. Mitra, R. Rajaraman, New results on systems with second-class constraints. \\ Annals of  Physics 203, 137 (1990)                 
\bibitem{Har74}  G. Curci, R. Ferrari, Slavnov transformations and supersummetry. \\Phys. Lett. B 63, 91 (1976)                 
                 
                 
                 
                 
                 
              



\end{thebibliography}
\end{document}